\PassOptionsToPackage{margin=1in}{geometry}
\documentclass[fleqn,usenatbib]{mnras}
\usepackage[utf8]{inputenc} 
\usepackage{lmodern,microtype}
\usepackage{multirow}
\usepackage{xcolor}
\usepackage[table]{xcolor}
\usepackage{newtxtext,newtxmath}
\usepackage{siunitx}

\usepackage[T1]{fontenc}
\usepackage[abs]{overpic}  
\usepackage{makecell}
\DeclareRobustCommand{\VAN}[3]{#2}
\let\VANthebibliography\thebibliography
\def\thebibliography{\DeclareRobustCommand{\VAN}[3]{##3}\VANthebibliography}

\usepackage{graphicx}	
\usepackage{subcaption}
\usepackage{amsmath}	
\usepackage{fix-cm}
\usepackage{booktabs}
\usepackage{graphicx}
\usepackage{tikz}
\usepackage{changepage}
\usepackage{orcidlink}
\usetikzlibrary{calc}
\usepackage{xcolor}

\definecolor{orcidlogocol}{HTML}{A6CE39}

\title[Multi--band Reconstruction of 16 PISCO Lens Systems]{{Multi--band Reconstruction of Sixteen Gravitational Lens Systems using PISCO data}}

\author[H. Qu et al.]{Huimin Qu\,\orcidlink{0009-0006-0299-0265}$^{1}$\thanks{E-mail: \href{mailto:huimin.qu@sydney.edu.au}{huimin.qu@sydney.edu.au}},
Daniel J. Ballard\,\orcidlink{0009-0003-3198-7151}$^{1}$,
Geraint F. Lewis\,\orcidlink{0000-0003-3081-9319}$^{1}$,
Karl Glazebrook\,\orcidlink{0000-0002-3254-9044}$^{2,3}$,
Antony Stark\,\orcidlink{0000-0002-2718-9996}$^{4}$, \and
Sarah M. Sweet\,\orcidlink{0000-0002-1576-2505}$^{5,3}$,
Colin Jacobs\,\orcidlink{0000-0003-4239-4055}$^{2,3}$,
Kim-Vy Tran\,\orcidlink{0000-0001-9208-2143}$^{4,3}$,
Brian Stalder\,\orcidlink{0000-0003-0973-4900}$^{6}$,
Tania M. Barone\,\orcidlink{0000-0002-2784-564X}$^{2,3}$, \and
Tucker Jones\,\orcidlink{0000-0001-5860-3419}$^{7}$,
Keerthi Vasan G.C.\,\orcidlink{0000-0002-2645-679X}$^{8}$,
Thomas E. Collett\,\orcidlink{0000-0001-5564-3140}$^{9}$,
Glenn G. Kacprzak\,\orcidlink{0000-0003-1362-9302}$^{2,3}$, \and
Dorota Bayer\,\orcidlink{0000-0003-2510-3017}$^{2,3}$\\
\\
$^{1}$Sydney Institute for Astronomy, School of Physics A28, The University of Sydney, NSW 2006, Australia\\
$^{2}$Centre for Astrophysics and Supercomputing, Swinburne University of Technology, PO Box 218, Hawthorn, VIC 3122, Australia\\
$^{3}$The ARC Centre of Excellence for All Sky Astrophysics in 3 Dimensions (ASTRO 3D), Australia\\
$^{4}$Center for Astrophysics, Harvard \& Smithsonian, Cambridge, MA 02138, USA\\
$^{5}$School of Mathematics and Physics, University of Queensland, Brisbane, QLD 4072, Australia\\
$^{6}$Vera C. Rubin Observatory Project Office, 950 N Cherry Ave, Tucson,
AZ, 85719, USA\\
$^{7}$Department of Physics and Astronomy, University of California, Davis, 1 Shields Avenue, Davis, CA 95616, USA\\
$^{8}$The Observatories of the Carnegie Institution for Science, 813 Santa Barbara Street, Pasadena, CA 91101, USA\\
$^{9}$Institute of Cosmology and Gravitation, University of Portsmouth, Burnaby Road, Portsmouth, PO1 3FX, UK\\}

\date{Accepted XXX. Received YYY; in original form ZZZ}

\pubyear{\the\year{}}

\begin{document}
\label{firstpage}
\pagerange{\pageref{firstpage}--\pageref{lastpage}}
\maketitle

\begin{abstract}
Next-generation surveys such as the Euclid survey, the Legacy Survey of Space and Time (LSST), and the China Space Station Telescope (CSST) survey are expected to discover $\sim\!10^5$ galaxy–galaxy scale strong gravitational lenses.
This motivates the development of scalable and robust lens modeling approaches that can efficiently and reliably learn from wide-field survey datasets before high-resolution follow-up. 
We design a scalable, Bayesian, \textsc{Lenstronomy}-based pipeline and apply it to a sample of sixteen lens candidates observed with the Parallel Imager for Southern Cosmology Observations (PISCO) on the Magellan telescope. PISCO provides four-band imaging ($z$, $i$, $r$, $g$) with colours, depth and seeing conditions comparable to LSST.
To fully exploit the constraining power of this dataset, our pipeline performs simultaneous multi--band modeling, using a common mass profile across all four bands while allowing independent light profiles in each. This approach leverages color information to provide joint constraints on the lens mass and yields reduced uncertainties compared to single-band analyses.
Fifteen out of sixteen PISCO lens candidates are successfully recovered with interpretable lensing configurations, including DESJ0533–2536, the first reported hyperbolic-umbilic galaxy–galaxy scale strong lensing candidate.
We further assess how much model complexity can be reliably constrained given the resolution and seeing of PISCO-like data.
Overall, our results demonstrate that scalable, multi--band lens modeling of ground-based data can extract meaningful constraints on mass and source morphology, providing a practical pathway to maximize the scientific return from large samples in upcoming surveys.
\end{abstract}



\section{Introduction}
Gravitational lensing is a phenomenon in which the gravitational field of a massive object, such as a galaxy or galaxy cluster, bends the light from a distant luminous source. In the regime of strong gravitational lensing, where the alignment between the source, lens, and observer is near--perfect, this effect 
produces multiple arc-shaped images or even complete Einstein rings \citep[see e.g.][]{1992grle.book.....S}.

By reconstructing the lensing signatures found in galaxy-scale lenses, one can perform detailed studies into the astrophysics of the lenses themselves -- typically massive early-type galaxies. Lensing measures the combined dark and baryonic content of halos on these mass scales \citep{sahuAGELConflictReal2024, shajibDarkMatterHalos2021, shajibMassiveEllipticalGalaxies2020, treu_initial_2010}, as well as potentially the circumgalactic medium around their line of sight \citep{2018Natur.554..493L, 2021ApJ...914...92M, baroneGravitationalLensingReveals2024}.  On sub-galactic scales, they probe a lesser--explored low mass regime of the dark matter halo mass function through detections and non--detections of substructures, which behave as perturbative lenses close to lensed arcs \citep[see e.g.][]{vegettiDetectionDarkSubstructure2010, vegettiGravitationalDetectionLowmass2012, nierenbergDetectionSubstructureAdaptive2014, hezavehDetectionLensingSubstructure2016, despaliModellingLineofsightContribution2018, gilmanWarmDarkMatter2020}.

Strong lenses also enable magnified observations of the higher-redshift background sources, extending the redshift reach of galaxy evolution studies \citep[see e.g.][]{mcleodNewRedshift92015, zavala_dusty_2018, vasanSpatiallyResolvedGalactic2025, amvrosiadisOnsetBarFormation2025}. Additionally, time--varying components of background sources can be used for time--delay cosmography, which measures the Hubble constant, $H_{0}$ \citep{wong_h0licow_2020, shajib_strides_2020, kelly_constraints_2023, pascale_sn_2024}. And, in more elusive, multiple source plane systems, or with a statistically significant sample of systems with good stellar dynamics data in the lens plane, strong lensing can constrain the dark energy equation of state \citep{sahuCosmographyDoubleSource2025, collettCosmologicalConstraintsDouble2014, caminhaGalaxyClusterStrong2022, liCosmologyLargePopulations2024}. Stellar kinematics in combination with strong lensing have also been used to test the validity general relativity on extragalactic scales \citep{collettPreciseExtragalacticTest2018}.

For decades, however, the scientific potential of strong gravitational lensing has been limited by the relatively small number of known systems. The advent of next-generation surveys such as Euclid, the Legacy Survey of Space and Time \citep[LSST;][]{ivezicLSSTScienceDrivers2019, 2009arXiv0912.0201L}, and the China Space Station Telescope \citep[CSST;][]{csstcollaboration2025introductionchinesespacestation, li2024csststronglensingpreparation, cao2025csststronglensingpreparation} is expected to increase the number of identified lenses by several orders of magnitude \citep{collettPOPULATIONGALAXYGALAXY2015, 2025}. There are therefore two main challenges facing the strong lensing community in the coming years. The first is discovering large quantities of clear lensing signals in these forthcoming vast data sets; the second is obtaining robust lens reconstructions for them all.

Attempts to confront the first of these challenges involve machine learning algorithms trained on simulations of strong lenses. Convolutional neural networks (CNNs) have been used to discover lens candidates in Dark Energy Survey \citep[DES;][]{2005astro.ph.10346T, 2018ApJS..239...18A} data and the Dark Energy Camera Legacy Survey (DECaLS; \citealt{2019AJ....157..168D}) fields \citep{2019ApJS..243...17J, 2019MNRAS.484.5330J}, for example. Building on this work, the ASTRO 3D Galaxy Evolution with Lenses (AGEL) survey \citep{tranAGELSurveyStrong2023, baroneAGELSurveyData2025} is enabling us to build large, diverse samples of confirmed lenses by combining high-resolution imaging with spectroscopic follow-up on these CNN--discovered candidate catalogues. CNNs, however, rely on resemblences to simulated images only, and are blind to real lensing physics. Fitting gravitational lens models to lens candidates therefore helps to remove false positive detections, and is an important stage to implement into the lens discovery process.

To combat the second of these challenges, the lens modelling procedure needs to be scalable to large samples of lensing data from upcoming surveys, as since we ultimately wish to extract science from a statistically significant sample size of new gravitational lenses, individual analyses of all the new data would require an untractable amount of investigator time.

The lens models required to fit the data in an automated manner can be selected from a plethora of physically motivated analytic profiles for mass distributions and source light profiles.
\textsc{Lenstronomy}\footnote{\url{https://lenstronomy.readthedocs.io/en/latest/}} \citep{2018PDU....22..189B, 2021JOSS....6.3283B} offers a Bayesian framework to sample over the parameters of such profiles using Markov Chain Monte Carlo 
(MCMC, e.g. \textsc{emcee} \citealt{2013PASP..125..306F}; \textsc{zeus} \citealt{2021MNRAS.508.3589K}), to obtain physically meaningful posterior distributions that faithfully reproduce the observed data. 
For cases where the data demands more complexity to reproduce the source morphology, the flexibility afforded by a basis set source composition \citep[shapelets;][]{2003MNRAS.338...35R, 2003MNRAS.338...48R, 2004MNRAS.348..214M, Massey_2005}, the semi--linear inversion approach of \citet{Warren_2003} with an adaptively pixelated source grid as in \textsc{PyAutoLens} \citep{Nightingale_2015, Nightingale_2018, Nightingale_2021}, or forward modelled pixelised sources as in e.g. \cite{enzi2024overconcentrateddarkhalostrong} may be favoured over a simplistic analytic light profile. These approaches are significantly more computationally expensive and may therefore not be ideal for automated analyses. However, machine learning techniques being developed to handle automated lens modelling with complex source reconstruction appear promising \citep[see e.g.][]{karchevStrongLensingSourceReconstruction2022, Gentile_2023, stone_caustics_2024, 2025AJ....169..254A} 

Previous work such as that of \citet{2019MNRAS.483.5649S} and \citet{shajib2025dolphinfullyautomatedforward} used \textsc{Lenstronomy} to attempt to select mass profiles systematically and determine the level of lens model complexity that can feasibly be applied to high resolution Hubble Space Telescope (HST) data, given analytic or shapelet basis set source profiles. However, in the future we may wish to apply automated lens modelling pipelines immediately after data has been taken, and it is currently unclear how much constraining power data like e.g. LSST has \textit{before} high--resolution follow--up.

In this work, we construct a \textsc{Lenstronomy} pipeline to fit gravitational lens models to ground--based data, which we apply to sixteen lens candidates observed with the Parallel Imager for Southern Cosmology Observations \citep[PISCO;][]{2014SPIE.9147E..3YS} on the \SI{6.5}{m} Magellan telescope, obtained by the AGEL collaboration. This data serves as an effective analogy to future LSST data, as it is ground--based and observes in similar, multiple colours.

Several studies have previously considered modelling lensed sources across multiple colour bands. \citet{Dye2014MNRAS} first extended the semi-linear inversion framework \citep{Warren_2003} to simultaneously reconstruct multiple wavebands under a single lens mass model, in which a best-fit lens light profile is first determined and then subtracted. \citet{lange2025galaxymassmodellingmultiwavelength} performed independent lens model fits across three JWST filters to study dark matter substructure, while \citet{Nightingale_2025} modelled four JWST wavebands for lens discovery. 
For strong lenses with data in $\mathcal{O}(1000)$ wavelength channels—such as in an Integral Field Unit (IFU) spectroscopy datacube—wavelength-regularised source reconstruction becomes particularly well-suited \citep{Young_2022}, as it introduces correlations in the source morphology across neighbouring wavelengths. A similar idea is employed by \citet{Galan_2024}, who modelled six JWST/NIRCam filters with a multi-band 3D correlated-field source model. 
These works highlight a growing trend towards exploiting multi-band information to improve the scientific reach of strong-lens modelling, which is expected to become increasingly ubiquitous in the near future. In this study, we perform simultaneous multi-band modeling of all four PISCO bands, fitting a single consistent lens mass distribution while allowing the lens and source light profiles to vary independently in each band. Our pipeline is designed to be insensitive to band-specific differences in coordinate grids or point spread functions (PSFs), meaning that it could also be scaled to data from multiple instruments simultaneously.

Based on the above design, our framework enables comprehensive multi-band modelling in strong lens surveys such as SLACS\citep{2008ApJ...682..964B} and BELLS \citep{BELLS2012ApJ...744...41B}, where rich multi-wavelength datasets have been available for years but rarely modeled jointly, likely due to substantial computational cost.
Using this framework, we assess whether information across colours can be effectively combined to mitigate the limitations imposed by poor angular resolution and seeing, and whether the benefits justify the additional computational cost. By determining how well the mass profiles of $z\approx0.6-0.9$ deflectors and their potential associated structural complexities can be constrained in this PISCO sample, we further indicate how much science can be immediately extracted from lenses in survey data. Additionally, by measuring how feasibly we can apply well--established lensing mass models to this sample, our pipeline helps to ascertain whether these systems, originally detected by CNN in DES data, are in fact plausible strong lenses or not.

The structure of the paper is as follows: Section~\ref{sec:PISCO data} details the data preparation process. In Section~\ref{sec:Pipeline}, we develop a comprehensive pipeline for handling and reconstructing the multi--band data, where the astronomical shift and rotation effects are considered and modeled. Section~\ref{sec:results} presents the key results obtained from our reconstructions. In Section~\ref{sec:discussion}, we discuss the further the insights gained from modelling this dataset using our pipeline, before summarising in Section~\ref{sec:conclusion}.

\section{PISCO data preparation}
\label{sec:PISCO data} 
\begin{figure}
	\includegraphics[width=\columnwidth]{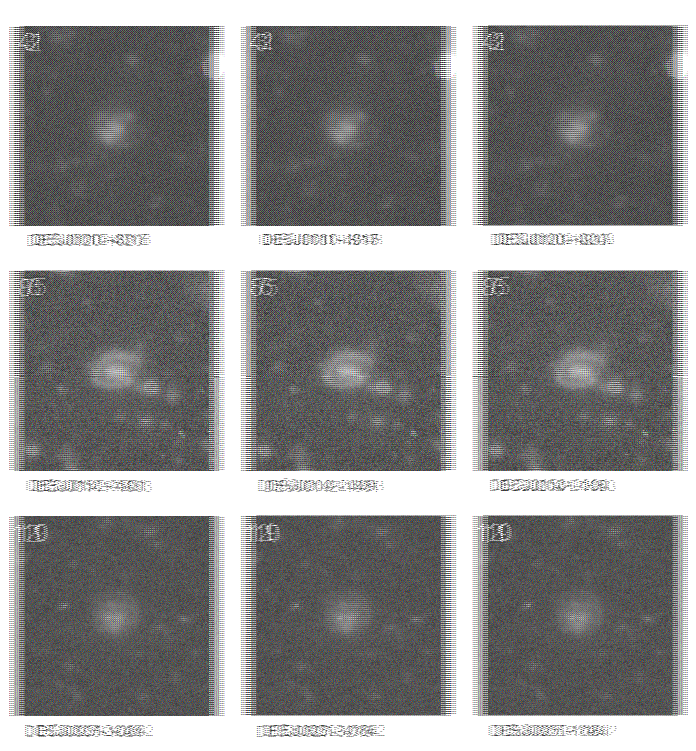}
    \caption{Dark Energy Survey (DES) data of the 16 lens candidates identified  using a Convolutional Neural Network (CNN)
    \citep{2019ApJS..243...17J, 2019MNRAS.484.5330J}. In most cases, the central lens galaxies appear red or orange, consistent with their quenched nature, while the lensed background sources are blue, reflecting ongoing star formation.}
    \label{fig:DES_image}
\end{figure}

Images of the 16 lens candidates were obtained using PISCO \citep{2014SPIE.9147E..3YS} on the \SI{6.5}{m} Magellan telescope. These candidates were originally identified in DES imaging using convolutional neural networks \citep{2019ApJS..243...17J, 2019MNRAS.484.5330J}; the DES imaging of our candidates is presented in Fig.~\ref{fig:DES_image}. 

PISCO employs a dichroic cube beam splitter to split the light into four optical channels, with the resulting bands (\SIrange{400}{550}{nm}, \SIrange{550}{700}{nm}, \SIrange{700}{850}{nm}, and \SIrange{820}{920}{nm}) defined after passing through "cleanup" glass filters placed before the CCDs, over a 9-arcminute field of view, enabling efficient multi--band follow-up. 
These bands correspond to the LSST/Sloan Digital Sky Survey \citep[SDSS;][]{1998AJ....116.3040G} filter set $g$, $r$, $i$, and the $z$ (`z-short' with a cutoff at 920~nm, similar to the LSST $z$ filter but different from the SDSS $z$ filter).
Unlike DES, DECaLS, LSST, or SDSS, where different filters are obtained sequentially under varying conditions, PISCO provides truly simultaneous multi--band imaging under similar observing conditions.
This configuration is well-suited to confirm strong lenses at follow-up depths deeper than survey discovery images, and guided the construction of our observed sample, which prioritizes higher-mass, higher-redshift systems, diverse environments, and exotic sources.

In addition, for a subset of the systems, the AGEL survey provides spectroscopic redshift measurements—primarily obtained with Keck/ESI and VLT/X-shooter for deflectors, and Keck/NIRES and VLT/X-shooter for sources—and high-resolution HST imaging, enabling a diverse range of scientific analyses.

\section{Data Handling and Reconstruction Pipeline}
\label{sec:Pipeline}

In this section, we establish the pipeline to handle and reconstruct the 16 PISCO lens systems. This involves extracting them from four--band PISCO data, obtaining data specifics such as background noise and full width at half maximum (FWHM) of the PSF, choosing the parametric models, and performing MCMC sampling whilst taking into account possible misalignments between the imaging data across the different colour bands.

\subsection{Multi--Band Extraction of Lens Systems}
\label{sec:Extraction} 

We utilize the computer vision tool \textsc{CV2}\footnote{\url{https://pypi.org/project/opencv-python/}}
to locate lens systems from multi--band data.
Each system was first visually identified in one of the four PISCO bands, using the DES data (Fig.~\ref{fig:DES_image}) as a reference, and this band served as a template for normalized cross-correlation to match regions across the PISCO dataset.
Background subtraction is then performed with \textsc{photutils}\footnote{\url{https://photutils.readthedocs.io/}} (v1.8.0) to remove sky light, yielding clean multi--band cutouts for subsequent steps.

\subsection{Data specifics}
\label{sec:data_specifics}

To enable realistic modeling, we extract key data specifics from the PISCO datasets, including the pixel scale, image noise characteristics, and point spread function (PSF) properties. The pixel scale of 0.218~arcsec per pixel is used to configure the coordinate grid for each system.

The observational noise in each pixel has two main contributions: Poisson noise from photon counts and Gaussian background noise. For pixel $i$, the Poisson noise is $\sigma_{\text{poisson}, i} = \sqrt{d_i / t_{\text{exp}}}$,
where $d_i$ is the pixel data value (counts per second) and $t_{\text{exp}} = 200~\si{\second}$ is the uniform exposure time for all bands. 
The background noise $\sigma_\text{bkgd} \sim \mathcal{N}(0, \sigma_\text{rms})$, with $\sigma_{\text{rms}}$ the root-mean-square (RMS) of the background and thus $\sigma_\text{bkgd} = \sigma_\text{rms}$.
The total per-pixel noise is then given by $\sigma_i = \sqrt{\sigma_{\text{poisson}, i}^2 + \sigma_\text{bkgd}^2}$.

We use the \textsc{photutils} package to obtain initial estimates of the $\sigma_{\text{rms}}$ values, configured with a 10$\times$10 pixel box size, a 3$\times$3 median filter, and a sigma-clipping threshold of 3. These estimations are performed on 400$\times$400 pixel, background-subtracted regions—larger than the typical $\sim$50$\times$50 cutout used later in our analyses -- to ensure statistical robustness. The resulting RMS values are further refined by fitting the pixel value distributions of each cutout to a Gaussian profile.

\begin{table*}
\centering
\caption{Summary of model configurations adopted for each system. The mass profile is consistent across all bands, while the light profiles of the deflector(s) and source(s) may either share the same model across bands or adopt different components, with separate parameters for each band. If present, satellite components are modeled as stated in the table.}
\label{tab:model_summary}
\resizebox{\textwidth}{!}{%
\begin{tabular}{cllllc}
\toprule
\textbf{Label} & \textbf{Target Name (model)} & \textbf{Mass Profiles} & \textbf{Light profiles of Deflector(s)} & \textbf{Light profiles of source} & \textbf{Number of parameters}  \\ \hline \midrule 
1 & DESJ0003--3348 (1) & EPL, Shear & \makecell[l]{Main deflector: two Elliptical Sérsic ($z$, $i$, $r$ bands), \\ single Elliptical Sérsic ($g$ band);\\
Satellite: Sérsic} & Elliptical Sérsic  & 99 \\ \midrule

1 & DESJ0003--3348 (2) & EPL, Shear, SIS & \makecell[l]{Main deflector: two Elliptical Sérsic ($z$, $i$, $r$ bands), \\ single Elliptical Sérsic ($g$ band);\\
Satellite: Sérsic}  & Elliptical Sérsic  & 102 \\ \midrule

1 & DESJ0003--3348 (3) & \makecell[l]{EPL, Shear,\\ SIS (location fixed)}  & \makecell[l]{Main deflector: two Elliptical Sérsic ($z$, $i$, $r$ bands), \\ single Elliptical Sérsic ($g$ band);\\
Satellite: Sérsic}  & Elliptical Sérsic &  100 \\ \midrule

2 & DESJ0010--4315 & EPL, Shear & \makecell[l]{Main deflector: Elliptical Sérsic \\ 2 Satellite: 2 Sersic} & Elliptical Sérsic  & 97 \\ \midrule

3 & DESJ0101--4917 & EPL, Shear & Elliptical Sérsic & Elliptical Sérsic  & 65 \\ \midrule

4 & DESJ0120--1820 (1) & EPL, Shear & Elliptical Sérsic & Elliptical Sérsic  & 65\\ \midrule

4 & DESJ0120--1820 (2) & EPL, Shear & two Elliptical Sérsic & Elliptical Sérsic  & 89\\ \midrule

5 & DESJ0141--1303 & EPL, Shear & \makecell[l]{Main deflector: Elliptical Sérsic \\ Satellite: Elliptical Sersic}& Elliptical Sérsic  & 89\\ \midrule

6 & DESJ0142--1831 (1) & EPL, Shear & Elliptical Sérsic & Elliptical Sérsic  & 65 \\ \midrule

6 & DESJ0142--1831 (2) & EPL, Shear & \makecell[l]{Main deflector: 1 Elliptical Sérsic + 1 Sersic,\\ Satellite(s): Elliptical Sersic (r, g bands)} & Elliptical Sérsic  & 93  \\ \midrule

7 & DESJ0150--0304 (1) & \makecell[l]{EPL, Shear,\\ SIS (location fixed)} & \makecell[l]{Deflector inside: Elliptical Sérsic, \\Satellite 1: Elliptical Sérsic, Satellite 2: Gaussian} & \makecell[l]{Elliptical Sérsic ($z$, $i$ bands), \\ Shapelets ($r$, $g$ bands)}  & 99 \\ \midrule

7 & DESJ0150--0304 (2) & \makecell[l]{EPL, Shear,\\ SIS (location fixed)} & \makecell[l]{Deflector inside: Elliptical Sérsic, \\Satellite 1: Elliptical Sérsic, Satellite 2: Gaussian} & Elliptical Sérsic  & 105 \\ \midrule

8 & DESJ0202--2445 (1) & EPL, Shear & \makecell[l]{Main deflector: two Elliptical Sérsic ($z$, $i$, $r$ bands), \\single Elliptical Sérsic ($g$ band)} & Elliptical Sérsic  & 83 \\ \midrule

8 & DESJ0202--2445 (2) & EPL, Shear, Flexion & \makecell[l]{Main deflector: two Elliptical Sérsic ($z$, $i$, $r$ bands), \\single Elliptical Sérsic ($g$ band)} & Elliptical Sérsic  & 87 \\ \midrule

9 & DESJ0212--0852 & EPL, Shear & Elliptical Sérsic & Elliptical Sérsic  & 65\\ \midrule

10 & DESJ0250--4104 (1) & EPL, Shear & \makecell[l]{Main deflector: Elliptical Sérsic \\ Satellite: Sersic} & Elliptical Sérsic & 79 \\ \midrule

10 & DESJ0250--4104 (2) & \makecell[l]{EPL, Shear,\\ SIS (location fixed)} & \makecell[l]{Main deflector: Elliptical Sérsic \\ Satellite: Sersic} & Elliptical Sérsic & 80 \\ \midrule

11 & DESJ0305--1024 & EPL, Shear & Elliptical Sérsic & Elliptical Sérsic  & 65 \\ \midrule

12 & DESJ0327--3246 & EPL, Shear & \makecell[l]{Main deflector: Elliptical Sérsic, \\ 2 Satellite: 2 Elliptical Sersic} & Elliptical Sérsic  & 113 \\ \midrule

13 & DESJ0354--1609 & EPL, Shear & Elliptical Sérsic & Elliptical Sérsic  & 65 \\ \midrule

14 & DESJ0533--2536 & EPL, Shear & \makecell[l]{Main deflector: Elliptical Sérsic, \\ Satellite: Elliptical Sersic (r band)}& Elliptical Sérsic  & 71\\ \midrule

15 & DESJ2032--5658 & EPL, Shear & \makecell[l]{Main deflector: Elliptical Sérsic, \\ 2 Satellite: 2 Elliptical Sersic} & Elliptical Sérsic & 113\\ \midrule

16 & DESJ2125--6504 (1) & EPL, Shear & \makecell[l]{Main deflector: Elliptical Sérsic, \\ Satellite: Elliptical Sersic} & Elliptical Sérsic & 89 \\ \midrule 

16 & DESJ2125--6504 (2) & EPL, Shear, SIS & \makecell[l]{Main deflector: Elliptical Sérsic, \\ Satellite: Elliptical Sersic} & Elliptical Sérsic & 92 \\ \midrule 

16 & DESJ2125--6504 (3) & \makecell[l]{EPL, Shear,\\ SIS (location fixed)} & \makecell[l]{Main deflector: Elliptical Sérsic, \\ Satellite: Elliptical Sersic} & Elliptical Sérsic & 84 \\ \midrule 

\bottomrule
\end{tabular}
}
\end{table*}

To model the observational point spread function (PSF), we adopt a Gaussian PSF kernel. We use the \textsc{Source Extractor}\footnote{\url{https://sextractor.readthedocs.io}} package to identify and select star-like objects in a larger field of each image. The FWHM is then estimated by averaging the \texttt{FWHM\_IMAGE} values of these objects. Across the 16 lens systems, typical FWHM values range from 0.6 to 1.1 $\mathrm{arcsec}$, corresponding to approximately 2.8–5.0 pixels.

\subsection{Selecting parametric models}

In this study, we adopt parametric models to describe the three key components of a gravitational lens system: lens mass, source light, and lens light. Parametric modelling is chosen to avoid the computational cost of performing multiple pixel-based source reconstructions, which are unnecessary given the seeing-limited nature of the data and the inability to resolve complex source morphology. To reconstruct the multi--band data, we use a consistent lens mass model across all bands, while allowing distinct source and lens light models for each band, as the lensing geometry remains fixed but the emission profiles can vary significantly. Note that the number of free parameters increases substantially in multi-band models ($\ge 65$ for all four-band PISCO systems in this sample, see Table \ref{tab:model_summary} for details), compared to single-band models in much of the previous literature (typically $N \lesssim 20$). To assure that our pipeline can retrieve an appropriate posterior under this many degrees of freedom, we perform a test on mock data, described in Appendix \ref{app:mock_data_test}.

We adopt a fiducial model consisting of an elliptical power-law (EPL) mass profile with external shear, and elliptical Sérsic profiles to describe the surface brightness of the main deflector and background source in each band, without assuming any correlation between these profiles across the different bands. Small inter-band shifts and rotations are further modelled to account for uncertainties in cutout alignment and minor chromatic-aberration-like effects, using affine transformations applied to the image coordinates.

Most systems are well-described by this fiducial model; however, when significant residuals remain in the lensed arcs or deflector light, we introduce additional complexity, as detailed in our analyses later. The final configurations for all 16 systems are summarized in Table \ref{tab:model_summary}. Details of the parameterised component profiles used to build our fiducial and extended models are described throughout the remainder of this section.

\subsubsection{The EPL Model}
The most significant component to our lensing mass in our fiducial model setup is the Elliptical Power Law (EPL) model \citep{2015A&A...580A..79T}. This describes an ellipsoidal mass profile with a power-law dependence on the radial distance. The dimensionless surface mass density of the EPL model is given by:

\begin{equation}
\kappa(x, y) = \frac{3-\gamma}{2} \left(\frac{\theta_{E}}{\sqrt{q x^2 + y^2/q}} \right)^{\gamma-1}
\end{equation}
where $\theta_{E}$ is the Einstein radius, $\gamma$ is the negative power-law slope, with 
$\gamma=2$ corresponding to an isothermal profile, and $q$ is the axis ratio (minor-to-major) of the mass distribution.

This profile allows for the so--called "bulge–halo conspiracy" to take effect, where the total mass profile of elliptical galaxies follows a power law, despite the baryonic and dark matter components not individually doing so \citep{2004ApJ...611..739T}. This makes it a convenient, albeit not the most physically motivated choice for gravitational lens modeling, since neither the baryonic nor dark components of the lensing mass are explicitly constrained.

\subsubsection{External Shear}

The other component in our fiducial mass model is the external shear, which accounts for a weak lensing signal inflicted by other nearby objects. This is characterized by a distortion of the images without adding mass to the main deflector itself. The lens potential for an external shear is:

\begin{equation}
\psi(x, y) = \frac{\gamma}{2} (x^2 - y^2) \cos(2\phi_\gamma) + \gamma xy \sin(2\phi_\gamma)
\end{equation}
where \( \gamma \) is the shear strength and \( \phi_\gamma \) is its orientation.

We model $(\gamma_{1}, \gamma_{2})$ as free parameters for external shear, where $(\gamma_{1}, \gamma_{2}) = \gamma\times(\sin2\phi_{\gamma}, \cos2\phi_{\gamma})$.

\subsubsection{The SIS Model}
To extend on our fiducial model to account for visible satellite objects that could contribute a strong lensing signal, we use the Singular Isothermal Sphere (SIS) model. This is a special case of the elliptical power-law (EPL) mass profile with $\gamma = 2$, assuming a spherically symmetric mass distribution with a constant velocity dispersion. Its surface mass density is:
\begin{equation}
\kappa(r) = \frac{\theta_E}{2r}
\end{equation}
where \( r \) is the radial distance and \(\theta_E\)
is the Einstein radius. 

\subsubsection{External flexion}
As a higher-order lensing signal beyond external shear, we also test for the presence of measurable external flexion in some of our systems. This signal captures asymmetric skewing and arc-like bending, typically induced by small-scale mass perturbations like dark matter substructures \citep{2005ApJ...619..741G, 2006MNRAS.365..414B, 2008A&A...485..363S}. It introduces third-order derivatives of the lensing potential, modeled in \textsc{Lenstronomy} as:

\begin{equation}
    f(x, y) = \frac{1}{6} \left( g_1 x^3 + 3g_2 x^2 y + 3g_3 x y^2 + g_4 y^3 \right)
\label{eq4}
\end{equation}
where the four flexion terms represents third-order derivatives of the lensing potential along different directions ($g_1 = \frac{\partial^3 \psi}{\partial x^3}$, $g_2 = \frac{\partial^3 \psi}{\partial x^2 \partial y}$, $g_3 = \frac{\partial^3 \psi}{\partial x \partial y^2}$, and $g_4 = \frac{\partial^3 \psi}{\partial y^3}$).

According to \citet{2022iglp.book.....M}, the flexion terms can be associated with the spin-1 flexion \( \mathcal{F} \) and spin-3 flexion \( \mathcal{G} \), as originally defined in \citet{2006MNRAS.365..414B}, with the following components:
\begin{align}
\mathcal{F}_1 &= \frac{1}{2} (g_1 + g_3), \\
\mathcal{F}_2 &= \frac{1}{2} (g_2 + g_4), \\
\mathcal{G}_1 &= \frac{1}{2} (g_1 - 3 g_3), \\
\mathcal{G}_2 &= \frac{1}{2} (g_2 - 3 g_4).
\end{align}

We sample over $g_{1}$, $g_{2}$, $g_{3}$ and $g_{4}$ when modelling external flexion in our analyses.

\subsubsection{Elliptical Sérsic Profile}
Our fiducial model for the surface mass distributions of our lenses and sources is an elliptical Sérsic profile. This describes how the intensity $I$ of a galaxy varies with distance $R$ from its center. Its surface brightness profile is:
\begin{equation}
    I(x, y) = I_e \exp \left[ -b_n \left( \left( \frac{\sqrt{q x^2 + y^2/q}}{R_e} \right)^{1/n} - 1 \right) \right]
\end{equation}
where \( I_e \) is the intensity at the effective radius \( R_e \), \( n \) is the Sérsic index controling the degree of steepness of the profile, and \( b_n \) is a constant given by the gamma function relation $\gamma(2n; b_{n})=\frac{1}{2}\Gamma(2n)$, which we approximate as $b_{n}=2n-1/3$.

\subsection{Multi--Band Lens Model Reconstruction Considering Shift and Rotation Effects}
After selecting our mass and light profiles, we reconstruct our data by simultaneously fitting its four bands ($z$, $i$, $r$, $g$) using the Particle Swarm Optimisation (PSO) algorithm within \textsc{lenstronomy} \citep{2018PDU....22..189B, 2021JOSS....6.3283B}, and Markov Chain Monte Carlo (MCMC) algorithm within \textsc{zeus} \citep{Karamanis_2021}, also wrapped within \textsc{lenstronomy}. 

\subsubsection{Our likelihood function}
For multi--band datasets, it is possible to reconstruct each band independently under an identical model form \citep{10.1093/mnras/stad587}, but the inferred mass parameters are not guaranteed to remain identical when the model is allowed to vary freely. \textit{A priori}, we believe that one mass model should be able to constrain the data with four independent source reconstructions simulateously, since lensing is an achromatic effect. We therefore construct our likelihood function for our fitting procedures to consider the data in all four PISCO bands simultaneously. It is expressed as:
\begin{equation}
    \log p(\mathcal{D}\mid\mathcal{M}) =
    \sum_{b \in \{z, i, r, g\}} \log p(\mathcal{D}_{b}\mid\mathcal{M}_{b}),
\end{equation}
where $\mathcal{D}$ and $\mathcal{M}$ are the true data and reconstructed data from the model, respectively, written specifically for band $b$ as $\mathcal{D}_{b}$ and $\mathcal{M}_{b}$. The likelihood function for each individual band follows:
\begin{equation}
    \log p(\mathcal{D}_{b} \mid \mathcal{M}_{b}) =
    - \sum_{i} \frac{(\mathcal{D}_{b, i}- \mathcal{M}_{b, i})^2}{2\sigma_i^2} + \text{const.}
\end{equation}
where $i$ represents each pixel. The uncertainty of each pixel, $\sigma_i$, is as described in Section \ref{sec:data_specifics}.

\begin{figure*}
\vfill
\centering

\hspace*{-8mm}
\begin{tikzpicture}[
  every node/.style={anchor=south west,inner sep=0pt},
  x=1mm, y=1mm
]

\def\rows{2}
\def\cols{8}
\def\xsep{24}
\def\ysep{32}
\def\scale{0.24}
\def\imgHeight{23.5}

\node[anchor=south west, font=\bfseries\LARGE]
  at (2, \ysep + \imgHeight + 2) {(1) Observed Data:};

\node at (0*\xsep,1*\ysep) {\includegraphics[scale=\scale]{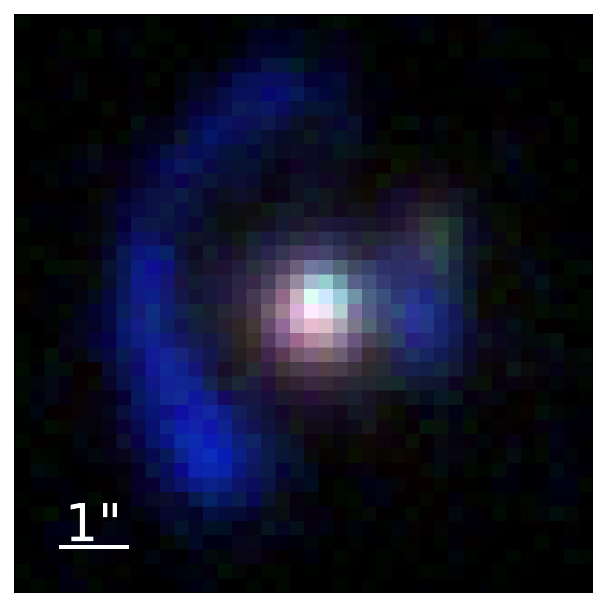}};
\node[anchor=north west, text=white, fill=black, rounded corners=1pt,
      font=\bfseries\footnotesize, inner sep=1pt]
      at (0*\xsep+1.5,1*\ysep+\imgHeight-0.5) {1};
\node[anchor=north, font=\footnotesize]
      at (0*\xsep+\imgHeight/2,1*\ysep-1) {DESJ0003--3348};

\node at (1*\xsep,1*\ysep) {\includegraphics[scale=\scale]{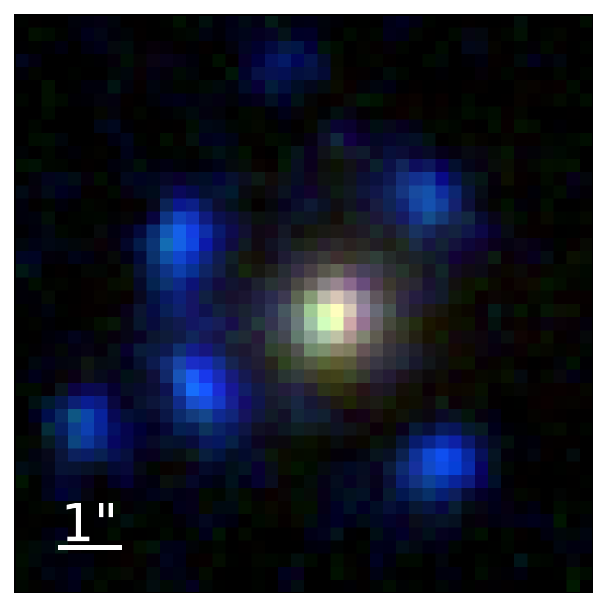}};
\node[anchor=north west, text=white, fill=black, rounded corners=1pt,
      font=\bfseries\footnotesize, inner sep=1pt]
      at (1*\xsep+1.5,1*\ysep+\imgHeight-0.5) {2};
\node[anchor=north, font=\footnotesize]
      at (1*\xsep+\imgHeight/2,1*\ysep-1) {DESJ0010--4315};

\node at (2*\xsep,1*\ysep) {\includegraphics[scale=\scale]{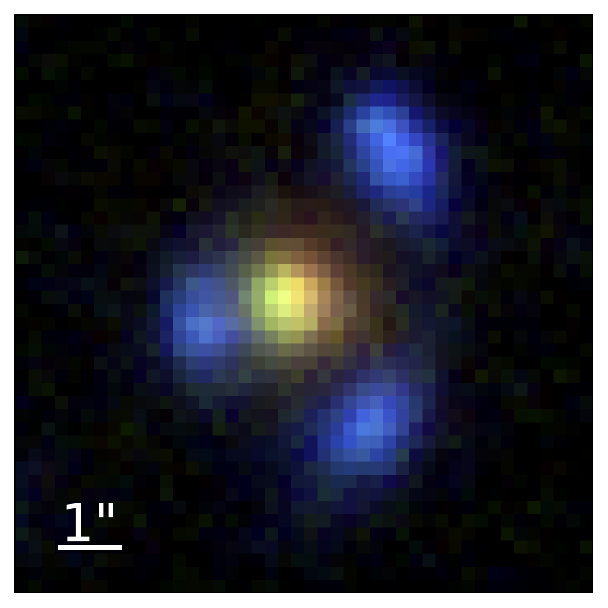}};
\node[anchor=north west, text=white, fill=black, rounded corners=1pt,
      font=\bfseries\footnotesize, inner sep=1pt]
      at (2*\xsep+1.5,1*\ysep+\imgHeight-0.5) {3};
\node[anchor=north, font=\footnotesize]
      at (2*\xsep+\imgHeight/2,1*\ysep-1) {DESJ0101--4917};

\node at (3*\xsep,1*\ysep) {\includegraphics[scale=\scale]{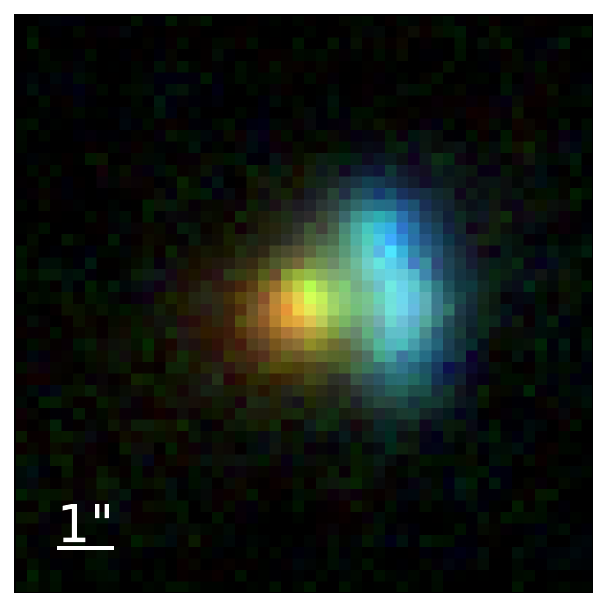}};
\node[anchor=north west, text=white, fill=black, rounded corners=1pt,
      font=\bfseries\footnotesize, inner sep=1pt]
      at (3*\xsep+1.5,1*\ysep+\imgHeight-0.5) {4};
\node[anchor=north, font=\footnotesize]
      at (3*\xsep+\imgHeight/2,1*\ysep-1) {DESJ0120--1820};

\node at (4*\xsep,1*\ysep) {\includegraphics[scale=\scale]{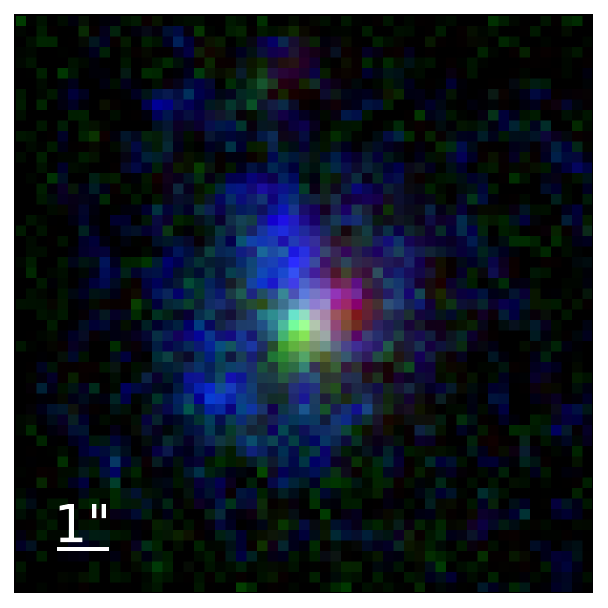}};
\node[anchor=north west, text=white, fill=black, rounded corners=1pt,
      font=\bfseries\footnotesize, inner sep=1pt]
      at (4*\xsep+1.5,1*\ysep+\imgHeight-0.5) {5};
\node[anchor=north, font=\footnotesize]
      at (4*\xsep+\imgHeight/2,1*\ysep-1) {DESJ0141--1303};

\node at (5*\xsep,1*\ysep) {\includegraphics[scale=\scale]{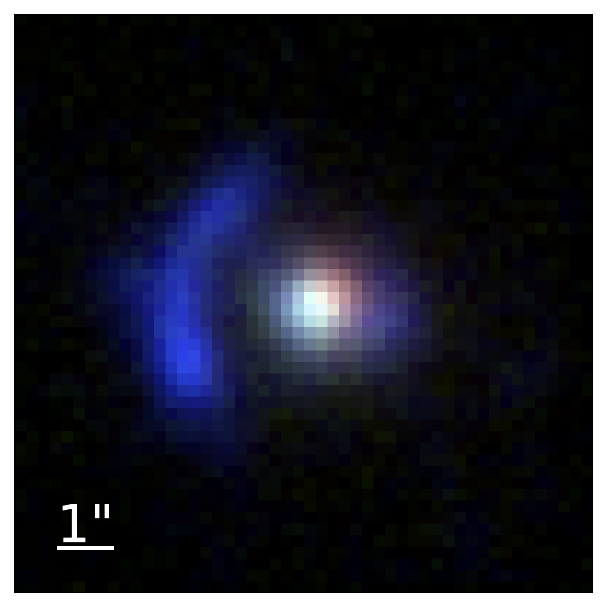}};
\node[anchor=north west, text=white, fill=black, rounded corners=1pt,
      font=\bfseries\footnotesize, inner sep=1pt]
      at (5*\xsep+1.5,1*\ysep+\imgHeight-0.5) {6};
\node[anchor=north, font=\footnotesize]
      at (5*\xsep+\imgHeight/2,1*\ysep-1) {DESJ0142--1831};

\node at (6*\xsep,1*\ysep) {\includegraphics[scale=\scale]{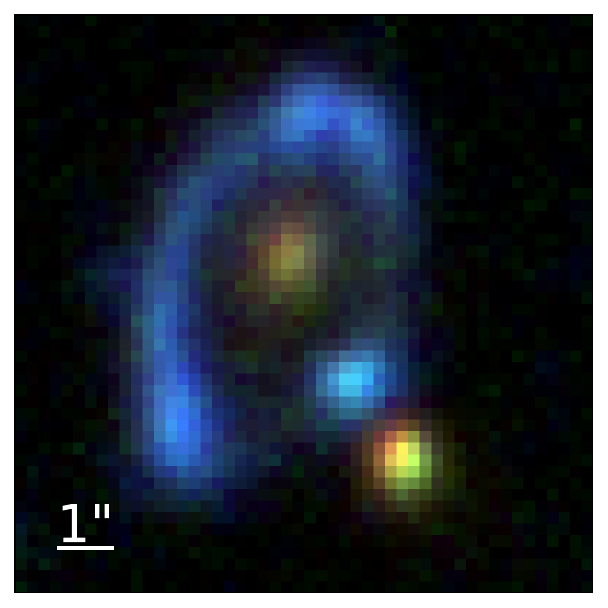}};
\node[anchor=north west, text=white, fill=black, rounded corners=1pt,
      font=\bfseries\footnotesize, inner sep=1pt]
      at (6*\xsep+1.5,1*\ysep+\imgHeight-0.5) {7};
\node[anchor=north, font=\footnotesize]
      at (6*\xsep+\imgHeight/2,1*\ysep-1) {DESJ0150--0304};

\node at (7*\xsep,1*\ysep) {\includegraphics[scale=\scale]{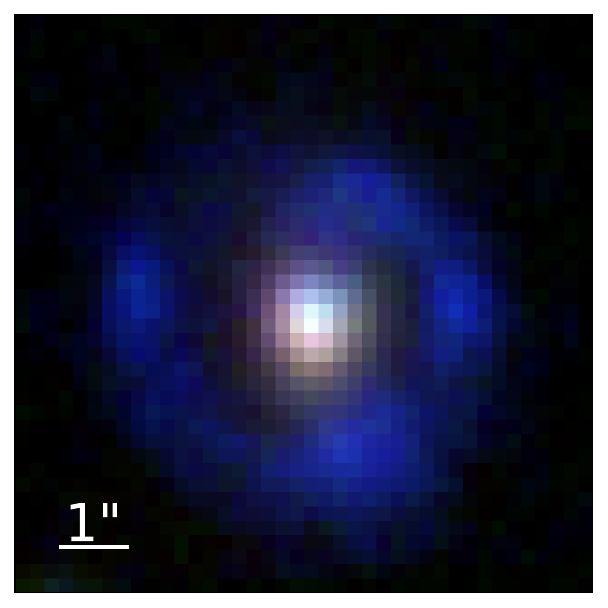}};
\node[anchor=north west, text=white, fill=black, rounded corners=1pt,
      font=\bfseries\footnotesize, inner sep=1pt]
      at (7*\xsep+1.5,1*\ysep+\imgHeight-0.5) {8};
\node[anchor=north, font=\footnotesize]
      at (7*\xsep+\imgHeight/2,1*\ysep-1) {DESJ0202--2445};

\node at (0*\xsep,0*\ysep) {\includegraphics[scale=\scale]{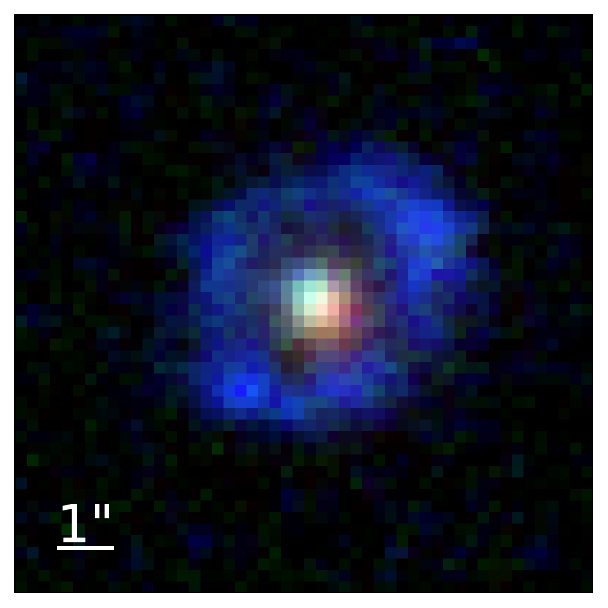}};
\node[anchor=north west, text=white, fill=black, rounded corners=1pt,
      font=\bfseries\footnotesize, inner sep=1pt]
      at (0*\xsep+1.5,0*\ysep+\imgHeight-0.5) {9};
\node[anchor=north, font=\footnotesize]
      at (0*\xsep+\imgHeight/2,0*\ysep-1) {DESJ0212--0852};

\node at (1*\xsep,0*\ysep) {\includegraphics[scale=\scale]{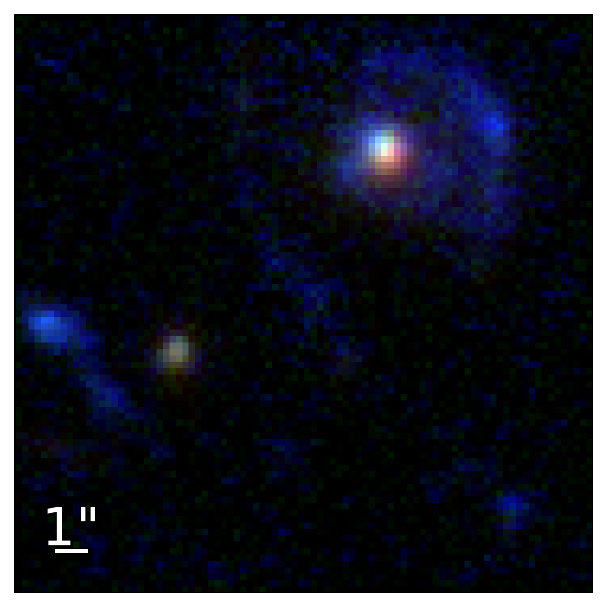}};
\node[anchor=north west, text=white, fill=black, rounded corners=1pt,
      font=\bfseries\footnotesize, inner sep=1pt]
      at (1*\xsep+1.5,0*\ysep+\imgHeight-0.5) {10};
\node[anchor=north, font=\footnotesize]
      at (1*\xsep+\imgHeight/2,0*\ysep-1) {DESJ0250--4104};

\node at (2*\xsep,0*\ysep) {\includegraphics[scale=\scale]{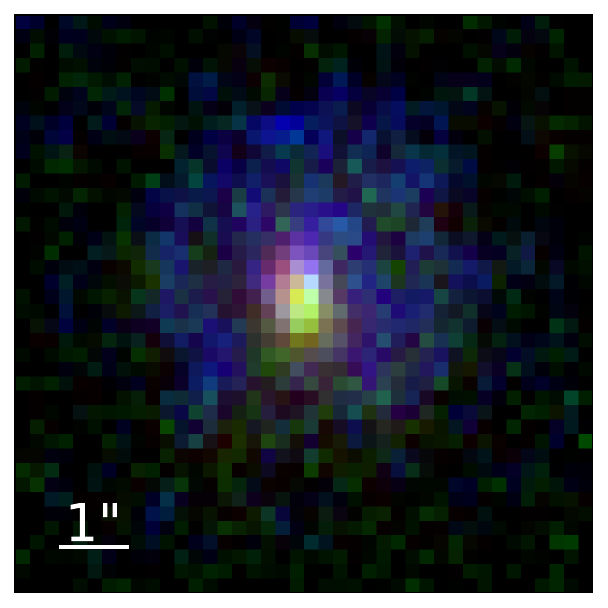}};
\node[anchor=north west, text=white, fill=black, rounded corners=1pt,
      font=\bfseries\footnotesize, inner sep=1pt]
      at (2*\xsep+1.5,0*\ysep+\imgHeight-0.5) {11};
\node[anchor=north, font=\footnotesize]
      at (2*\xsep+\imgHeight/2,0*\ysep-1) {DESJ0305--1024};

\node at (3*\xsep,0*\ysep) {\includegraphics[scale=\scale]{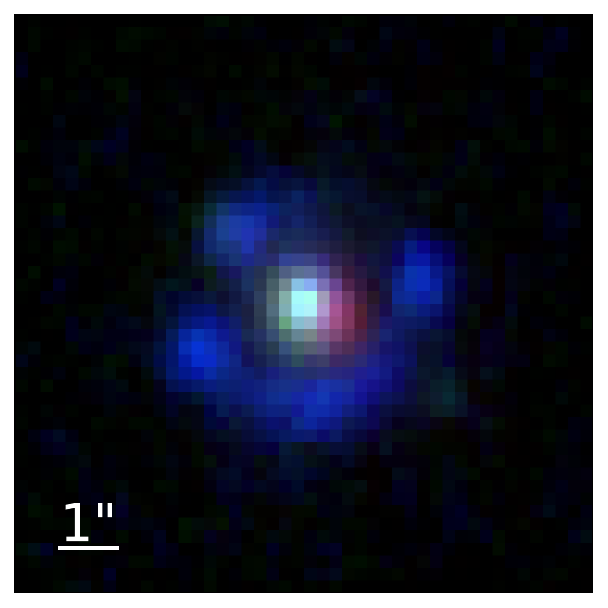}};
\node[anchor=north west, text=white, fill=black, rounded corners=1pt,
      font=\bfseries\footnotesize, inner sep=1pt]
      at (3*\xsep+1.5,0*\ysep+\imgHeight-0.5) {12};
\node[anchor=north, font=\footnotesize]
      at (3*\xsep+\imgHeight/2,0*\ysep-1) {DESJ0327--3246};

\node at (4*\xsep,0*\ysep) {\includegraphics[scale=\scale]{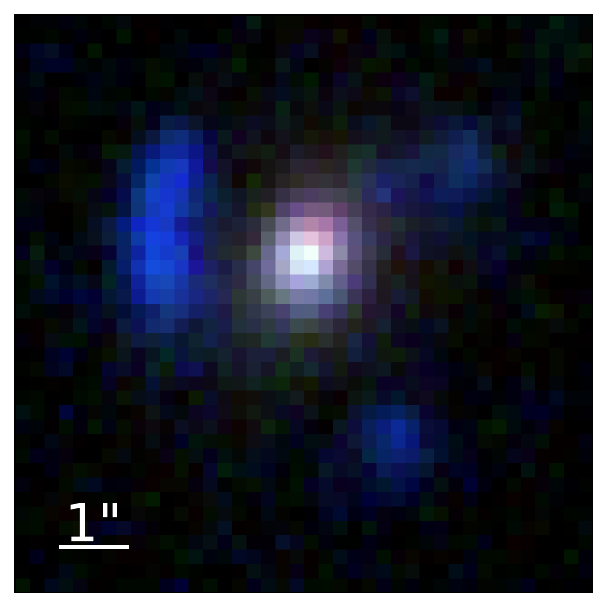}};
\node[anchor=north west, text=white, fill=black, rounded corners=1pt,
      font=\bfseries\footnotesize, inner sep=1pt]
      at (4*\xsep+1.5,0*\ysep+\imgHeight-0.5) {13};
\node[anchor=north, font=\footnotesize]
      at (4*\xsep+\imgHeight/2,0*\ysep-1) {DESJ0354--1609};

\node at (5*\xsep,0*\ysep) {\includegraphics[scale=\scale]{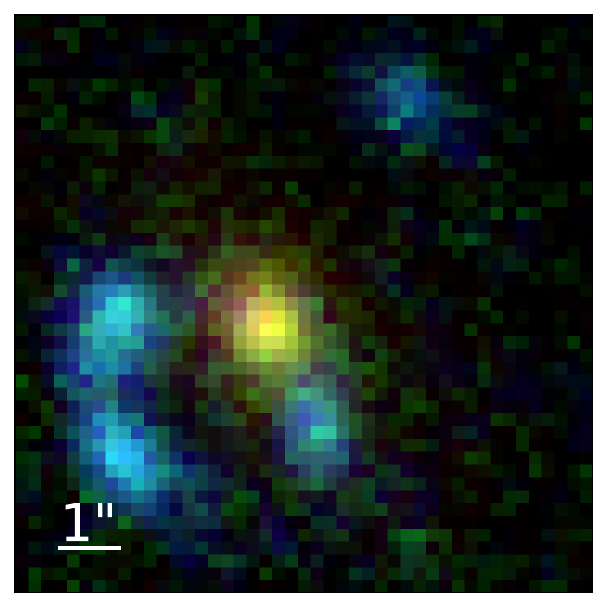}};
\node[anchor=north west, text=white, fill=black, rounded corners=1pt,
      font=\bfseries\footnotesize, inner sep=1pt]
      at (5*\xsep+1.5,0*\ysep+\imgHeight-0.5) {14};
\node[anchor=north, font=\footnotesize]
      at (5*\xsep+\imgHeight/2,0*\ysep-1) {DESJ0533--2536};

\node at (6*\xsep,0*\ysep) {\includegraphics[scale=\scale]{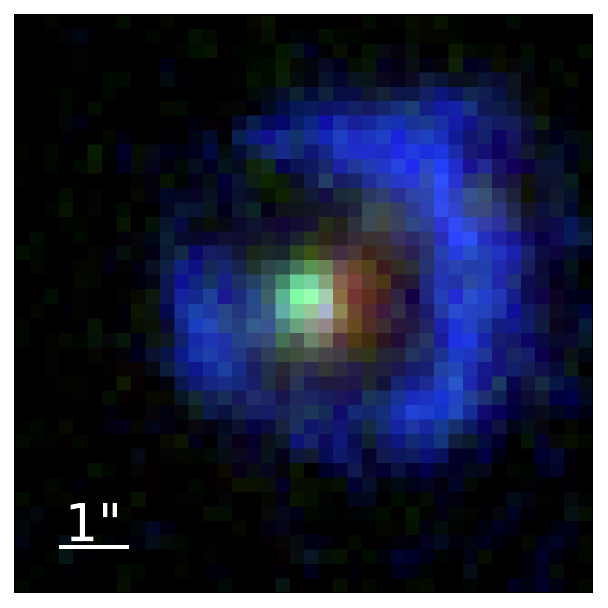}};
\node[anchor=north west, text=white, fill=black, rounded corners=1pt,
      font=\bfseries\footnotesize, inner sep=1pt]
      at (6*\xsep+1.5,0*\ysep+\imgHeight-0.5) {15};
\node[anchor=north, font=\footnotesize]
      at (6*\xsep+\imgHeight/2,0*\ysep-1) {DESJ2032--5658};

\node at (7*\xsep,0*\ysep) {\includegraphics[scale=\scale]{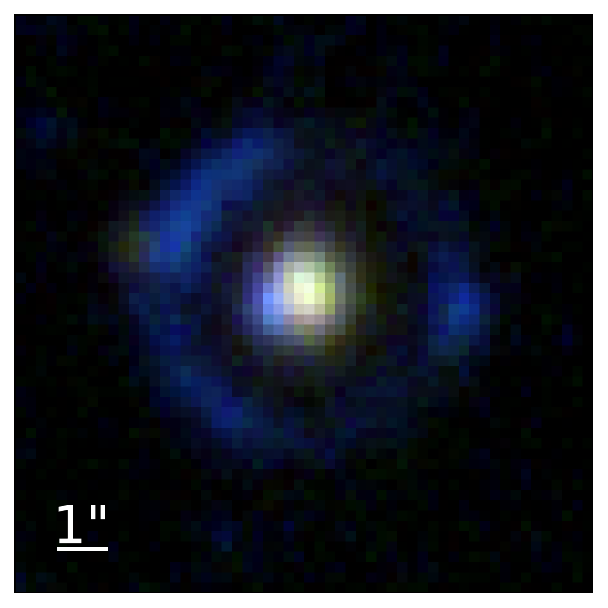}};
\node[anchor=north west, text=white, fill=black, rounded corners=1pt,
      font=\bfseries\footnotesize, inner sep=1pt]
      at (7*\xsep+1.5,0*\ysep+\imgHeight-0.5) {16};
\node[anchor=north, font=\footnotesize]
      at (7*\xsep+\imgHeight/2,0*\ysep-1) {DESJ2125--6504};

\end{tikzpicture}

\vspace{3em}  

\hspace*{-8mm}
\begin{tikzpicture}[every node/.style={anchor=south west,inner sep=0pt}, x=1mm, y=1mm]
  \def\rows{2}
  \def\cols{8}
  \def\xsep{24}
  \def\ysep{24}
  \def\scale{0.24}
  \def\imgHeight{23.5}

  \node[anchor=south west, font=\bfseries\LARGE] 
    at (2, \ysep + \imgHeight + 2) {(2) Model Reconstruction:};

  \node at (0* \xsep, 1 * \ysep) {\includegraphics[scale=\scale]{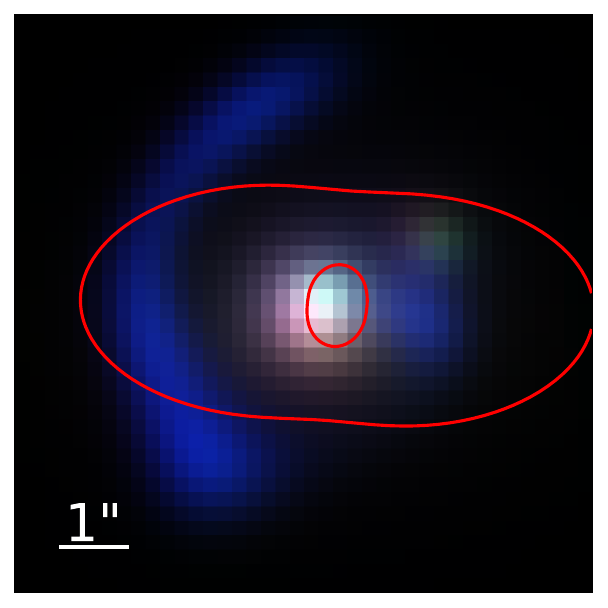}};
  \node at (1* \xsep, 1 * \ysep) {\includegraphics[scale=\scale]{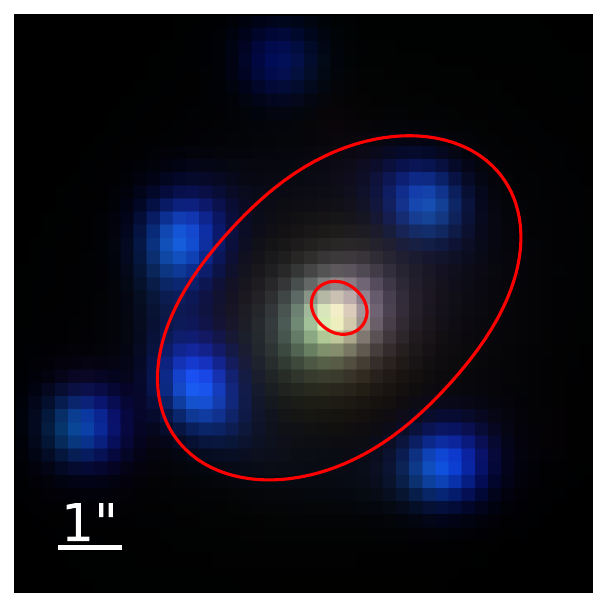}};
  \node at (2* \xsep, 1 * \ysep) {\includegraphics[scale=\scale]{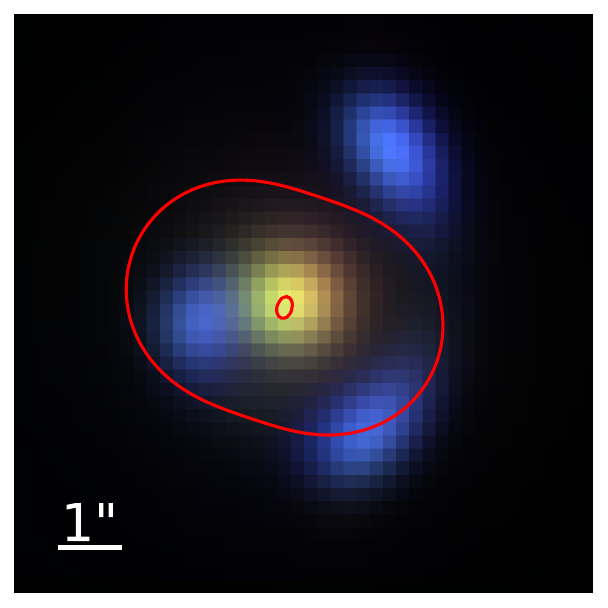}};
  \node at (3* \xsep, 1 * \ysep) {\includegraphics[scale=\scale]{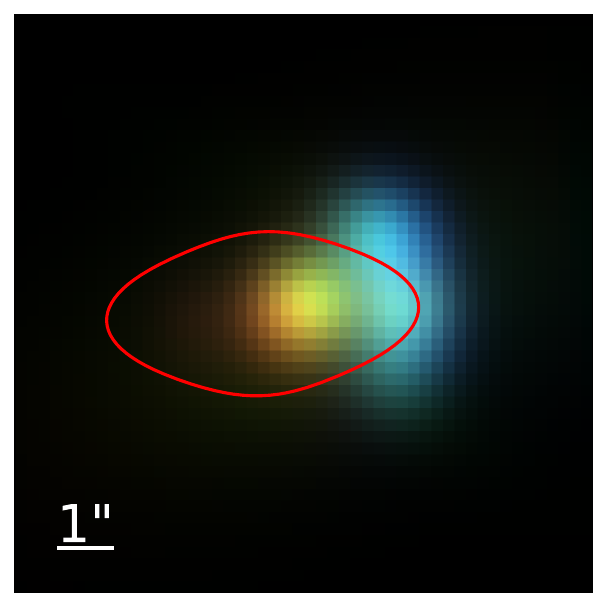}};
  \node at (4* \xsep, 1 * \ysep) {\includegraphics[scale=\scale]{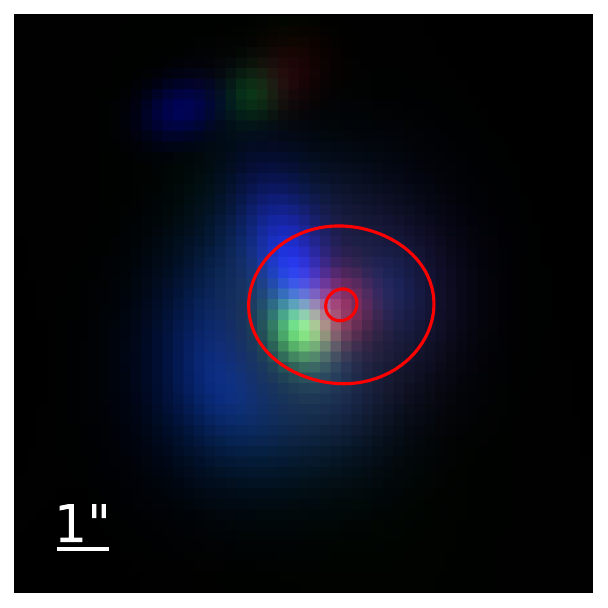}};
  \node at (5* \xsep, 1 * \ysep) {\includegraphics[scale=\scale]{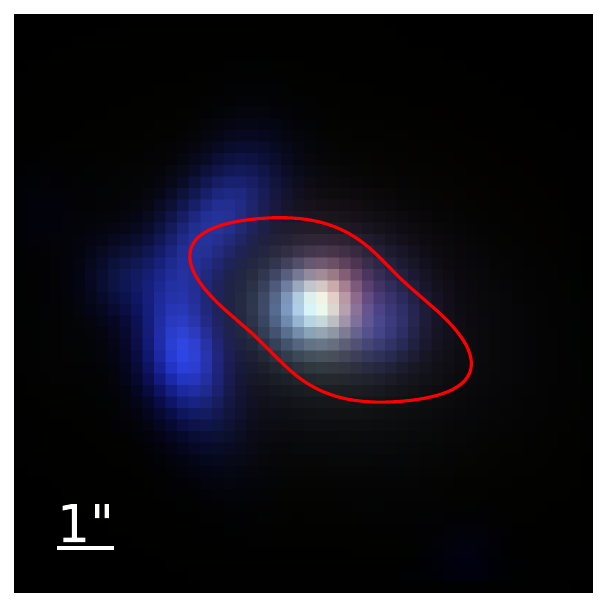}};
  \node at (6* \xsep, 1 * \ysep) {\includegraphics[scale=\scale]{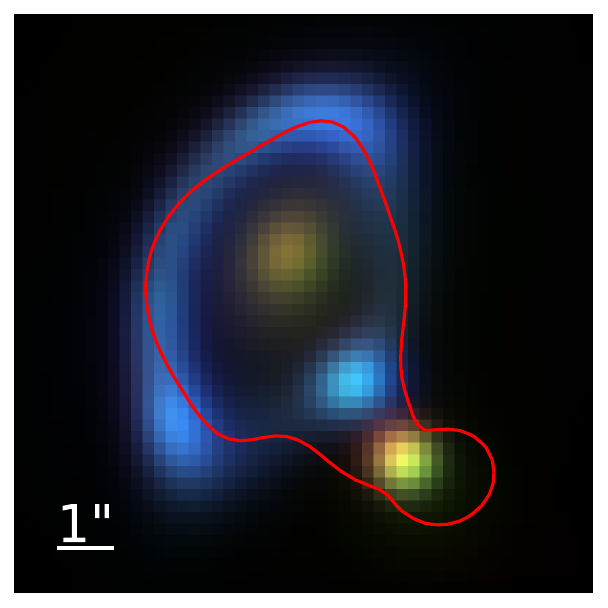}};
  \node at (7* \xsep, 1 * \ysep) {\includegraphics[scale=\scale]{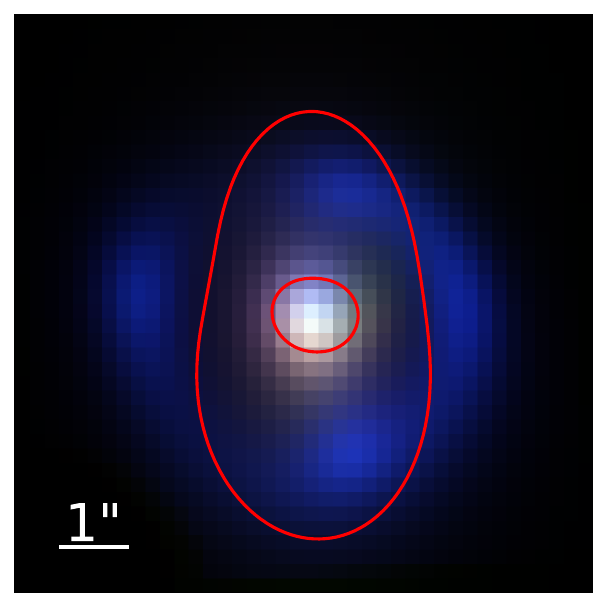}};

  \node at (0* \xsep, 0 * \ysep) {\includegraphics[scale=\scale]{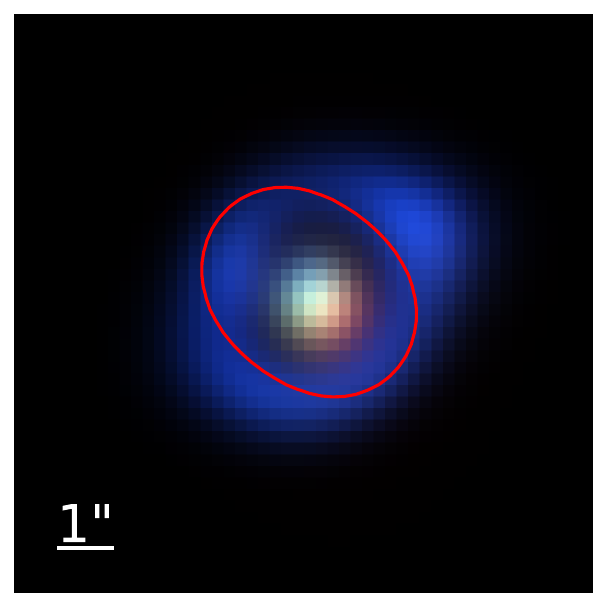}};
  \node at (1* \xsep, 0 * \ysep) {\includegraphics[scale=\scale]{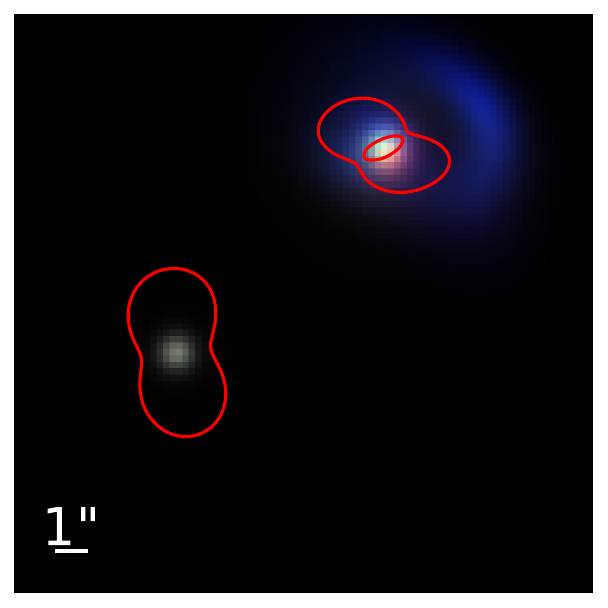}};
  \node at (2* \xsep, 0 * \ysep) {\includegraphics[scale=\scale]{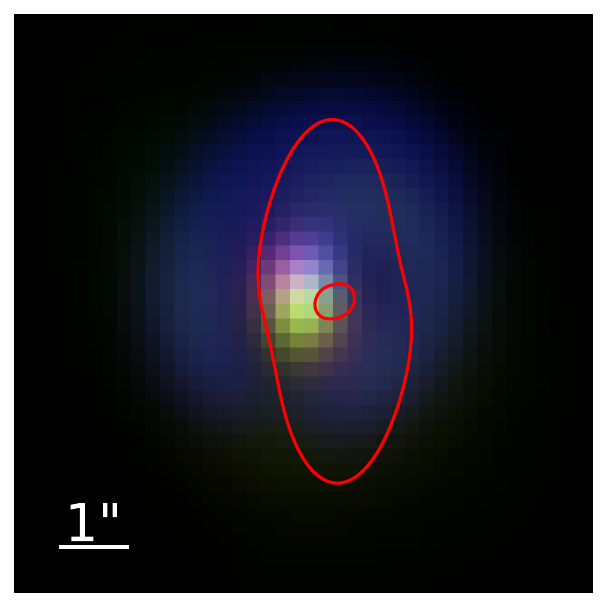}};
  \node at (3* \xsep, 0 * \ysep) {\includegraphics[scale=\scale]{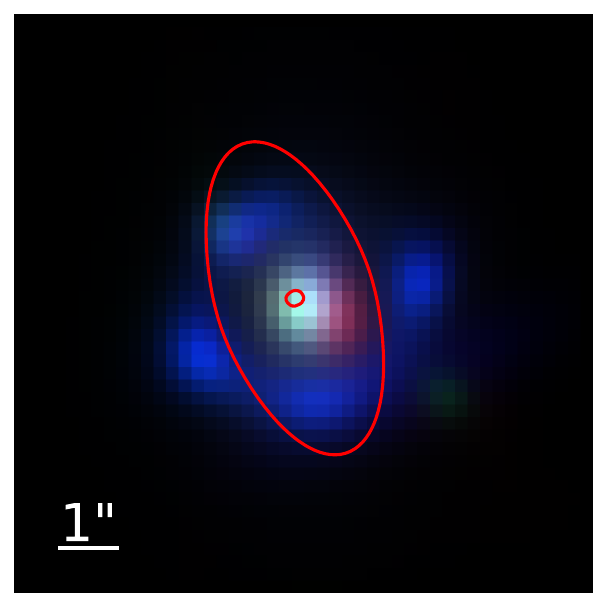}};
  \node at (4* \xsep, 0 * \ysep) {\includegraphics[scale=\scale]{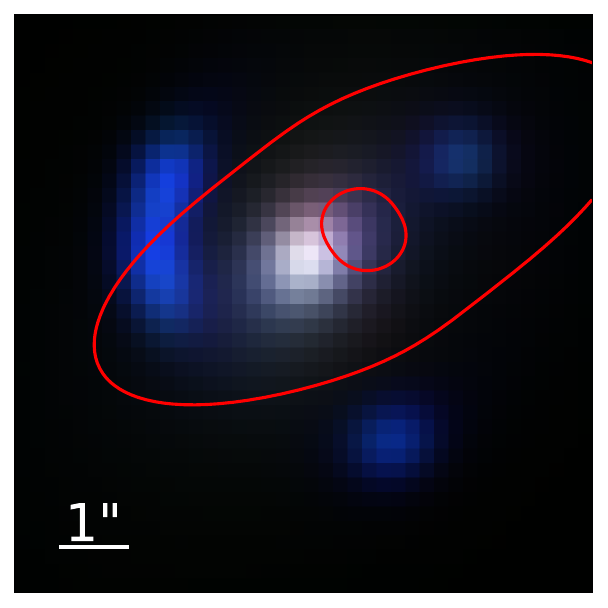}};
  \node at (5* \xsep, 0 * \ysep) {\includegraphics[scale=\scale]{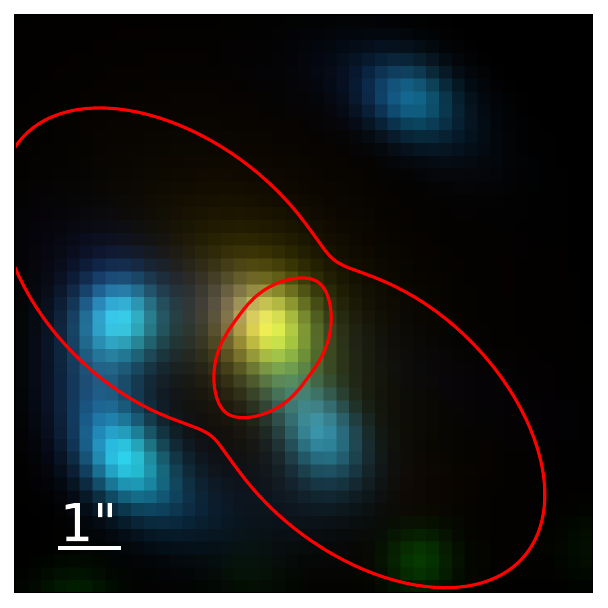}};
  \node at (6* \xsep, 0 * \ysep) {\includegraphics[scale=\scale]{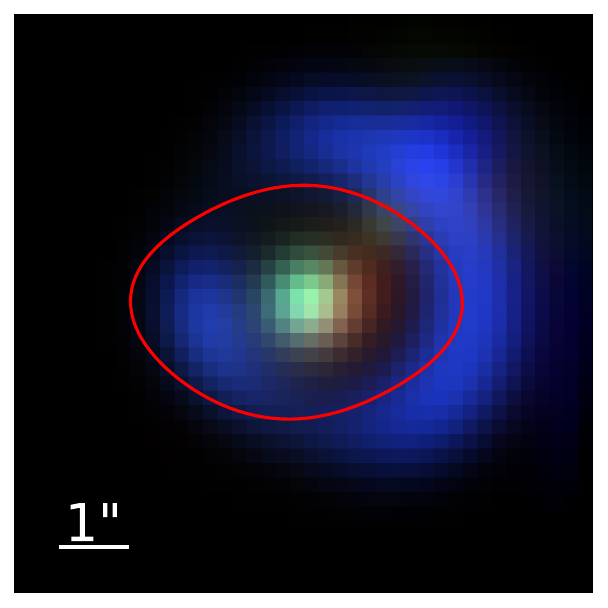}};
  \node at (7* \xsep, 0 * \ysep) {\includegraphics[scale=\scale]{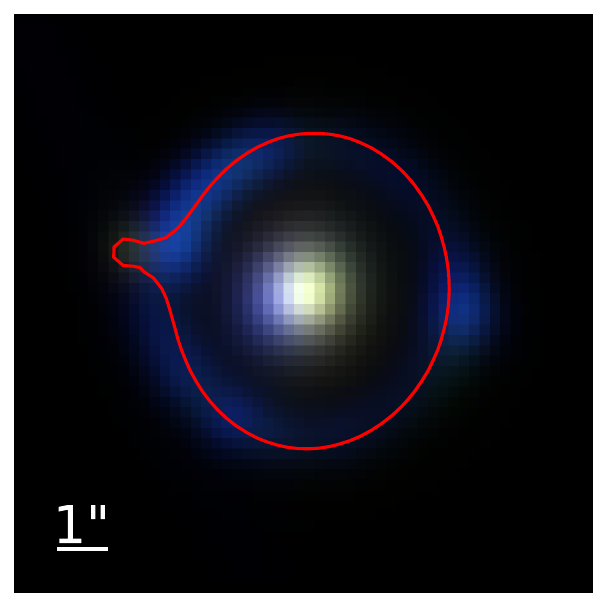}};

  \node[anchor=north west, text=white, fill=black, rounded corners=1pt,
        font=\bfseries\footnotesize, inner sep=1pt] at (0* \xsep + 1.5, 1 * \ysep + \imgHeight - 0.5) {1};
  \node[anchor=north west, text=white, fill=black, rounded corners=1pt,
        font=\bfseries\footnotesize, inner sep=1pt] at (1* \xsep + 1.5, 1 * \ysep + \imgHeight - 0.5) {2};
  \node[anchor=north west, text=white, fill=black, rounded corners=1pt,
        font=\bfseries\footnotesize, inner sep=1pt] at (2* \xsep + 1.5, 1 * \ysep + \imgHeight - 0.5) {3};
  \node[anchor=north west, text=white, fill=black, rounded corners=1pt,
        font=\bfseries\footnotesize, inner sep=1pt] at (3* \xsep + 1.5, 1 * \ysep + \imgHeight - 0.5) {4};
  \node[anchor=north west, text=white, fill=black, rounded corners=1pt,
        font=\bfseries\footnotesize, inner sep=1pt] at (4* \xsep + 1.5, 1 * \ysep + \imgHeight - 0.5) {5};
  \node[anchor=north west, text=white, fill=black, rounded corners=1pt,
        font=\bfseries\footnotesize, inner sep=1pt] at (5* \xsep + 1.5, 1 * \ysep + \imgHeight - 0.5) {6};
  \node[anchor=north west, text=white, fill=black, rounded corners=1pt,
        font=\bfseries\footnotesize, inner sep=1pt] at (6* \xsep + 1.5, 1 * \ysep + \imgHeight - 0.5) {7};
  \node[anchor=north west, text=white, fill=black, rounded corners=1pt,
        font=\bfseries\footnotesize, inner sep=1pt] at (7* \xsep + 1.5, 1 * \ysep + \imgHeight - 0.5) {8};

  \node[anchor=north west, text=white, fill=black, rounded corners=1pt,
        font=\bfseries\footnotesize, inner sep=1pt] at (0* \xsep + 1.5, 0 * \ysep + \imgHeight - 0.5) {9};
  \node[anchor=north west, text=white, fill=black, rounded corners=1pt,
        font=\bfseries\footnotesize, inner sep=1pt] at (1* \xsep + 1.5, 0 * \ysep + \imgHeight - 0.5) {10};
  \node[anchor=north west, text=white, fill=black, rounded corners=1pt,
        font=\bfseries\footnotesize, inner sep=1pt] at (2* \xsep + 1.5, 0 * \ysep + \imgHeight - 0.5) {11};
  \node[anchor=north west, text=white, fill=black, rounded corners=1pt,
        font=\bfseries\footnotesize, inner sep=1pt] at (3* \xsep + 1.5, 0 * \ysep + \imgHeight - 0.5) {12};
  \node[anchor=north west, text=white, fill=black, rounded corners=1pt,
        font=\bfseries\footnotesize, inner sep=1pt] at (4* \xsep + 1.5, 0 * \ysep + \imgHeight - 0.5) {13};
  \node[anchor=north west, text=white, fill=black, rounded corners=1pt,
        font=\bfseries\footnotesize, inner sep=1pt] at (5* \xsep + 1.5, 0 * \ysep + \imgHeight - 0.5) {14};
  \node[anchor=north west, text=white, fill=black, rounded corners=1pt,
        font=\bfseries\footnotesize, inner sep=1pt] at (6* \xsep + 1.5, 0 * \ysep + 1* \imgHeight - 0.5) {15};
  \node[anchor=north west, text=white, fill=black, rounded corners=1pt,
        font=\bfseries\footnotesize, inner sep=1pt] at (7* \xsep + 1.5, 0 * \ysep + \imgHeight - 0.5) {16};
\end{tikzpicture}

\vspace{3em}  

\hspace*{-8mm}
\begin{tikzpicture}[every node/.style={anchor=south west,inner sep=0pt}, x=1mm, y=1mm]
  \def\rows{2}
  \def\cols{8}
  \def\xsep{24}
  \def\ysep{24}
  \def\scale{0.24}
  \def\imgHeight{23.5}

  \node[anchor=south west, font=\bfseries\LARGE] 
    at (2, \ysep + \imgHeight + 2) {(3) Source--plane Reconstruction:};

  \node at (0* \xsep, 1 * \ysep) {\includegraphics[scale=\scale]{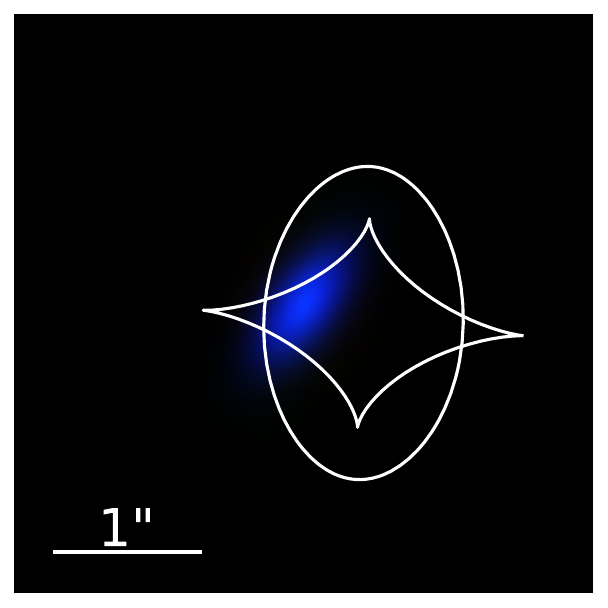}};
  \node at (1* \xsep, 1 * \ysep) {\includegraphics[scale=\scale]{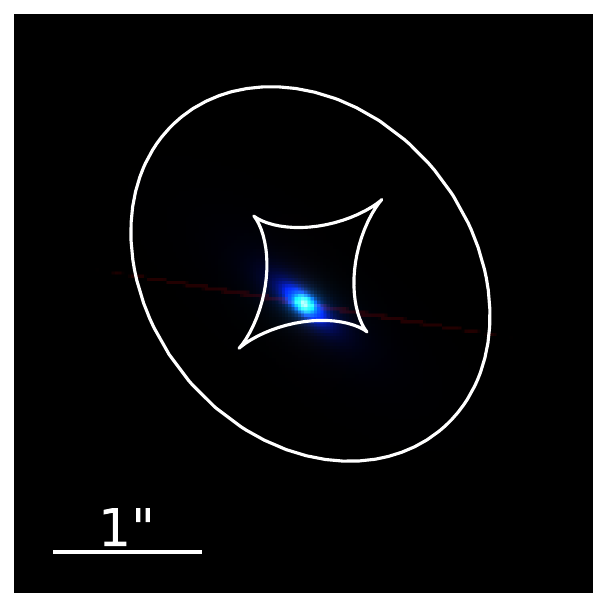}};
  \node at (2* \xsep, 1 * \ysep) {\includegraphics[scale=\scale]{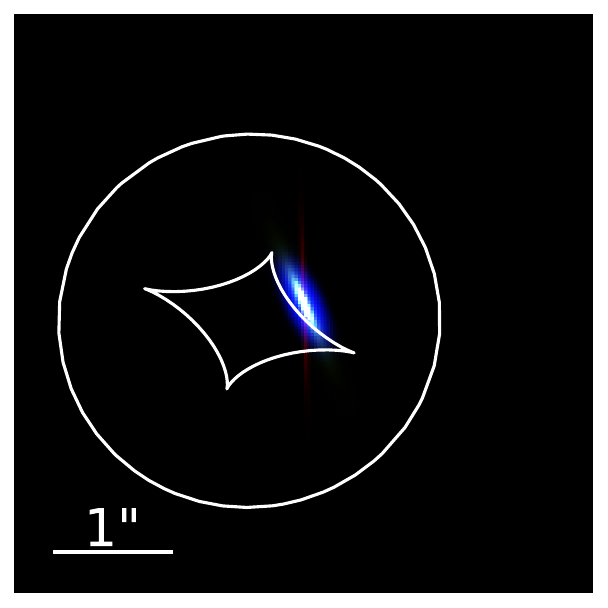}};
  \node at (3* \xsep, 1 * \ysep) {\includegraphics[scale=\scale]{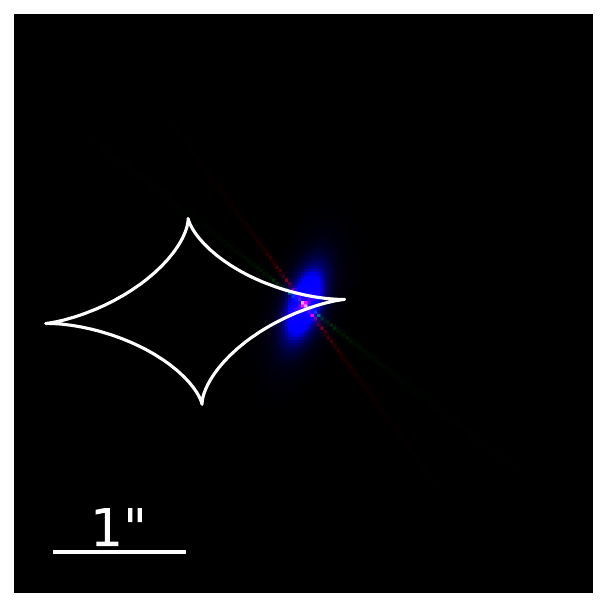}};
  \node at (4* \xsep, 1 * \ysep) {\includegraphics[scale=\scale]{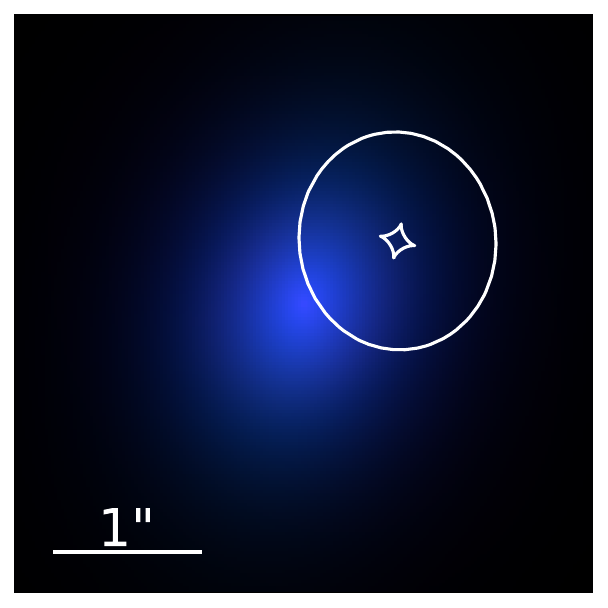}};
  \node at (5* \xsep, 1 * \ysep) {\includegraphics[scale=\scale]{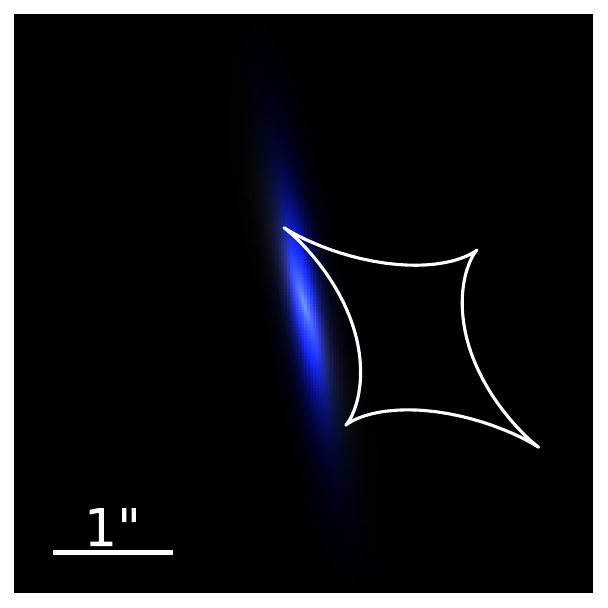}};
  \node at (6* \xsep, 1 * \ysep) {\includegraphics[scale=\scale]{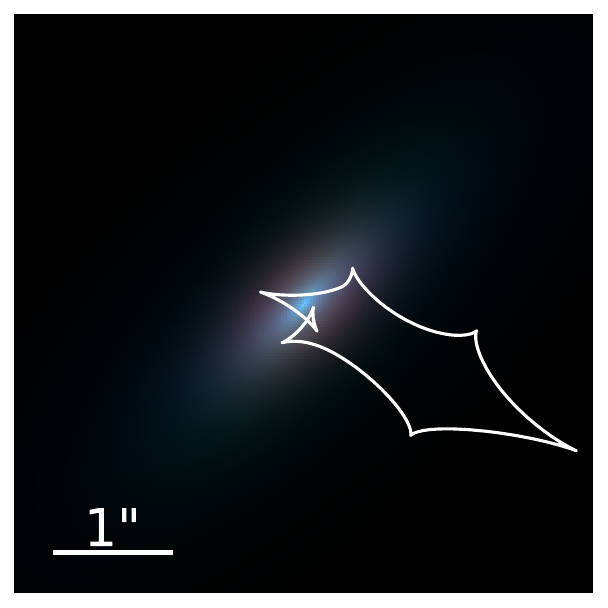}};
  \node at (7* \xsep, 1 * \ysep) {\includegraphics[scale=\scale]{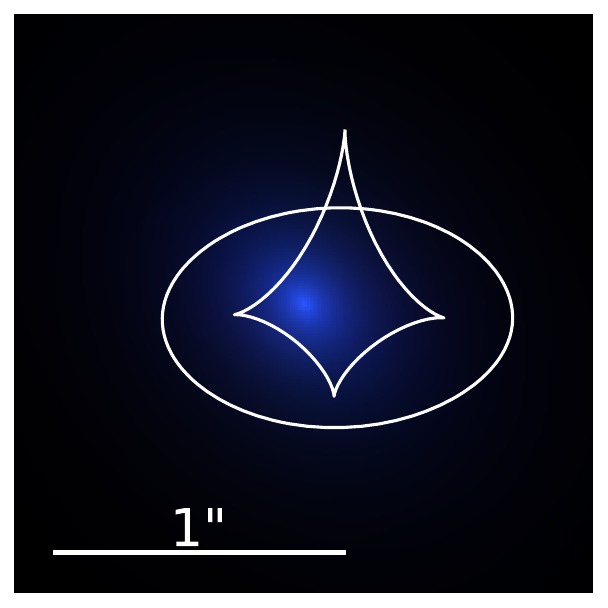}};

  \node at (0* \xsep, 0 * \ysep) {\includegraphics[scale=\scale]{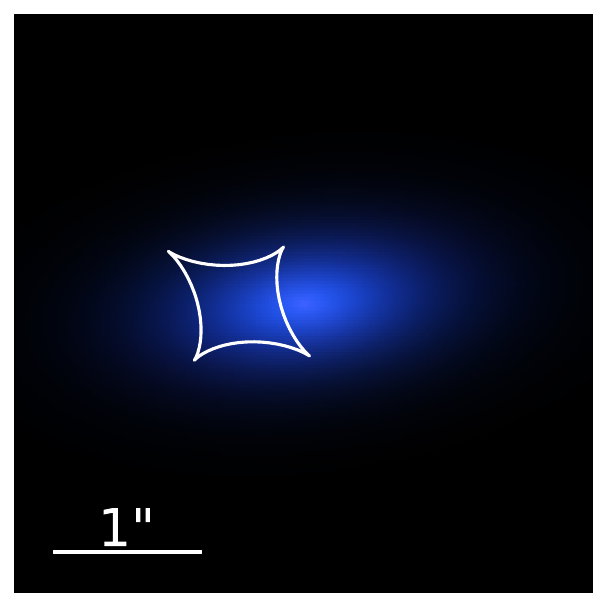}};
  \node at (1* \xsep, 0 * \ysep) {\includegraphics[scale=\scale]{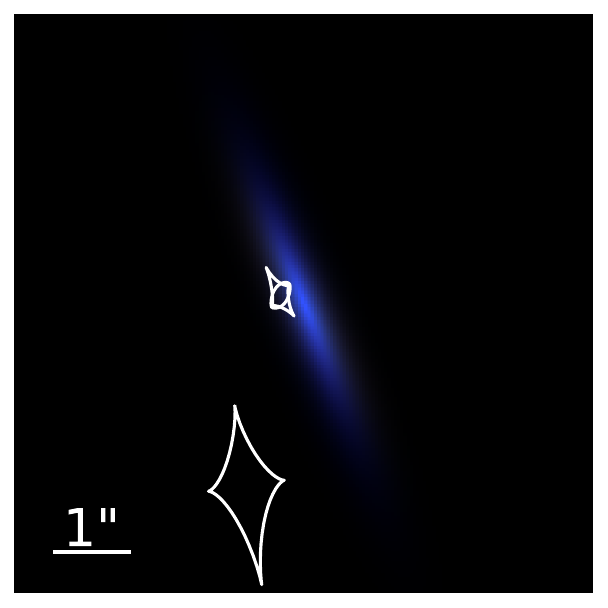}};
  \node at (2* \xsep, 0 * \ysep) {\includegraphics[scale=\scale]{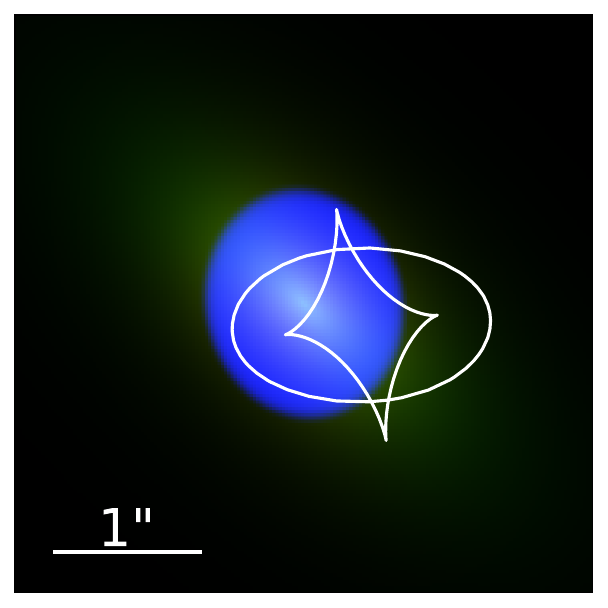}};
  \node at (3* \xsep, 0 * \ysep) {\includegraphics[scale=\scale]{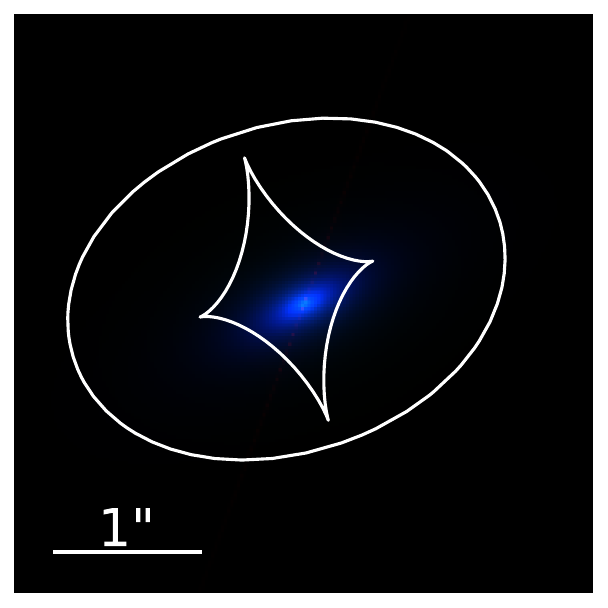}};
  \node at (4* \xsep, 0 * \ysep) {\includegraphics[scale=\scale]{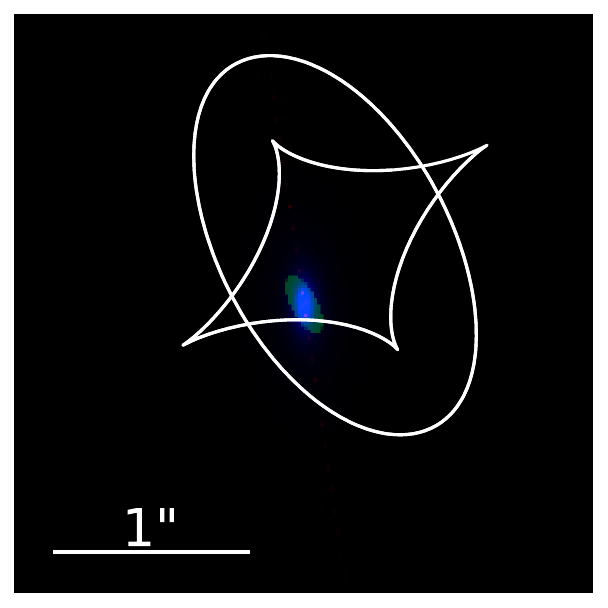}};
  \node at (5* \xsep, 0 * \ysep) {\includegraphics[scale=\scale]{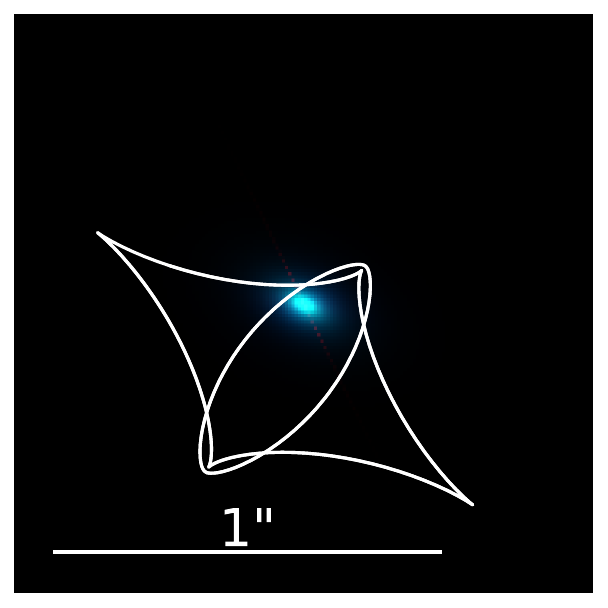}};
  \node at (6* \xsep, 0 * \ysep) {\includegraphics[scale=\scale]{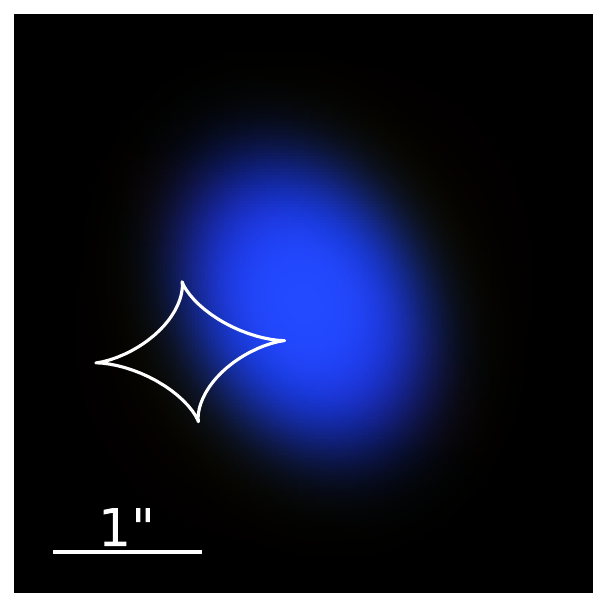}};
  \node at (7* \xsep, 0 * \ysep) {\includegraphics[scale=\scale]{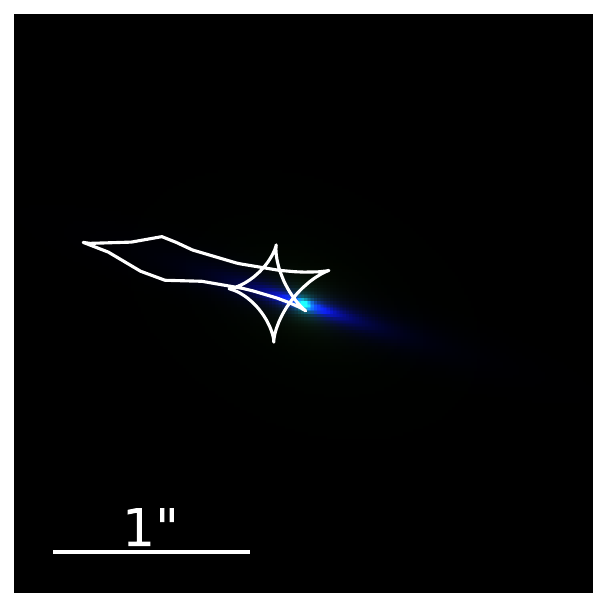}};

  \node[anchor=north west, text=white, fill=black, rounded corners=1pt,
        font=\bfseries\footnotesize, inner sep=1pt] at (0* \xsep + 1.5, 1 * \ysep + \imgHeight - 0.5) {1};
  \node[anchor=north west, text=white, fill=black, rounded corners=1pt,
        font=\bfseries\footnotesize, inner sep=1pt] at (1* \xsep + 1.5, 1 * \ysep + \imgHeight - 0.5) {2};
  \node[anchor=north west, text=white, fill=black, rounded corners=1pt,
        font=\bfseries\footnotesize, inner sep=1pt] at (2* \xsep + 1.5, 1 * \ysep + \imgHeight - 0.5) {3};
  \node[anchor=north west, text=white, fill=black, rounded corners=1pt,
        font=\bfseries\footnotesize, inner sep=1pt] at (3* \xsep + 1.5, 1 * \ysep + \imgHeight - 0.5) {4};
  \node[anchor=north west, text=white, fill=black, rounded corners=1pt,
        font=\bfseries\footnotesize, inner sep=1pt] at (4* \xsep + 1.5, 1 * \ysep + \imgHeight - 0.5) {5};
  \node[anchor=north west, text=white, fill=black, rounded corners=1pt,
        font=\bfseries\footnotesize, inner sep=1pt] at (5* \xsep + 1.5, 1 * \ysep + \imgHeight - 0.5) {6};
  \node[anchor=north west, text=white, fill=black, rounded corners=1pt,
        font=\bfseries\footnotesize, inner sep=1pt] at (6* \xsep + 1.5, 1 * \ysep + \imgHeight - 0.5) {7};
  \node[anchor=north west, text=white, fill=black, rounded corners=1pt,
        font=\bfseries\footnotesize, inner sep=1pt] at (7* \xsep + 1.5, 1 * \ysep + \imgHeight - 0.5) {8};

  \node[anchor=north west, text=white, fill=black, rounded corners=1pt,
        font=\bfseries\footnotesize, inner sep=1pt] at (0* \xsep + 1.5, 0 * \ysep + \imgHeight - 0.5) {9};
  \node[anchor=north west, text=white, fill=black, rounded corners=1pt,
        font=\bfseries\footnotesize, inner sep=1pt] at (1* \xsep + 1.5, 0 * \ysep + \imgHeight - 0.5) {10};
  \node[anchor=north west, text=white, fill=black, rounded corners=1pt,
        font=\bfseries\footnotesize, inner sep=1pt] at (2* \xsep + 1.5, 0 * \ysep + \imgHeight - 0.5) {11};
  \node[anchor=north west, text=white, fill=black, rounded corners=1pt,
        font=\bfseries\footnotesize, inner sep=1pt] at (3* \xsep + 1.5, 0 * \ysep + \imgHeight - 0.5) {12};
  \node[anchor=north west, text=white, fill=black, rounded corners=1pt,
        font=\bfseries\footnotesize, inner sep=1pt] at (4* \xsep + 1.5, 0 * \ysep + \imgHeight - 0.5) {13};
  \node[anchor=north west, text=white, fill=black, rounded corners=1pt,
        font=\bfseries\footnotesize, inner sep=1pt] at (5* \xsep + 1.5, 0 * \ysep + \imgHeight - 0.5) {14};
  \node[anchor=north west, text=white, fill=black, rounded corners=1pt,
        font=\bfseries\footnotesize, inner sep=1pt] at (6* \xsep + 1.5, 0 * \ysep + \imgHeight - 0.5) {15};
  \node[anchor=north west, text=white, fill=black, rounded corners=1pt,
        font=\bfseries\footnotesize, inner sep=1pt] at (7* \xsep + 1.5, 0 * \ysep + \imgHeight - 0.5) {16};
\end{tikzpicture}
\caption{Three-panel overview of all 16 lens systems. Each panel shows 2×8 images arranged by candidate ID. From top to bottom: (1) observed data, (2) best-fit lens model reconstructions, and (3) source-plane reconstructions. Each image is created by stacking the $z$, $i$, $r$, and $g$ bands after registering them to the $r$-band frame, with a label in the top-left corner. 
The reconstructions represent the highest likelihood lens configurations in the MCMC chain, under the best-performing lens models (grey-highlighted row in Table \ref{tab:physical_quantities}), with red critical curves overlaid in the lens plane and white caustic curves shown in the source plane. The presence (or absence) of an inner critical curve (corresponding to the outer caustic in the source plane) reflects the slope of the mass density profile, appearing only when the logarithmic slope is shallower than isothermal ($\gamma < 2$).}
\label{fig:overview_16}
\vfill
\end{figure*}
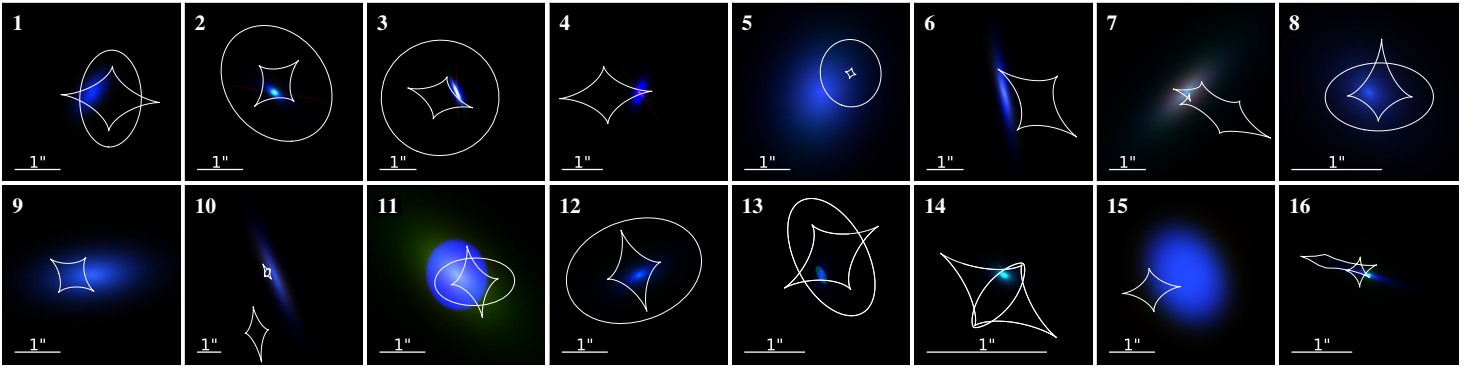

\begin{figure*} 
    \centering
    \includegraphics[width=0.9\textwidth,height=0.9\textheight,keepaspectratio]{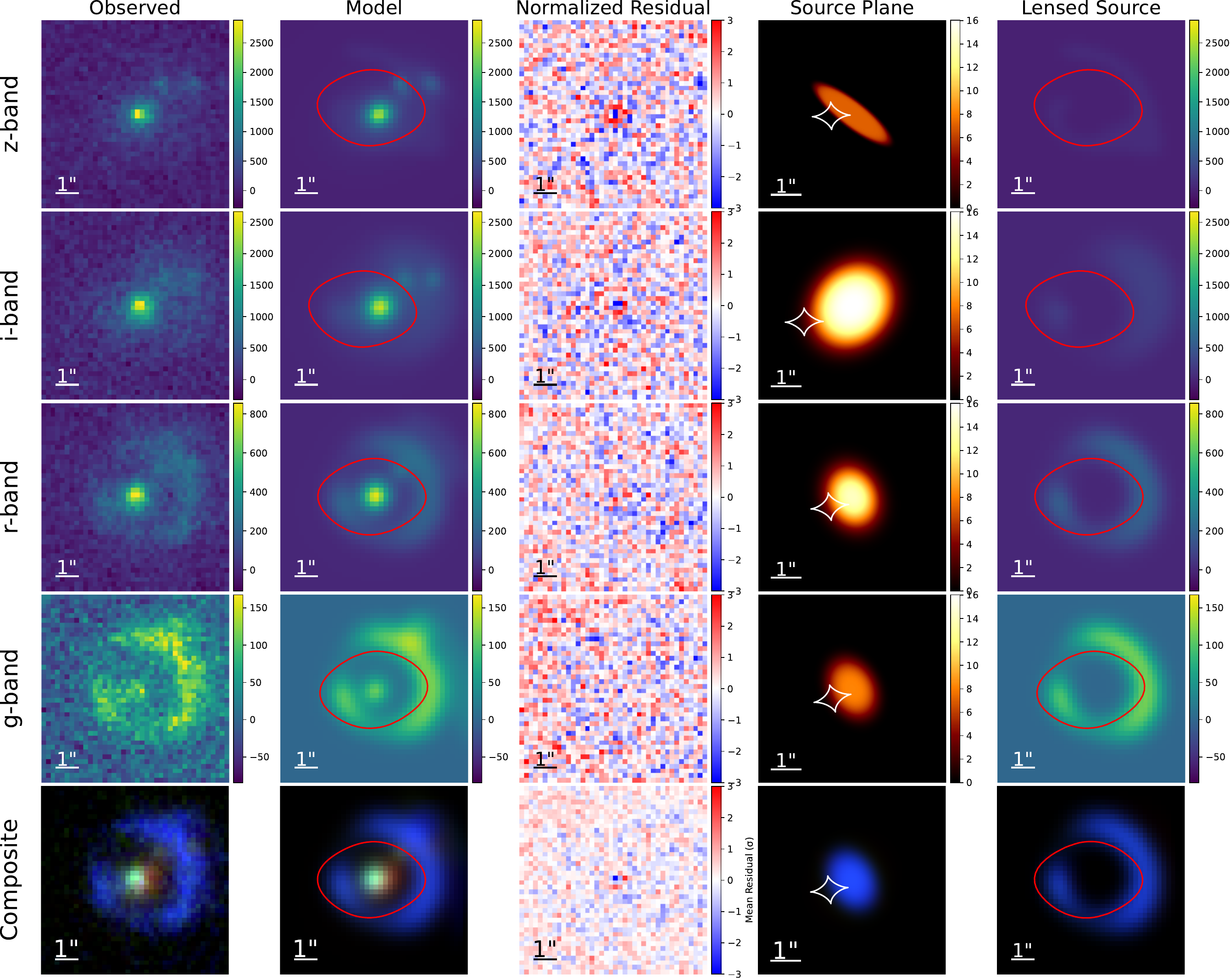}
    \caption{Example of multi--band gravitational lens reconstruction for the system DESJ2032-5658. 
    Each row corresponds to a different filter—$z$, $i$, $r$, and $g$ from top to bottom—followed by a composite color image based on band-normalized data in the final row. 
    The columns display: (i) observed data; (ii) reconstructed model image with red critical curves; (iii) normalized residuals, defined as $(\mathrm{model} - \mathrm{data})/\sigma$; (iv) source-plane reconstruction with white caustics; and (v) lensed source light in the image plane, also with red critical curves. 
    Within each band, a consistent color scale is used for the data, model, and lensed source image, revealing the increasing prominence of the lensed source in bluer bands. 
    Critical curves, computed from a shared mass model, are transformed into each band’s frame using the inferred relative shifts and rotations with respect to the r-band. 
    For the composite row, all images are aligned to the r-band frame, and the residuals are averaged across bands.} 
    \label{fig:recon_example}
\end{figure*}

\begin{figure}
    \centering
    \includegraphics[width=1\linewidth]{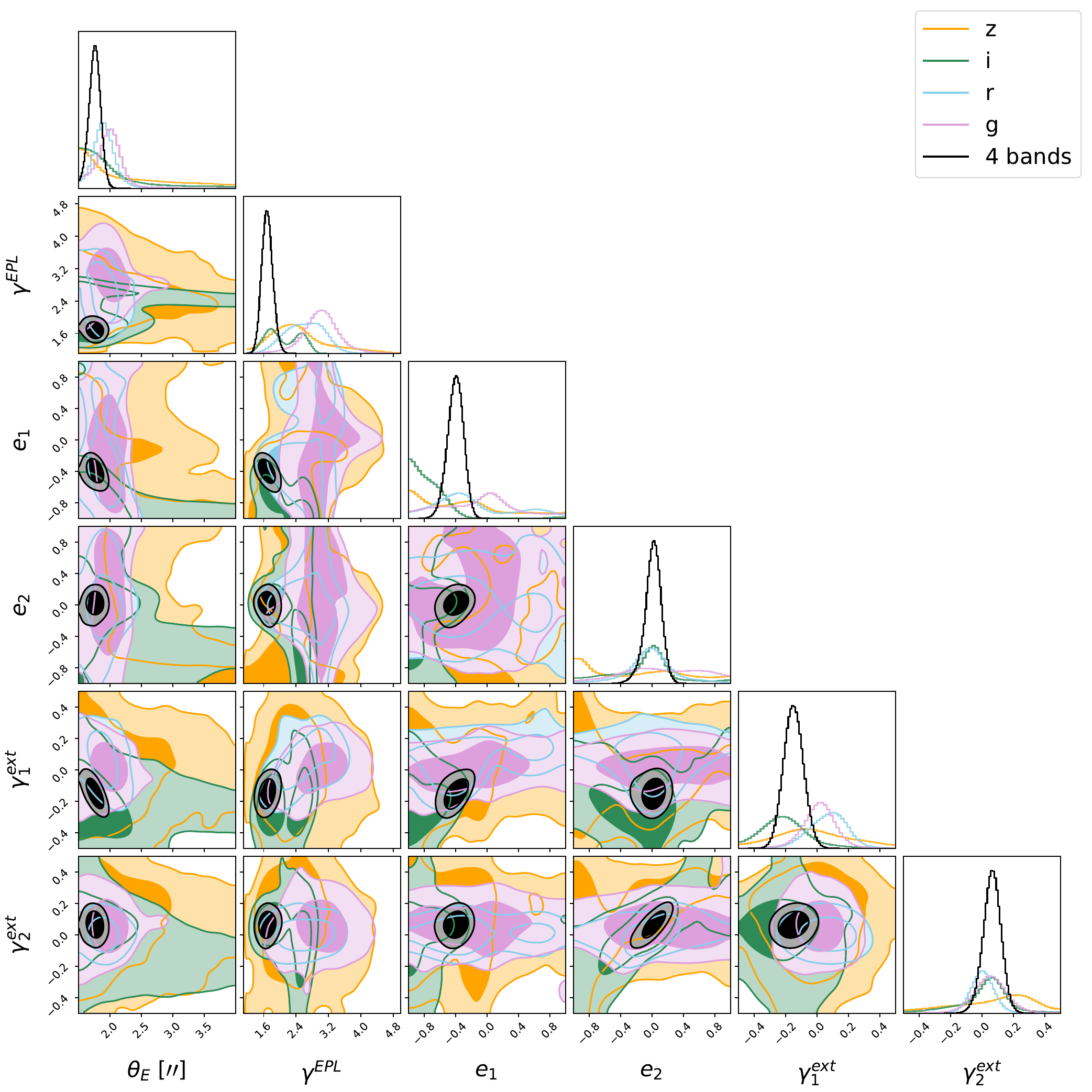}
    \caption{The $1\sigma$ and $2\sigma$ posterior distributions of Einstein radius, power law slope, ellipticity parameters, external shear parameters, for DESJ0305-1024. Colored contours show single-band reconstructions ($z$, $i$, $r$, $g$), which exhibit broad and inconsistent posteriors. The black line shows the joint four-band reconstruction, demonstrating tighter and more stable posteriors.}
    \label{fig:singleVS4}
\end{figure}

\subsubsection{Shifts and rotations between colour bands}
When one enforces a shared mass model across all reconstructed bands, pre-registration can be conducted \citep{2019MNRAS.483.5649S, Ballard_2024} to account for astrometric offsets between bands and enable joint reconstruction on multi--band datasets with a single mass model. However, the lens and source light in this PISCO sample are sufficiently fuzzy that precise morphological alignment between bands is particularly challenging, and small rotational offsets, likely due to minor chromatic effects, are observed. To address these issues, we implement the explicit modelling of inter-band shifts and rotations. Specifically, we introduce three nuisance parameters per band—two for translation and one for rotation—relative to the r-band data which offers the clearest visibility of both lens and source light. This approach not only guarantees robust multi--band reconstructions based on seeing-limited, ground-based observations, but also facilitates future reconstructions of datasets from different instruments, such as the stubborn misalignment between HST and ALMA datasets in \citet{Zhang_2023}.

\subsubsection{Sampling our posterior}
We initially optimise for a high-likelihood solution using Particle Swarm Optimisation (PSO), which offers fast convergence and computational efficiency, making it well-suited for this preliminary stage. In high-dimensional parameter spaces (as in this work, $\geq 65$ parameters), the optima found by PSO may be local rather than global. To mitigate this, we run the PSO multiple times in parallel and select the result with the highest likelihood. The number of PSO runs depends on the complexity of the lens; through experimentation, we find $\sim30-100$ runs to be sufficient.

Following PSO optimisation, we use \textsc{zeus} to sample the posterior distribution and quantify parameter uncertainties. Compared to traditional samplers like \textsc{emcee}, \textsc{zeus} offers faster convergence and lower memory usage, requiring fewer iterations for stability.
We use $2 \text{--} 4 \times N$ walkers (where N is the number of free parameters). To verify convergence, we conducted autocorrelation analyses \citep{gelman2013bayesian, Hogg_2018}, ensuring that the autocorrelation time stabilizes as the sample size increases and that the autocorrelation function reaches an asymptote for each parameter.

Once we have obtained converged chains, we additionally compute the Bayesian Information Criterion (BIC), which helps to identify well-performing models while discarding overly complex ones, especially in lens systems that demand more complexity when we come to perform our analyses. The BIC is given by:

\begin{equation}
    \text{BIC} = N_{\text{params}}\log{N_{\text{pixels}}}-2\log{ p(\mathcal{D}\mid\mathcal{M})_{\text{best}}},
\end{equation}

where $N_{\text{params}}$ is the number of free parameters in the model, $N_{\text{pixels}}$ is the number of pixels in total across all four bands, and $p(\mathcal{D}\mid\mathcal{M})_{\text{best}}$ is the best--fitting likelihood, which in this work we define as the highest likelihood in our chain.

\section{Results}
\label{sec:results}
In this section, we present the modelling results for our sample of 16 strong lens candidates. This includes an overview of the best-fit model reconstructions; a detailed example of a multi--band lens reconstruction; a summary table of the key inferred physical quantities; a comparison between single-band and multi--band modelling; and the discovery of a possible exotic image configuration.

A detailed presentation of the modeling for each system is available in Appendix~\ref{app:individual_lenses}.

\subsection{Overview of Lens Reconstructions}

Figure \ref{fig:overview_16} presents the observational data, best-fit reconstructions in the image plane, and source-plane reconstructions for the full sample of 16 strong lens candidates. The color composite images are created by stacking normalized images from each band, all aligned to the r-band reference frame using the inferred shifts and rotations. These visualizations showcase the morphological diversity of our sample, which our pipeline has successfully reconstructed.

\subsubsection{Demonstration with DESJ2032-5658}

In Figure~\ref{fig:recon_example}, we highlight system DESJ2032-5658 as a representative example to demonstrate the quality and consistency of our multi--band modeling. We show the observed data, best fit model, the residuals, the corresponding source plane reconstructions, and the lensed arcs with the modelled lens light component subtracted. We present these for each band individually, followed by the final row presenting a composite of all bands.

Across all bands, the residual maps show low and spatially uniform values, confirming the model’s excellent performance. In the bluer bands ($r$ and $g$), the lensed source becomes increasingly prominent, consistent with the intrinsically bluer light distribution revealed in the source-plane reconstruction column.

\subsection{Posteriors on physical quantities}
A selection of key physical quantities inferred from the posterior distributions for all systems in our sample is shown in Table \ref{tab:physical_quantities}. For each lens system, we report the Einstein radius of the main deflector, the total projected mass enclosed within that radius, and the external shear strength, with the 68\% credible interval as uncertainties.
The mass enclosed within Einstein radius \( R_E \) under the EPL model is computed by: 
\begin{equation}
    M(< \theta_E) = \frac{c^2}{4 G} \frac{D_L D_S}{D_{LS}} \theta_E^{2}.
\end{equation}
where $\theta_E$ is the Einstein radius, and $D_L$, $D_S$, and $D_{LS}$ are the angular diameter distances to the lens, the source, and between the lens and the source, respectively. We adopt the \texttt{Planck18} cosmology \citep{planckcollaborationPlanck2018Results2020} as implemented in the default settings of the \textsc{astropy}\footnote{\url{https://docs.astropy.org/}} package.

Spectroscopic redshifts from the AGEL survey are available for some systems and are listed in Table~\ref{tab:physical_quantities} (marked with $\dagger$); when unavailable, deflector redshifts are taken from photometry, and source redshifts are assumed to be $z_s = 2$ for estimating enclosed masses, consistent with the peak of the source redshift distribution in the AGEL survey (see Fig.~11 of \citealt{2025arXiv250308041B}). 
We note that the systems with spectroscopic measurements in this sample have relatively high-redshift deflectors ($\langle z_\mathrm{lens}^\dagger \rangle \sim 0.68$) and sources ($\langle z_s^\dagger \rangle \sim 2.14$), broadly representative of those found in the AGEL survey but higher than in other surveys such as SLACS \citep{2008ApJ...682..964B}. This may reflect that DES is mostly sensitive to large Einstein radii.

\subsection{Single vs multi--band}

In order to obtain more powerful constraints on the parameters of our mass models, we use all the available data and reconstruct the source light distributions in all four bands ($z$, $i$, $r$, $g$) simultaneously, rather than selecting only one preferred band throughout our analysis.

With the \textsc{zeus} sampler, we find that a similar number of MCMC steps are required to obtain a fully converged posterior ($\sim\!10^5$), though the four-band reconstruction involves many more parameters and demands more MCMC ``walkers'' as a result. 
While this increase could become a limitation for very large samples \citep[e.g., $\mathcal{O}(10^3)$ lenses or more; see][]{shajib2025dolphinfullyautomatedforward}, our results demonstrate that the scientific gain from multi--band modelling---tighter, more consistent parameter constraints and more robust lens models---clearly justifies the computational expense. Moreover, GPU-accelerated tools such as \textsc{Herculens} \citep{Galan_2022}, \textsc{GIGA-Lens} \citep{Gu_2022}, \textsc{JaxTronomy}\footnote{\url{https://github.com/lenstronomy/jaxtronomy}}, \textsc{TinyLensGPU} \citep{cao2025csststronglensingpreparation} will likely make such approaches feasible for survey-scale applications.

\subsubsection{Demonstration with DESJ0305-1024}
To assess the impact of multi--band modelling, we reconstructed system 11 (DESJ0305–1024) using the four bands individually (henceforth ``single--band''), as well as simultaneously (henceforth ``multi--band''). As shown in Figure \ref{fig:singleVS4}, single-band fits yield broad, inconsistent posteriors for key mass parameters, whereas the multi--band reconstruction (black) produces tight, stable posteriors, demonstrating that multi--band information suppresses lens–source degeneracies and mitigates ground-based seeing limitations.

We also examined the likelihoods of the best-fit reconstructions. While the likelihood for each individual band in the multi--band fit is generally lower than in the single--band fits—reflecting the trade-off of jointly fitting all bands—the single-band fits can produce inconsistent lens mass models across bands and carry the risk of overfitting noise. By contrast, the multi--band fit yields a more consistent and robust mass model overall.

\begin{table*} 
\vfill
\centering
\caption{Summary of key physics quantities for the full lens system sample. Listed are the target label, deflector redshift $z_\mathrm{lens}$ (with $^\dagger$ indicating spectroscopic values and unmarked values from photometric estimates), right ascension and declination (in degrees), Einstein radius $\theta_E^{\text{\tiny EPL}}$, mass profile slope $\gamma^{\text{\tiny EPL}}$, total projected mass enclosed within the Einstein radius assuming either a fiducial source redshift of $z_s = 2$ or the system's spectroscopic $z_s^\dagger$ when available, external shear strength $\gamma^{\text{\small ext}}$, and position angles of the lens and shear components. Values with uncertainties are either directly obtained or derived from the posterior distributions, with 68\% credible intervals. Spectroscopic redshifts are drawn from the ASTRO 3D Galaxy Evolution with Lenses (AGEL) survey
\citep{tranAGELSurveyStrong2023, baroneAGELSurveyData2025}. The row highlighted in grey corresponds to the lens model shown in Figure \ref{fig:overview_16}.}
\label{tab:physical_quantities}
\renewcommand{\arraystretch}{1.5}

\resizebox{\textwidth}{!}{
\begin{tabular}{clccccccccc}
\hline
\hline
Label & Target Name 
& \(z_\mathrm{lens}\)
& \makecell{RA\\ (deg)} 
& \makecell{Dec\\ (deg)} 
& \makecell{$\theta_E^{\text{\tiny EPL}}$ \\ \((^{\prime\prime})\)} 
& $\gamma^{\text{\tiny EPL}}$
& \makecell[c]{%
\(\log_{10}\left[\frac{M(<\theta_{\rm E}^{\text{\tiny EPL}})}{M_\odot}\right]\) \\
(\(z_s = 2\) or true \(z_s^\dagger\))
}
& $\gamma^{\text{\small ext}}$
& \makecell{$\phi^{\text{\tiny EPL}}$ \\ ($^\circ$)}
& \makecell{$\phi^{\text{\small ext}}$ \\ ($^\circ$)}
\\
\hline
\multirow{3}{*}{1} & DESJ0003--3348 (1) \cellcolor[gray]{0.9} & \multirow{3}{*}{0.659$^\dagger$} & \multirow{3}{*}{0.818255} & \multirow{3}{*}{-33.8012} 
& $2.65^{+0.04}_{-0.04}$  \cellcolor[gray]{0.9}
& $1.74^{+0.07}_{-0.07}$  \cellcolor[gray]{0.9}
& \makecell[c]{$12.40^{+0.01}_{-0.01}\ (z_s^\dagger=1.834)$}  \cellcolor[gray]{0.9}
& $0.05^{+0.03}_{-0.02}$  \cellcolor[gray]{0.9}
& $-1.24^{+2.26}_{-2.23}$ \cellcolor[gray]{0.9}
& $11.52^{+18.76}_{-9.13}$  \cellcolor[gray]{0.9}
\\ 
                   & DESJ0003--3348 (2) &                        &                            &                            
                   & $2.35^{+0.07}_{-0.06}$
                   & $1.89^{+0.12}_{-0.10}$ 
                   & $12.29^{+0.03}_{-0.02}\ (z_s^\dagger=1.834)$
                   & $0.05^{+0.02}_{-0.02}$
                   & $6.33^{+3.30}_{-3.42}$
                   & $62.69^{+15.54}_{-29.75}$
                   \\
                   & DESJ0003--3348 (3) &                        &                            &                            & $2.61^{+0.05}_{-0.06}$ & $1.75^{+0.08}_{-0.07}$ 
                   & \makecell[c]{$12.38^{+0.02}_{-0.02}\ (z_s^\dagger=1.834)$} 
                   & $0.05^{+0.03}_{-0.02}$ 
                   & $-1.85^{+2.29}_{-2.28}$
                   & $11.25^{+21.51}_{-9.74}$
                   \\

\hline
2 \cellcolor[gray]{0.9} & DESJ0010--4315 \cellcolor[gray]{0.9} & 0.84 \cellcolor[gray]{0.9} & 2.626778 \cellcolor[gray]{0.9} & -43.254127 \cellcolor[gray]{0.9}
& $2.78^{+0.01}_{-0.03}$ \cellcolor[gray]{0.9}
& $1.82^{+0.18}_{-0.13}$ \cellcolor[gray]{0.9}
& $12.54^{+0.00}_{-0.01}$ \cellcolor[gray]{0.9}
& $0.05^{+0.02}_{-0.01}$ \cellcolor[gray]{0.9}
& $36.16^{+1.66}_{-2.37}$ \cellcolor[gray]{0.9}
& $7.01^{+13.33}_{-27.92}$ \cellcolor[gray]{0.9}
\\
\hline
3  \cellcolor[gray]{0.9}
& DESJ0101--4917  \cellcolor[gray]{0.9}
& 0.77  \cellcolor[gray]{0.9}
& 15.491818  \cellcolor[gray]{0.9}
& -49.293942  \cellcolor[gray]{0.9}
& $2.22^{+0.01}_{-0.01}$  \cellcolor[gray]{0.9}
& $1.98^{+0.06}_{-0.06}$  \cellcolor[gray]{0.9}
& $12.24^{+0.00}_{-0.00}\ (z_s^\dagger = 2.632)$  \cellcolor[gray]{0.9}
& $0.22^{+0.02}_{-0.02}$  \cellcolor[gray]{0.9}
& $80.16^{+5.60}_{-109.33}$  \cellcolor[gray]{0.9}
& $74.66^{+0.97}_{-1.14}$  \cellcolor[gray]{0.9}
\\
\hline
\multirow{2}{*}{4} & DESJ0120--1820 (1)
& \multirow{2}{*}{0.56} & \multirow{2}{*}{20.107359} & \multirow{2}{*}{-18.333811} 
& $1.78^{+0.06}_{-0.09}$ 
& $1.70^{+0.04}_{-0.04}$ 
& $11.95^{+0.03}_{-0.05}$ 
& $0.18^{+0.02}_{-0.03}$ 
& $3.72^{+1.23}_{-0.81}$ 
& $-6.40^{+3.05}_{-5.43}$
\\
                   & DESJ0120--1820 (2)  \cellcolor[gray]{0.9}
                   & & & 
                   & $2.10^{+0.02}_{-0.01}$ \cellcolor[gray]{0.9}
                   & $2.18^{+0.08}_{-0.05}$ \cellcolor[gray]{0.9}
                   & $12.10^{+0.01}_{-0.01}$ \cellcolor[gray]{0.9}
                   & $0.04^{+0.02}_{-0.01}$ \cellcolor[gray]{0.9}
                   & $2.58^{+0.86}_{-0.68}$ \cellcolor[gray]{0.9}
                   & $-8.87^{+5.78}_{-13.77}$ \cellcolor[gray]{0.9}
\\
\hline                   
5 \cellcolor[gray]{0.9}
& DESJ0141--1303 \cellcolor[gray]{0.9}& 0.69 \cellcolor[gray]{0.9}& 25.254148 \cellcolor[gray]{0.9}& -13.050894 \cellcolor[gray]{0.9}
& $1.77^{+0.07}_{-0.07}$ \cellcolor[gray]{0.9}
& $1.81^{+0.67}_{-0.18}$ \cellcolor[gray]{0.9}
& $12.05^{+0.03}_{-0.04}$ \cellcolor[gray]{0.9}
& $0.12^{+0.07}_{-0.06}$ \cellcolor[gray]{0.9}
& $13.36^{+29.33}_{-38.22}$ \cellcolor[gray]{0.9}
& $28.14^{+20.52}_{-29.05}$ \cellcolor[gray]{0.9}
\\
\hline
\multirow{2}{*}{6} & DESJ0142-1831 (1) & \multirow{2}{*}{0.637$^\dagger$} & \multirow{2}{*}{25.720295} & \multirow{2}{*}{-18.521051} 
& $2.21^{+0.01}_{-0.02}$ 
& $2.52^{+0.10}_{-0.18}$
& $12.16^{+0.00}_{-0.01}\ (z_s^\dagger = 2.47)$
& $0.04^{+0.03}_{-0.02}$
& $-6.75^{+1.42}_{-1.77}$
& $-0.50^{+23.56}_{-8.88}$ 
\\
                   & DESJ0142--1831 (2) \cellcolor[gray]{0.9}
                   & & & 
                   & $1.95^{+0.02}_{-0.02}$ \cellcolor[gray]{0.9}
                   & $2.25^{+0.08}_{-0.08}$ \cellcolor[gray]{0.9}
                   & $12.06^{+0.01}_{-0.01}\ (z_s^\dagger = 2.47)$ \cellcolor[gray]{0.9}
                   & $0.15^{+0.01}_{-0.01}$ \cellcolor[gray]{0.9}
                   & $-21.20^{+1.68}_{-1.46}$ \cellcolor[gray]{0.9}
                   & $22.54^{+3.89}_{-3.76}$ \cellcolor[gray]{0.9}
                   \\
\hline
\multirow{2}{*}{7} & DESJ0150--0304 (1) & \multirow{2}{*}{0.63675$^\dagger$} & \multirow{2}{*}{27.537943} & \multirow{2}{*}{-3.077297} 
& $2.53^{+0.01}_{-0.01}$ 
& $2.13^{+0.09}_{-0.06}$
& $12.42^{+0.00}_{-0.00}\ (z_s^\dagger = 1.390)$
& $0.29^{+0.01}_{-0.01}$
& $81.10^{+1.46}_{-2.34}$
& $74.68^{+0.76}_{-0.66}$
\\
                   & DESJ0150--0304 (2) \cellcolor[gray]{0.9}
                   & & & 
                   & $2.47^{+0.01}_{-0.01}$ \cellcolor[gray]{0.9}
                   & $2.05^{+0.08}_{-0.07}$ \cellcolor[gray]{0.9}
                   & $12.39^{+0.00}_{-0.00}\ (z_s^\dagger = 1.390)$ \cellcolor[gray]{0.9}
                   & $0.30^{+0.00}_{-0.00}$ \cellcolor[gray]{0.9}
                   & $73.25^{+0.49}_{-0.50}$ \cellcolor[gray]{0.9}
                   & $75.86^{+0.41}_{-0.41}$ \cellcolor[gray]{0.9}
                   \\
\hline
\multirow{2}{*}{8} & DESJ0202--0445 (1) & \multirow{2}{*}{0.71} & \multirow{2}{*}{30.527702} & \multirow{2}{*}{-24.751058}
& $2.39^{+0.02}_{-0.02}$
& $1.37^{+0.09}_{-0.07}$
& $12.32^{+0.01}_{-0.01}$
& $0.15^{+0.02}_{-0.02}$
& $87.45^{+1.35}_{-1.61}$
& $87.36^{+1.63}_{-3.60}$
\\
& DESJ0202--0445 (2) \cellcolor[gray]{0.9}
& & & 
& $2.34^{+0.02}_{-0.02}$ \cellcolor[gray]{0.9}
& $1.53^{+0.10}_{-0.09}$ \cellcolor[gray]{0.9}
& $12.30^{+0.01}_{-0.01}$ \cellcolor[gray]{0.9}
& $0.13^{+0.02}_{-0.02}$ \cellcolor[gray]{0.9}
& $-85.90^{+1.46}_{-1.43}$ \cellcolor[gray]{0.9}
& $-81.41^{+3.20}_{-2.66}$ \cellcolor[gray]{0.9}
\\
\hline
9  \cellcolor[gray]{0.9}
& DESJ0212--0852  \cellcolor[gray]{0.9}
& 0.759$^\dagger$  \cellcolor[gray]{0.9}
& 33.10507  \cellcolor[gray]{0.9}
& -8.896672  \cellcolor[gray]{0.9}
& $1.95^{+0.05}_{-0.04}$ \cellcolor[gray]{0.9}
& $2.70^{+0.21}_{-0.28}$ \cellcolor[gray]{0.9}
& $12.16^{+0.02}_{-0.02}\ (z_s^\dagger = 2.202)$ \cellcolor[gray]{0.9}
& $0.15^{+0.04}_{-0.03}$ \cellcolor[gray]{0.9}
& $19.68^{+44.36}_{-78.83}$ \cellcolor[gray]{0.9}
& $54.21^{+4.78}_{-9.23}$ \cellcolor[gray]{0.9}
\\\hline
\multirow{2}{*}{10} & DESJ0250--4104 (1) & \multirow{2}{*}{0.722$^\dagger$} & \multirow{2}{*}{27.537943} & \multirow{2}{*}{-3.077297} 
& $3.89^{+0.17}_{-0.20}$ 
& $1.25^{+0.08}_{-0.07}$
& $12.71^{+0.04}_{-0.05}\ (z_s^\dagger = 2.478)$
& $0.16^{+0.06}_{-0.07}$
& $53.69^{+5.25}_{-17.64}$
& $59.23^{+9.69}_{-50.35}$
\\
                   & DESJ0250--4104 (2) \cellcolor[gray]{0.9}
                   & & & 
                   & $1.41^{+0.09}_{-0.15}$ \cellcolor[gray]{0.9}
                   & $1.31^{+0.08}_{-0.07}$ \cellcolor[gray]{0.9}
                   & $11.83^{+0.06}_{-0.10}\ (z_s^\dagger = 2.478)$ \cellcolor[gray]{0.9}
                   & $0.35^{+0.03}_{-0.09}$ \cellcolor[gray]{0.9}
                   & $-3.16^{+3.19}_{-3.01}$ \cellcolor[gray]{0.9}
                   & $9.05^{+4.63}_{-3.27}$ \cellcolor[gray]{0.9}
\\\hline
11  \cellcolor[gray]{0.9}
& DESJ0305--1609  \cellcolor[gray]{0.9}
& 0.75  \cellcolor[gray]{0.9}
& 46.270357  \cellcolor[gray]{0.9}
& -10.40325  \cellcolor[gray]{0.9}
& $1.76^{+0.08}_{-0.09}$ \cellcolor[gray]{0.9}
& $1.69^{+0.13}_{-0.12}$ \cellcolor[gray]{0.9}
& $12.08^{+0.04}_{-0.04}$ \cellcolor[gray]{0.9}
& $0.17^{+0.05}_{-0.05}$ \cellcolor[gray]{0.9}
& $79.07^{+8.33}_{-166.22}$ \cellcolor[gray]{0.9}
& $75.46^{+8.62}_{-16.72}$ \cellcolor[gray]{0.9}
\\\hline
12 \cellcolor[gray]{0.9}& DESJ0327--3246  \cellcolor[gray]{0.9}& 0.65 \cellcolor[gray]{0.9}& 51.797294 \cellcolor[gray]{0.9}& -32.776455 \cellcolor[gray]{0.9}
& $1.99^{+0.04}_{-0.03}$ \cellcolor[gray]{0.9}
& $1.92^{+0.10}_{-0.12}$ \cellcolor[gray]{0.9}
& $12.12^{+0.02}_{-0.01}$ \cellcolor[gray]{0.9}
& $0.08^{+0.04}_{-0.03}$ \cellcolor[gray]{0.9}
& $-72.07^{+2.28}_{-2.02}$ \cellcolor[gray]{0.9}
& $-70.39^{+7.44}_{-5.81}$ \cellcolor[gray]{0.9}
\\\hline
13  \cellcolor[gray]{0.9}
& DESJ0354--1609  \cellcolor[gray]{0.9}
& 0.574$^\dagger$  \cellcolor[gray]{0.9}
& 58.576126  \cellcolor[gray]{0.9}
& -16.16415  \cellcolor[gray]{0.9}
& $3.01^{+0.19}_{-0.03}$ \cellcolor[gray]{0.9}
& $1.58^{+0.05}_{-0.15}$ \cellcolor[gray]{0.9}
& $12.43^{+0.05}_{-0.01}\ (z_s^\dagger = 1.91)$ \cellcolor[gray]{0.9}
& $0.24^{+0.09}_{-0.02}$ \cellcolor[gray]{0.9}
& $26.87^{+0.52}_{-0.52}$ \cellcolor[gray]{0.9}
& $23.97^{+1.45}_{-1.01}$ \cellcolor[gray]{0.9}
\\\hline
14 \cellcolor[gray]{0.9}& DESJ0533--2536 \cellcolor[gray]{0.9}& 0.71 \cellcolor[gray]{0.9}& 83.455527 \cellcolor[gray]{0.9}& -25.615115 \cellcolor[gray]{0.9}
& $3.47^{+0.07}_{-0.07}$ \cellcolor[gray]{0.9}
& $1.21^{+0.03}_{-0.03}$ \cellcolor[gray]{0.9}
& $12.65^{+0.02}_{-0.02}$ \cellcolor[gray]{0.9}
& $0.21^{+0.02}_{-0.03}$ \cellcolor[gray]{0.9}
& $-40.89^{+1.34}_{-1.36}$ \cellcolor[gray]{0.9}
& $-42.39^{+1.71}_{-1.90}$ \cellcolor[gray]{0.9}
\\\hline
15  \cellcolor[gray]{0.9}
& DESJ2032--5658  \cellcolor[gray]{0.9}
& 0.92  \cellcolor[gray]{0.9}
& 345.130757  \cellcolor[gray]{0.9}
& -56.970349  \cellcolor[gray]{0.9}
& $2.11^{+0.07}_{-0.03}$ \cellcolor[gray]{0.9}
& $2.26^{+0.22}_{-0.27}$ \cellcolor[gray]{0.9}
& $12.36^{+0.03}_{-0.01}$ \cellcolor[gray]{0.9}
& $0.05^{+0.04}_{-0.02}$ \cellcolor[gray]{0.9}
& $6.61^{+8.47}_{-8.35}$ \cellcolor[gray]{0.9}
& $1.56^{+26.54}_{-51.98}$ \cellcolor[gray]{0.9}
\\\hline
\multirow{3}{*}{16} & DESJ2125--6504 (1)
& \multirow{3}{*}{0.779$^\dagger$} & \multirow{3}{*}{321.300117} & \multirow{3}{*}{-65.074076}
& $3.17^{+0.01}_{-0.02}$
& $2.46^{+0.08}_{-0.09}$
& $12.59^{+0.00}_{-0.01}\ (z_s^\dagger = 2.223)$
& $0.09^{+0.01}_{-0.01}$ 
& $21.77^{+1.83}_{-2.82}$ 
& $29.31^{+2.43}_{-2.95}$ 
\\
& DESJ2125--6504 (2) 
& & & 
& $3.08^{+0.01}_{-0.01}$
& $2.36^{+0.11}_{-0.16}$
& $12.57^{+0.00}_{-0.00}\ (z_s^\dagger = 2.223)$
& $0.02^{+0.01}_{-0.01}$
& $78.46^{+2.42}_{-2.22}$
& $59.72^{+7.21}_{-7.96}$
\\
& DESJ2125--6504 (3)  \cellcolor[gray]{0.9}
& & & 
& $3.08^{+0.01}_{-0.01}$ \cellcolor[gray]{0.9} \cellcolor[gray]{0.9}
& $2.33^{+0.11}_{-0.10}$ \cellcolor[gray]{0.9}
& $12.57^{+0.00}_{-0.00}\ (z_s^\dagger = 2.223)$ \cellcolor[gray]{0.9}
& $0.01^{+0.00}_{-0.00}$ \cellcolor[gray]{0.9} \cellcolor[gray]{0.9}
& $83.55^{+2.07}_{-2.24}$ \cellcolor[gray]{0.9}
& $62.13^{+16.32}_{-26.59}$ \cellcolor[gray]{0.9}
\\
\hline
\end{tabular}}
\end{table*}

\subsection{Discovering a hyperbolic umbilic candidate}
\label{subsec: HU lens}

Most of the sources reconstructed by our pipeline in Figure \ref{fig:overview_16} exhibit typical Einstein cross, cusp or fold--like image configurations. However, in system 14 (DESJ0533–2536), the tangential and radial caustic meet at a cusp in the source plane, very close to where our reconstructed source is located. In the image plane, this results in three images of the source around a ring on one side of the deflector, and a fourth counter-image on the opposite side of the deflector.

Systems where a fifth image is also observed around the ring of images, due to the source lying closer to the caustic interchange, are known as hyperbolic umbilic (HU) lenses, corresponding to a theoretical catastrophe \citep{1992grle.book.....S, 2001stgl.book.....P, Meena_2020, Meena_2023, Meena_2024}. These systems are well–primed for more detailed studies of the dark matter mass profile of the lens. They produce images both near the lens center and at much larger radii, allowing us to probe the mass profile slope over vastly different radii. They also provide extended image-plane coverage for potential dark matter substructure searches.

The galaxy cluster Abell 1703 was the first reported observation of this characteristic decentralized quad-ring morphology with an additional far away counter-image, which was later confirmed spectroscopically as belonging to the same source \citep{2009MNRAS.399....2O}. More such systems have since was discovered in \citep{Lagattuta_2023}. So far, no observed HU configuration has been reported in a galaxy-galaxy strong lens. However, here we report DESJ0533–2536 as the first candidate for a hyperbolic umbilic galaxy-galaxy lens.

\begin{table*} 
\centering
\setlength{\tabcolsep}{4pt}
\renewcommand{\arraystretch}{1.2}
\begin{tabular}{lccccccccc}
\toprule
Models for DESJ0003-3348 & $\theta_E^{\text{\tiny EPL}}$ & $\gamma^{\text{\tiny EPL}}$ 
& \makecell[c]{%
\(\log_{10}\left[\frac{M(<\theta_{\rm E}^{\text{\tiny EPL}})}{M_\odot}\right]\)
}
& $\theta_E^{\text{\tiny SIS}}$ 
& \makecell[c]{%
\(\log_{10}\left[\frac{M(<\theta_{\rm E}^{\text{\tiny SIS}})}{M_\odot}\right]\)
}
& $\gamma^{\text{\small ext}}$ 
& $R_\mathrm{eff}$ & $\mathrm{log}L$ & $\mathrm{BIC}$ \\ 
& $(^{\prime\prime})$ & &  & $(^{\prime\prime})$ 
&  & 
&$(^{\prime\prime})$ & & \\ 
\midrule
Model 1: No SIS & 2.66 
& 1.6 
& 12.40 
&  &  
& 0.07 
& 0.33 
& -3290.2 & 7448.0 \\
Model 2: With SIS (free) 
& 2.40 
& 1.8 
& 12.31 
& 0.4 
& 10.7 & 0.03 & 0.35 & -3272.3 & 7438.5 \\
Model 3: With SIS (constrained) 
& 2.64 
& 1.7 
& 12.39 
& 0.02 (84\% CL: <0.09) 
& <9.5 (84\% CL) 
& 0.06
& 0.32
& -3287.4 & 7451.2 \\
\bottomrule
\end{tabular}
\caption{The best-fit mass parameters for the DESJ0003 system under three different models, drawn from zeus samples. The number of decimal places for each parameter is chosen to reflect the precision suggested by the 68\% credible interval of its posterior distribution. For model with satellite mass (constrianed position), the SIS component's Einstein radius $\theta_E^{\text{\tiny SIS}}$ has a posterior distribution that peaks near zero, and we report the 84\% upper limit, derived from an exponential fit to the posterior.}
\label{tab:desj0003_best_fit}
\end{table*}
\begin{table*} 
\centering
\setlength{\tabcolsep}{4pt}
\renewcommand{\arraystretch}{1.2}
\begin{tabular}{lccccccccc}
\toprule
Models for DESJ2125-6504 & $\theta_E^{\text{\tiny EPL}}$ & $\gamma^{\text{\tiny EPL}}$ 
& \makecell[c]{%
\(\log_{10}\left[\frac{M(<\theta_{\rm E}^{\text{\tiny EPL}})}{M_\odot}\right]\)
}
& $\theta_E^{\text{\tiny SIS}}$ 
& \makecell[c]{%
\(\log_{10}\left[\frac{M(<\theta_{\rm E}^{\text{\tiny SIS}})}{M_\odot}\right]\)
}
& $\gamma^{\text{\small ext}}$ 
& $R_\mathrm{eff}$ & $\mathrm{log}L$ & $\mathrm{BIC}$ \\ 
& $(^{\prime\prime})$ & &  & $(^{\prime\prime})$ 
&  & 
&$(^{\prime\prime})$ & & \\ 
\midrule
Model 1: No SIS 
& 3.14 
& 2.6 
& 12.58 
&  &  
& 0.07 
& 2 
& -6103.1 & 13046.0 \\
Model 2: With SIS (free) 
& 3.08 
& 2.2 
& 12.57 
& 0.19 
& 10.1 
& 0.03 
& 0.6 
& -6062.7 
& 12993.5 \\
Model 3: With SIS (constrained) 
& 3.07 
& 2.2 
& 12.57 
& 0.18  
& 10.1 
& 0.01 
& 1.2 
& -6060.4 & 12970.1 \\
\bottomrule
\end{tabular}
\caption{Same as Table \ref{tab:desj0003_best_fit}, but for the DESJ2125 system. The inclusion of a SIS component significantly improves the fit, with BIC differences of approximately 52 and 76 for Model 2 and Model 3, respectively, relative to Model 1. The inferred external shear correspondingly decreases when the satellite is included, indicating that part of the perturbation previously absorbed by the shear is now attributed to the satellite mass.}
\label{tab:desj2125_best_fit}
\end{table*}

With an Einstein radius of $3.47\pm0.07''$ it has one of the most massive deflectors in our sample, also placing it in the galaxy–scale regime rather than the cluster–scale regime. Given the data quality of PISCO, we cannot observe a fifth image -- which could be faint and coincide with the observed location of the lens light. We are able to perform models using high resolution HST data, but we remained blind to this data during the work presented in this paper. We also speculate that in other wavelength regimes outside of optical, or in narrow-band emission features observable with IFU, there may exist visible features beyond the brightest part of our reconstructed sources which do lie in the correct position to form all five images. Regardless, with this data we are able to confirm that the correct caustic pattern to form a HU can occur in galaxy--scale lenses, possibly offering a unique opportunity to probe small-scale substructure and exploit the extreme magnification associated with HU lenses. 

\begin{figure}
    \centering
    \includegraphics[width=1\linewidth]{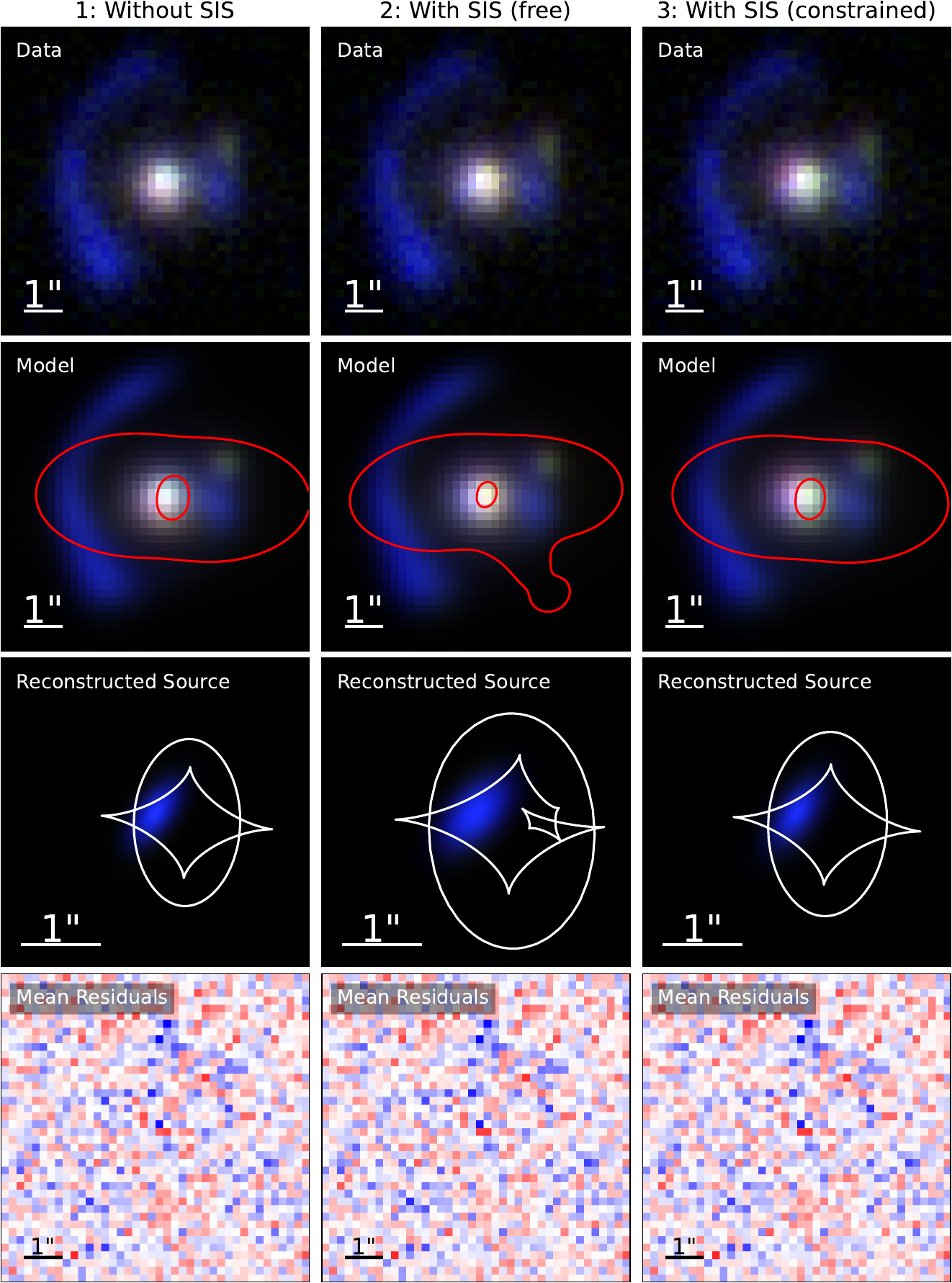}
    \caption{Best-fit reconstructions of system 1 (DESJ0003-3348) under three modeling assumptions. From left to right: Model 1 (no SIS component), Model 2 (SIS component with free position), and Model 3 (SIS component fixed at the satellite’s position). Each row displays the observed data, model prediction, reconstructed source, and mean normalized residual (averaged over 4 bands; 2 $\sigma$ level).}
    \label{fig:comparison_model_desj0003}
\end{figure}
\begin{figure}
    \centering
    \includegraphics[width=1\linewidth]{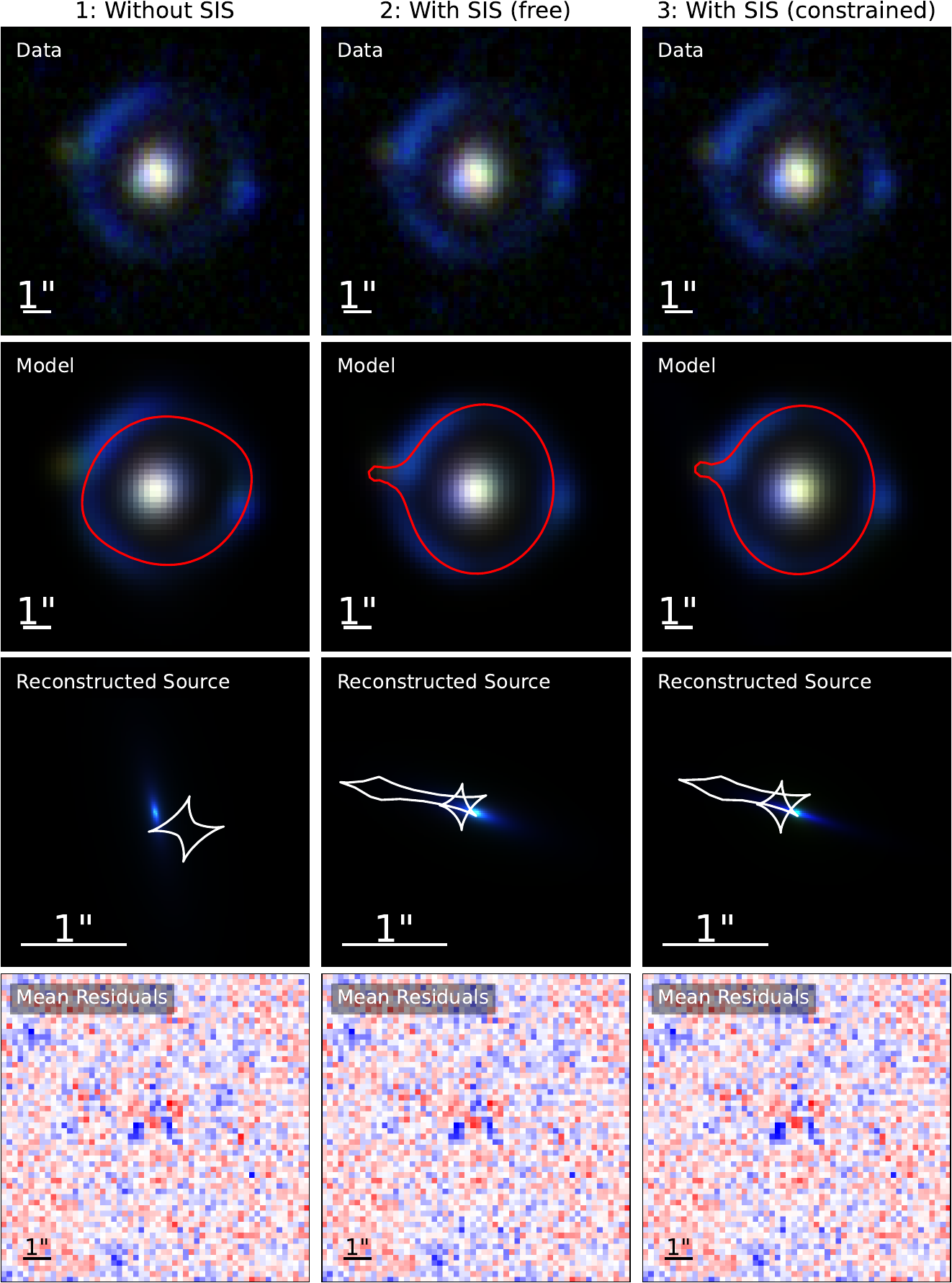}
    \caption{The same as Figure \ref{fig:comparison_model_desj0003}, but for system 16 (DESJ2125-6504). Models 2 and 3 include a SIS component associated with the satellite, leading to notably different de-lensed source morphologies and caustic structures compared to Model 1.}
    \label{fig:comparison_model_desj2125}
\end{figure}

\section{Discussion}
\label{sec:discussion}

In this section we evaluate how viably our pipeline can constrain the lens model in our sample, focusing on specific cases such as detecting potential satellite masses, improvements from including flexion, and challenges in modeling complex systems, then we discuss the implications of this for modelling future ground--based survey data of newly--discovered lenses in an automated fashion.

\subsection{How much model complexity can we constrain with PISCO--like data quality?}
\subsubsection{Satellite masses}
\label{subsec: substructure in DESJ0003 and DESJ2125}
We investigate the possible presence of additional lensing perturbations in systems that have a visible satellite galaxy. We tested three modelling assumptions to investigate how well our pipeline handles satellite masses in this dataset:
\begin{enumerate}
    \item Model 1: The fiducial mass model with a Sérsic elliptical profile added to account for the satellite light.
    \item Model 2: Based on Model 1, plus an SIS mass component with a free position.
    \item Model 3: Based on Model 1, plus an SIS mass which is forced to share the centroid of the satellite light (in the $r$-band).
\end{enumerate}
We tested these assumptions on systems 1 (DESJ0003-3348, where we constrain the satellite object close to the smaller of the two arcs) and 16 (DESJ2125-6504, where we constrain the satellite close to the larger of the two arcs). Their inferred highest-likelihood parameter values and BIC are presented in Tables \ref{tab:desj0003_best_fit} and \ref{tab:desj2125_best_fit} respectively.

We find no indication that a satellite mass is required in system 1 ($\Delta\mathrm{BIC} = \mathrm{BIC}(\mathrm{Model\,1}) - \mathrm{BIC}(\mathrm{Model\,3}) \approx -3$,with lower BIC indicating a better fit). In contrast, system 16 shows a strong preference for including the satellite ($\Delta\mathrm{BIC} \approx 76$), with Model 3 providing the better fit. 
While all models fit the data to the noise level (Figures \ref{fig:comparison_model_desj0003} and \ref{fig:comparison_model_desj2125}), 
the BIC and posterior distributions of key mass parameters (Appendix \ref{app:individual_lenses}) 
quantitatively support the conclusion that a satellite is unnecessary in system 1 but required in system 16.

We therefore find that \textbf{our pipeline is capable of constraining lower mass satellite deflectors, and that that for some systems (e.g., system 16), even ground--based data can reveal a satellite mass necessary for the best reconstruction of the arcs}. The strong evidence for a satellite mass in system 16 (DESJ2125–6504) also highlights the importance of accounting for mass perturbations from visible satellite galaxies.

\subsubsection{Asymmetric angular complexity}
\label{subsubsec:asymmetries}
\begin{figure*}
\centering
\begin{tikzpicture}
  \node[anchor=north west] (lefttop) at (0, 0) {
    \includegraphics[width=0.74\textwidth]{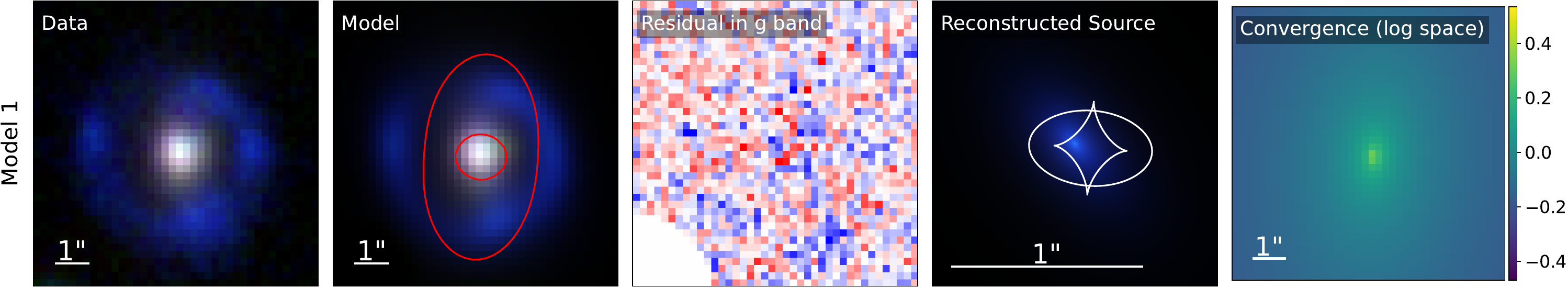}
  };
  \node[anchor=north west] (lefttop) at (0, -3) {
    \includegraphics[width=0.74\textwidth]{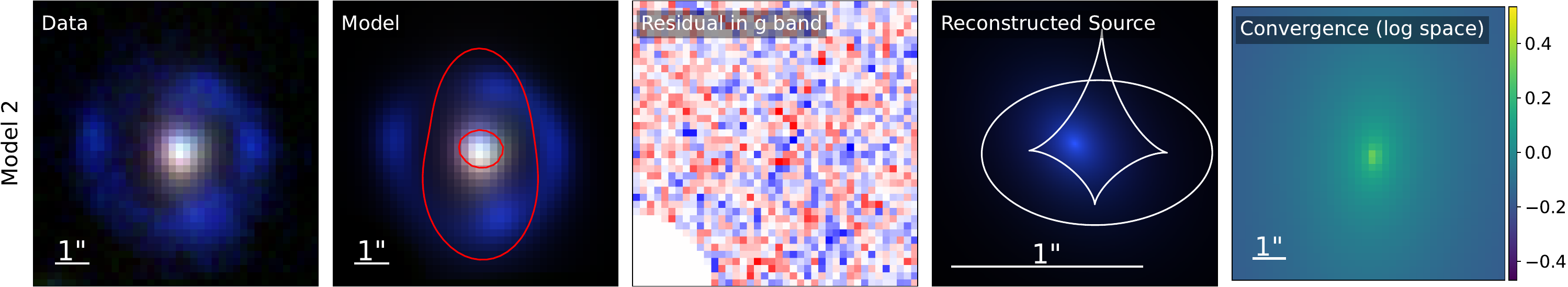}
  };
  \node[anchor=north west] (right) at (0.74\textwidth + 0.1cm, -0.4) {
    \includegraphics[height=0.23\textwidth]{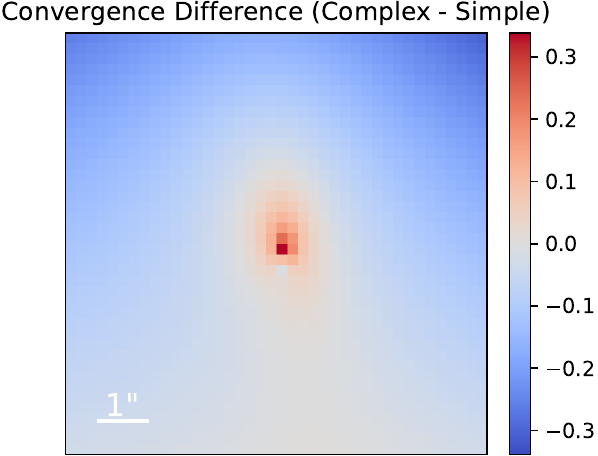}
  };
\end{tikzpicture}
\caption{Comparison of reconstructions for DESJ0202 (system 8). The left panel shows the best‐fit models under Model 1 (fiducial configuration) and Model 2 (including external flexion); Model 2 yields improved residuals in the $g$ band and enhances the observed light in the lower portion of the ring along a lopsided critical curve, while the right panel shows the convergence difference, revealing a gradient consistent with the added flexion component.}
\label{complexity_in_DESJ0202}
\end{figure*}
In some systems, the observed light distribution along the Einstein ring is notably uneven. We therefore additionally test whether our pipeline can measure multipolar deformations to the mass profile shape, or perturbation from an external flexion signal. When performing these tests on system 8 (DESJ0202-2445), we find that the external flexion-included model provided the best performance. Figure \ref{complexity_in_DESJ0202} presents a comparison of the best-fit reconstructions from the fiducial model and this model, with the flexion-included model better reproducing the asymmetric ring morphology. The convergence map illustrates the dimensionless surface mass density, while the difference between the two models' convergence maps reveals a directional gradient aligned with the inferred $g_4$ flexion mode (defined in Eq.~\ref{eq4}). Quantitatively, the flexion-included model is also strongly favored, with a BIC improvement of over 200.

We emphasise that this is the only system where adding an external flexion component led to a substantial improvement in BIC in our sample. While multipole perturbations can also improve the BIC for this system, the improvement is less pronounced than that from flexion. 
For the few other systems where flexion or multipoles were tested, no significant BIC gains were observed.
We therefore conclude that, \textbf{for PISCO–like ground-based imaging, we can constrain asymmetric mass profile shapes only when asymmetric signatures are very pronounced}. 
We caution, however, that reliably detecting genuine weak-lensing flexion is challenging due to intrinsic shape noise, PSF distortions, and shear–flexion coupling \citep{Massey_2007_shear_flexion, Viola_2011_WL_shear_flexion}. Consequently, the “flexion” model here should be interpreted primarily as a tool to capture asymmetric deviations from an elliptical mass distribution, rather than a secure measurement of physical flexion. Higher--resolution weak-lensing analyses would be required to robustly probe higher-order lensing signals.

\subsubsection{External shear}
\begin{figure}
    \centering
    \includegraphics[width=\linewidth]{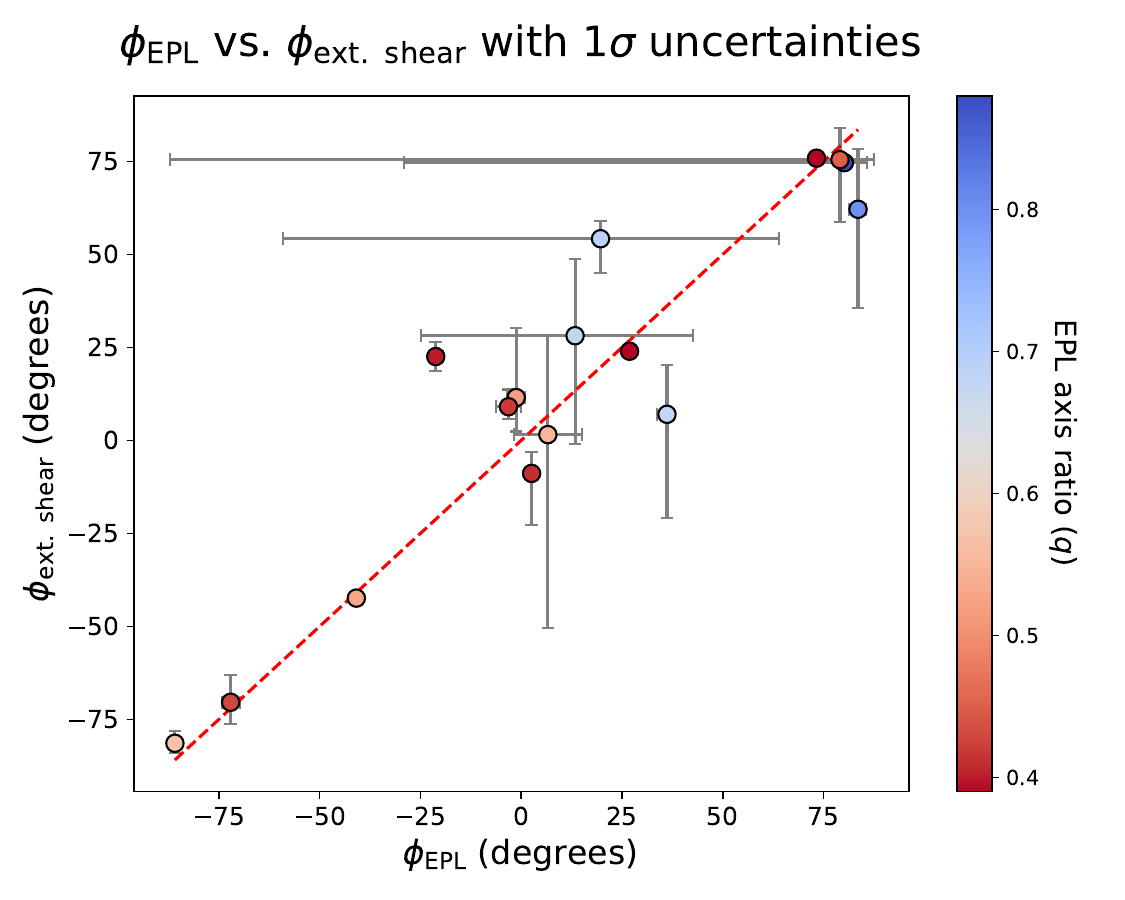}
    \caption{Position angle of the inferred EPL profile against the position angle of the inferred external shear, coloured according to the axis ratio of the EPL profile. The dashed line represents perfect alignment between these model components.}
    \label{fig:shear_alignment}
\end{figure}

We note that in the reconstructions presented in this work, most of the systems show external shear strengths of 0.05 or higher (see Table~\ref{tab:physical_quantities}), which is greater than expected \citep{treuSLACSSurveyVIII2009}.  \citet{etheringtonStrongGravitationalLensings2023} suggests, however, that the external shear commonly measured in strong lens modelling is not truly comparable to cosmic shear measurements in weak lensing, and is likely to be a compensation for our chosen mass profiles being insufficiently physical.

One way to identify where this is the case is by examining the aligment between the inferred power law and external shear components of the model. In Figure \ref{fig:shear_alignment}, we observe that our least spherical power law masses ($q<1$) have external shear position angle consistent with the EPL position angle. This suggests that the "external" shear in these cases is indicative of an \textit{internal} lensing signature that we have not modelled. We therefore concur that \textbf{the external shear posteriors we have presented in this work are likely not physically interpretable as true cosmic shear signals} comparable to those found in weak lensing studies.

\subsubsection{More complex source reconstruction}
\label{subsubsec:complexity in DESJ0150}

\begin{figure}
\centering
\begin{subfigure}{\linewidth}
  \centering
  \includegraphics[width=\linewidth]{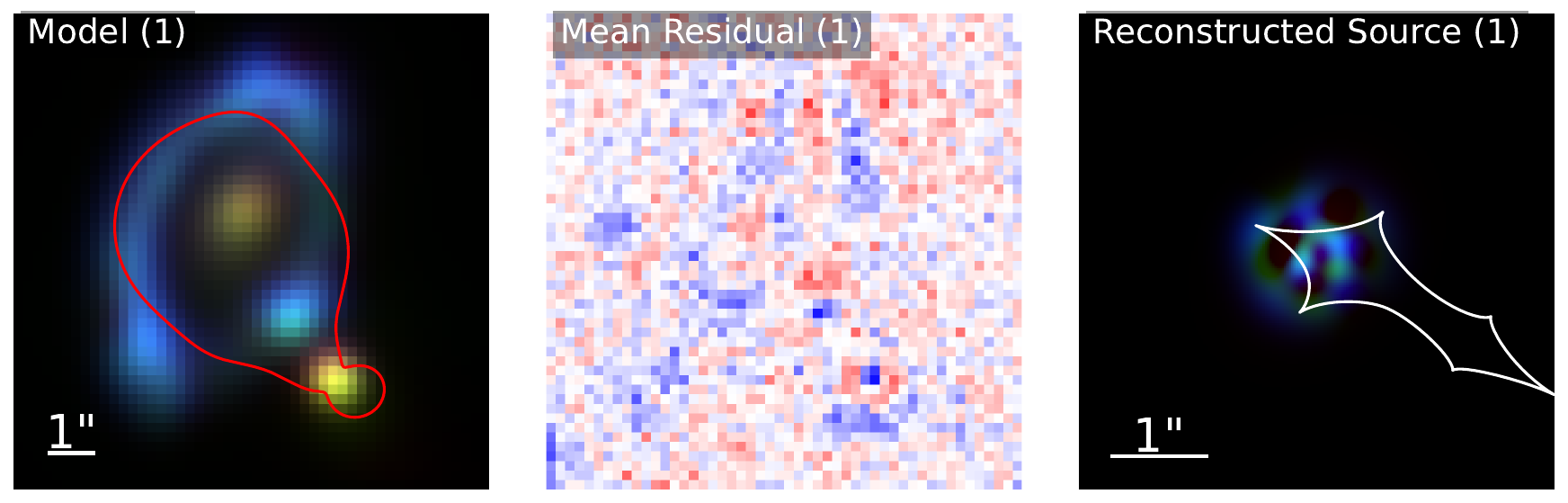}
  \vspace{-\baselineskip} 
\end{subfigure}
\begin{subfigure}{\linewidth}
  \centering
  \includegraphics[width=\linewidth]{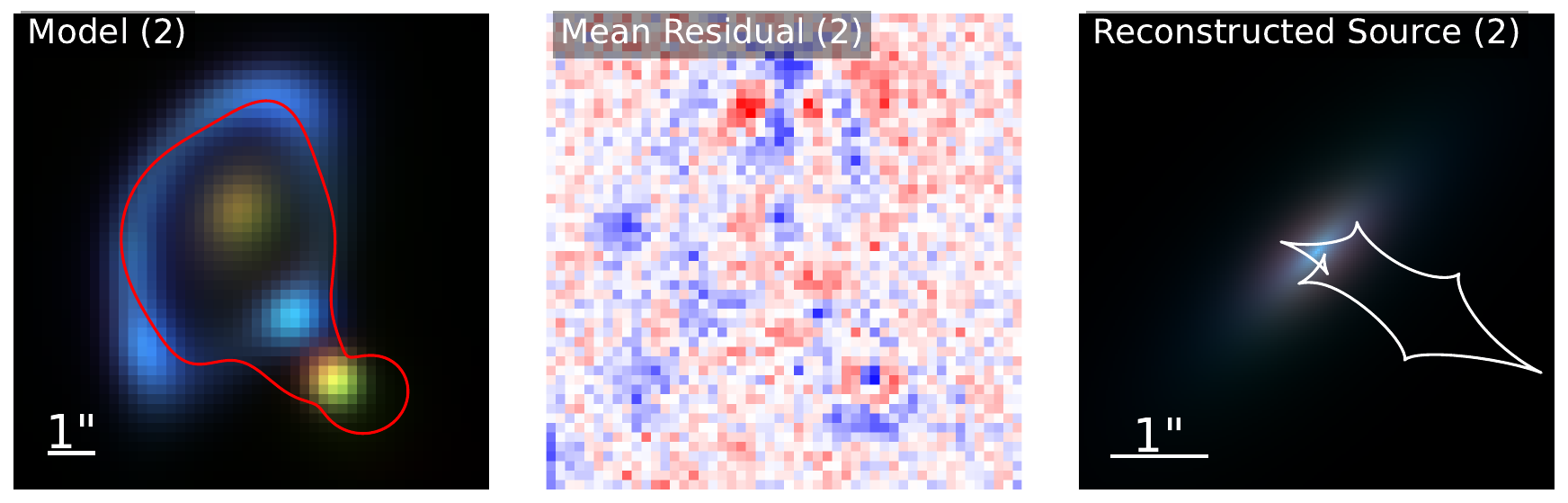}
  \vspace{-\baselineskip}
\end{subfigure}
\begin{subfigure}{\linewidth}
  \centering
  \includegraphics[width=\linewidth]{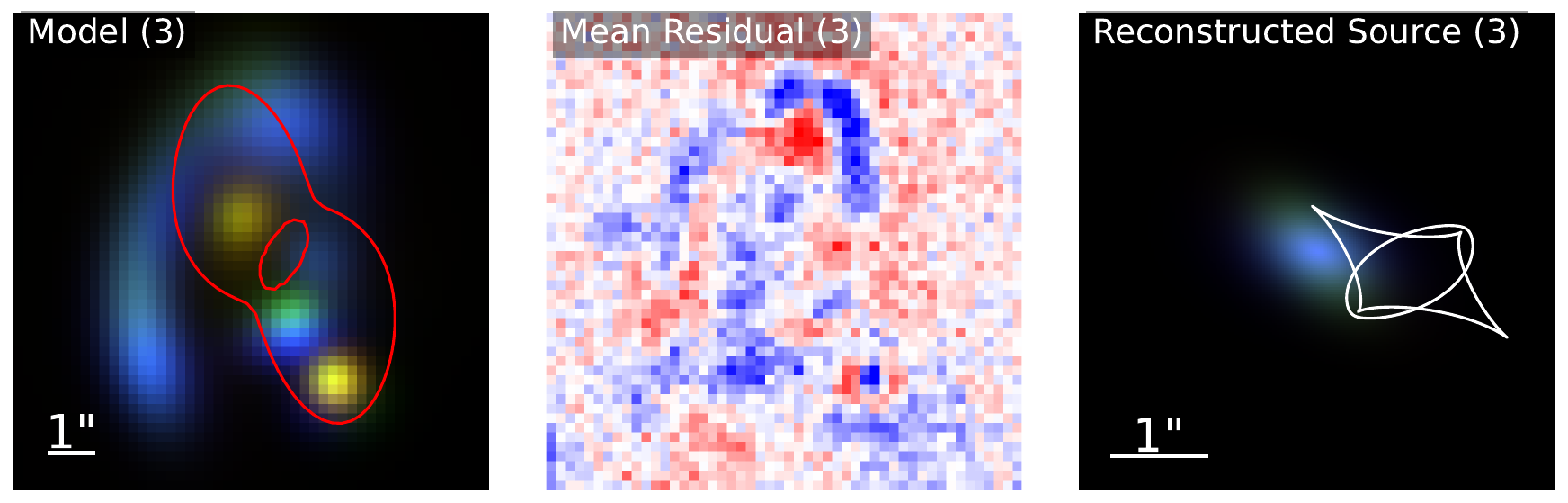}
  \vspace{-\baselineskip}
\end{subfigure}
\begin{subfigure}{\linewidth}
  \centering
  \includegraphics[width=\linewidth]{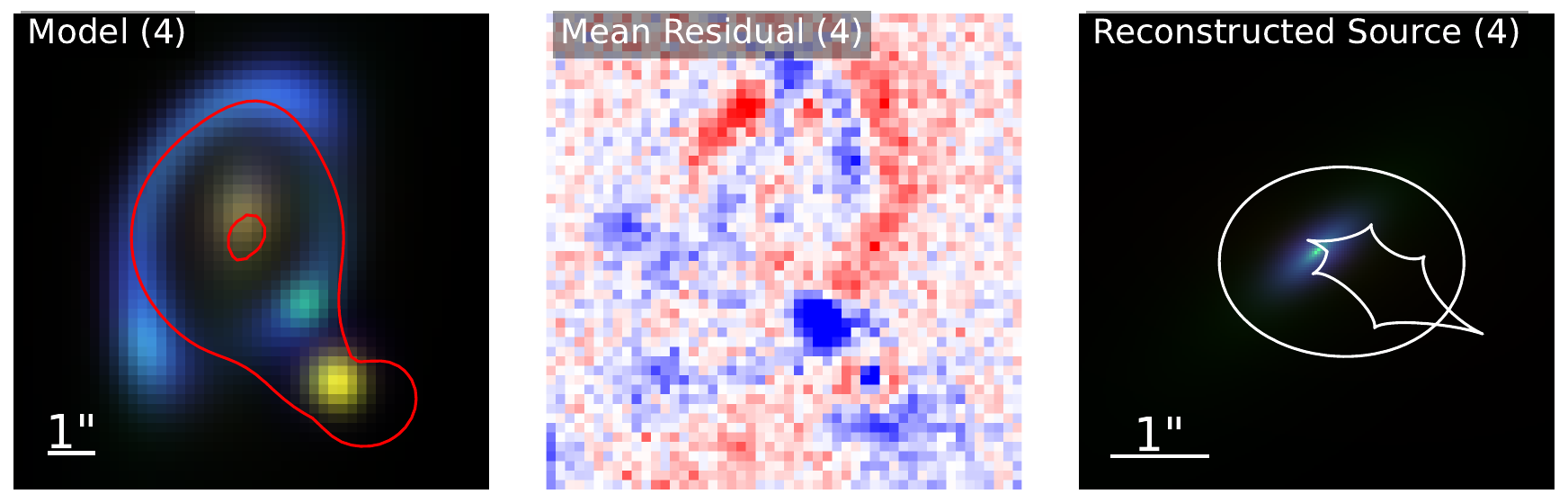}
  \vspace{-\baselineskip}
\end{subfigure}
\begin{subfigure}{\linewidth}
  \centering
  \includegraphics[width=\linewidth]{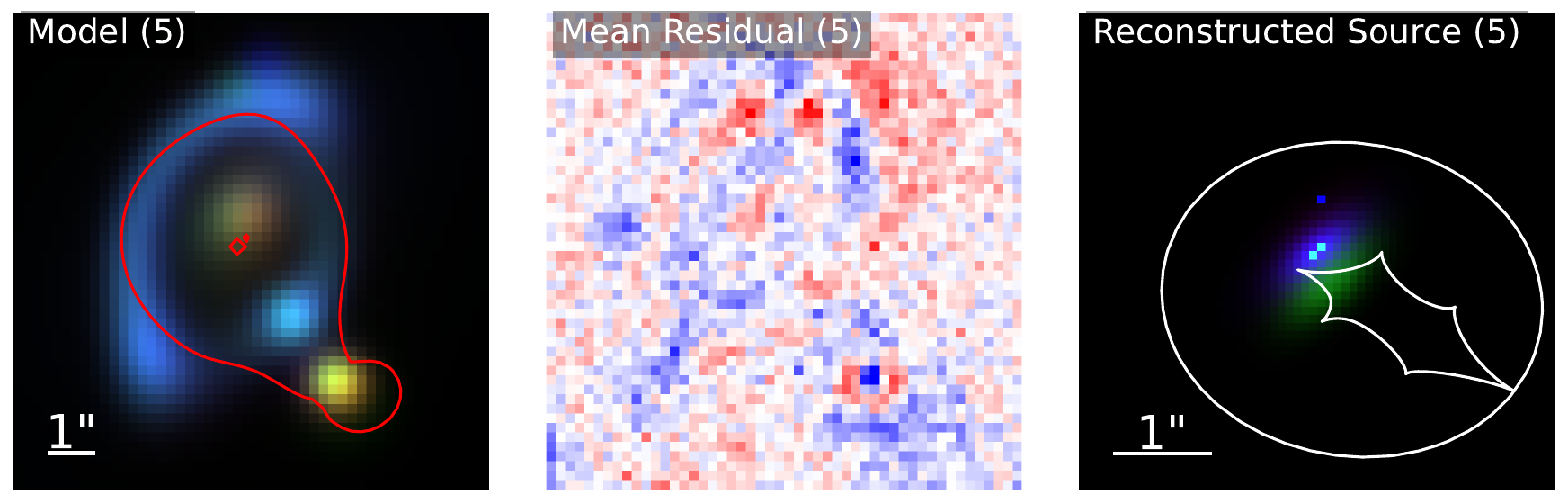}
\end{subfigure}
\begin{subfigure}{\linewidth}
  \centering
  \includegraphics[width=\linewidth]{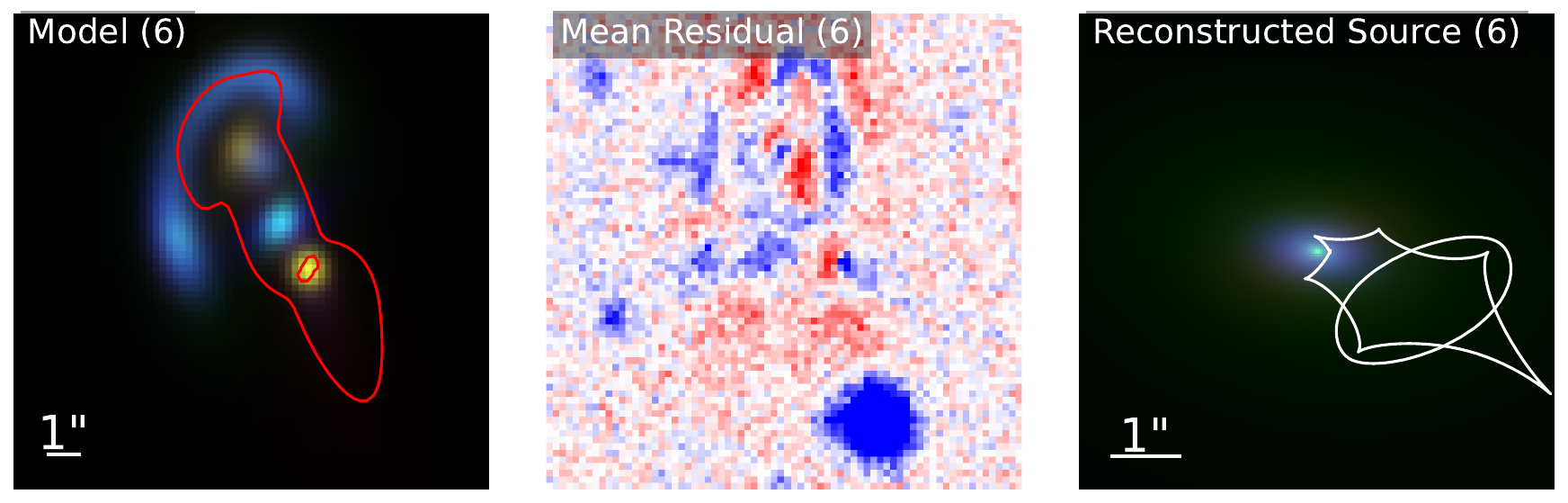}
\end{subfigure}
\caption{Reconstructions of DESJ0150-0304 under six lens models. Only the top two models (Models 1 and 2)—featuring an EPL+SIS+shear mass and three lens light components—provide more plausible reconstructions and are further refined via MCMC sampling. The remaining models (3–6) show the best-fit results from the PSO stage and fail to capture the ring morphology adequately.}
\label{fig:complexity-DESJ0150}
\end{figure}

An especially complex system in our sample is system 7 (DESJ0150--0304), which has an elongated ring–like morphology. In particular, a relatively bright region along the ring, located between two red blobs, is consistently brighter in the data than in our model. We applied our satellite mass and external flexion extensions to our pipeline to this system, and performed further tests. This included adding a foreground light profile whose position coincides with this problematic part of the ring (models 1, 2, 3, 5, 6), and the incorporation of blue objects in the further field of view as candidate fifth images for the source (model 6), as in a HU--like lens. We find that the latter scenario is implausible, while the former provides more satisfactory fits.

None of our models faithfully reproduces the observed brightness variations along the ring, despite having explored all mass model complexities that are plausibly constrainable with this dataset. We therefore tested whether added \textit{source} complexity could improve the fits by replacing the Sérsic profile with a shapelet basis in the $r$ and $g$ bands (model 1), where the source has the high signal--to--noise ratio. Varying $n_{\text{max}}$ with a PSO search showed that $n_{\text{max}}=6$ yielded the best likelihood.

However, running a fully converged PSO+MCMC pipeline on this system with $n_{\text{max}}=6$ gave us an unphysical appearance for the source, with gaps and disconnected features, despite the residuals in the $r$ and $g$ bands being improved. This could suggest either an unfocused lens model or an elaborate source plane morphology that is not resolvable within the resolution and seeing of this PISCO data. Representative reconstructions for these models are shown in Figure \ref{fig:complexity-DESJ0150}, with detailed descriptions of each model provided in Appendix \ref{app:individual_7}.

While shapelets can outperform Sérsic models for the source in likelihood, \textbf{our shapelet source reconstructions do not yield robust astrophysical interpretations}. Better priors that impose focused sources, or employing pixelated or exponential shapelet reconstructions that can capture cuspy features, may help, though higher-quality data would still be desirable in the latter case. Nevertheless, given the low-quality spectroscopic redshifts (see Appendix~\ref{app:individual_7} for details), there remains a small possibility that DESJ0150–0304 is not a genuine lens.

\subsection{Can our pipeline model strong lenses beyond galaxy--galaxy scale?}

One of the lenses in our sample, system 10 (DESJ0250--4104) is embedded within an environment populated by many other massive objects. Unsurprisingly, we observe an anomolously high external shear magnitude as a result, indicating that there is massive structure unaccounted for along the line of sight. Despite this, we are still able to measure the mass distribution of the main deflector, assuming that it can be represented by a power law. Our most reliable measurement for its mass is $6.8^{+0.9}_{-1.3} \times 10^{11} \mathrm{~M_\odot}$. This places the object on the lower end of masses expected for the bright central galaxy of a cluster (though is still feasible), or indicates that this is a group--scale lens system.

Computationally, our pipeline is not well--optimised to fit models to data spanning more pixels than the thumbnails in the top panel of Figure \ref{fig:overview_16}. These cutouts, excluding system 10, range from between $40-60$ pixels per side. System 10 has a thumbnail cutout which is 90 pixels per side, yet this still excludes a number of bright objects in the nearby environment. To make the modelling more tractable, we therefore apply a mask to this data, selecting the pixels that are most valuable to the lens model (see Appendix \ref{app:individual_10} for details).
However, note that for less computationally demanding lenses, including the fainter regions may still be beneficial, as regions that are absent of bright features also serve as valuable lens model constraints.

Our pipeline is not equipped to model the cluster--scale dark matter halo, thus interpretations of the inferred masses of such deflectors should be taken with caution. Nevertheless, for this specific system, with some manual effort invested to define an appropriate mask, the pipeline is able to successfully reproduce the shape of the observed arc. In the future, it could be possible to use automated tools to mask for the highest signal--to--noise ratio features, similar to the pipeline implemented in \textsc{dolphin} \citep{shajib2025dolphinfullyautomatedforward}.

\subsection{Is our pipeline automated enough for future survey data?}

Our current pipeline provides a semi-automated workflow for reconstructing strong lens systems, starting from a fiducial model and incrementally adding complexity as needed. All sixteen PISCO systems in this study were processed within this framework, with PSO+MCMC runs converging efficiently within $\mathcal{O}(10^{5})$ evaluations. The pipeline simultaneously fits multi--band data, naturally accounting for inter-band shifts and rotations, and can be readily extended to incorporate additional bands from different instruments without modifying the core workflow. 

Evaluation of residuals, for example through correlations between pixels, and the incremental addition of model components can be formalized into quantitative criteria, providing a systematic framework that supports automated application to larger datasets. 
Recent studies have demonstrated that machine-learning-assisted component recognition \citep[e.g., \textsc{DOLPHIN};][]{shajib2025dolphinfullyautomatedforward} and Bayesian neural network approaches \citep[e.g., \textsc{LEMON};][]{Gentile_2023} can significantly accelerate traditional lens modeling. Furthermore, GPU-accelerated modeling engines with automatic differentiation and advanced sampling methods, such as gradient-based Hamiltonian Monte Carlo (HMC), offer an efficient pathway to scale multi--band lens reconstruction to the large samples expected from upcoming surveys. In the future, our pipeline could be ported to such frameworks, further enhancing automation and computational efficiency while retaining flexibility for complex systems.

\section{Conclusions}
\label{sec:conclusion}

We have developed and applied a \textsc{Lenstronomy}--based pipeline to jointly model multi--band imaging of sixteen AGEL systems observed with Magellan--PISCO, as a demonstrative analogue for future LSST--quality data before high--resolution follow--up. Our conclusions are as follows:
\begin{enumerate}
    \item \textbf{Lens recovery} — Fifteen out of sixteen candidates in this sample can be reproduced with viable, well established lens modelling formalism, supporting their identification as genuine strong lenses. The system DESJ0150-0304 remains an outlier, with a mild chance of still being a false positive. We demonstrate that lens modelling of ground-based imaging thumbnails can help validate CNN-selected strong lens candidates and identify systems that merit higher--resolution follow--up. One particularly interesting such system is DESJ0533-2536, which we report as the first galaxy--galaxy scale system identified as a hyperbolic umbilic candidate.
    \item \textbf{Role of colour information} — Simultaneous modelling of four bands of ground--based data significantly  improves constraints on the lens mass distribution compared to single-band fits. This indicates that colour information is likely to carry genuine constraining power in future ground--based observations of lens candidates with e.g. LSST. By modelling all bands, we acheive an average uncertainty of about $2.2\%$ uncertainty on Einstein radius for the PISCO sample (estimated from Table \ref{tab:physical_quantities}).
    \item \textbf{Mass model complexity} — We assess the extent to which model complexity can be constrained with PISCO-quality data. Satellites can be included when visible, and priors should allow their contribution to lensing to vary. 
    Higher-order lensing signals may be detected with ground-based data in some cases—for instance, an external flexion term is favoured in DESJ0202–2445—but are not guaranteed for more typical systems. Overall, PISCO-quality imaging provides meaningful constraints on lens mass and source morphology, while more detailed modeling of complex morphologies remains limited by resolution and seeing.
\end{enumerate}

Our methodology and findings highlight the importance of fully exploiting multi--band lens modeling to maximize the scientific return from large upcoming surveys. 
For the $\mathcal{O}(10^4)$--$\mathcal{O}(10^5)$ lenses expected from Rubin, Euclid, CSST, and Roman, automated and uniform modeling with high success rates will enable large-scale statistical studies, from population-level analyses of lens properties \citep{Tan_2024, 2025arXiv250315329E, 2025AJ....170...44E} to forecasts for cosmological parameter inference \citep{10.1093/mnras/stad3514}. 
At the same time, a comparatively small but scientifically rich subset of systems will continue to require dedicated high-quality imaging and carefully hand-crafted modeling. To study these key systems in greater detail and quantitatively assess the robustness of inferences drawn from PISCO imaging, we will incorporate high-resolution HST follow--up data \citep{tranAGELSurveyStrong2023, baroneAGELSurveyData2025} in future work.

\section*{Acknowledgements}


We sincerely thank the referee for careful reading our manuscript and providing constructive comments. HQ gratefully acknowledges support from the International Research Training Program Scholarship, the ASA Student Travel Assistance Scheme, and funding from the Dark Matter under Gravitational Lensing workshop in Hong Kong and the Scaling up Lensing 2025 workshop in Liège. HQ also thanks the MaxEnt and Bayesian Association of Australia Inc. (MBAA) for sponsorship to attend MaxEnt 2025 in Auckland. 
DJB and GFL acknowledge the support of the Australian Research Council Discovery Project DP230101775.
SMS acknowledges funding from the Australian Research Council (DE220100003).
TJ gratefully acknowledges support from the National Science Foundation through grant AST-2108515, the Gordon and Betty Moore Foundation through Grant GBMF8549, and a UC Davis Chancellor's Fellowship.
KVGC was supported by NASA through the STScI grants JWST-GO-04265 and JWST-GO-03777.
Parts of this research were conducted by the Australian Research Council Centre of Excellence for All Sky Astrophysics in 3 Dimensions (ASTRO 3D), through project number CE170100013. DB acknowledges support by the Australian Research Council Centre of Excellence for All Sky Astrophysics in 3 Dimensions (ASTRO 3D), through project number CE170100013.

We thank Richard Massey, David Lagattuta, Daniel Gilman, Ashish K. Meena, Jasjeet Singh Bagla, and Tian Li for their insightful discussions and valuable advice. We also thank Wolfgang Enzi, Natalie Lines, and Rafaela Gsponer for their helpful comments on the abstract. We also thank Lana Eid, Laura Uronenfor, Brendon Brewer, and Yannis Kalaidzidis for helpful discussions during the workshops.

The authors acknowledge the facilities, and the scientific and technical assistance of the Sydney Informatics Hub at the University of Sydney, and in particular access to the high performance computing (HPC) facility \textit{Artemis}, which has contributed to the results reported in this paper. This research was also undertaken with the assistance of resources from the National Computational Infrastructure (NCI Australia), an NCRIS enabled capability supported by the Australian Government.  

Software: \textsc{Astropy}, \textsc{Lenstronomy}, \textsc{Photutils}, \textsc{OpenCV}, \textsc{Source Extractor}, \textsc{Corner}, \textsc{NumPy}, and \textsc{Matplotlib}.

\section*{Data Availability}

The imaging data underlying this article were obtained with PISCO on the Magellan telescope and are not publicly available. The reconstructed lens models generated in this study can be shared upon reasonable request to the corresponding author.




\bibliographystyle{mnras}
\bibliography{reference, dans_references} 




\appendix 

\section{Individual Lens System Descriptions}
\label{app:individual_lenses}

We provide system-by-system commentary to highlight the specific morphology and characteristics pertaining to each strong lens system and state the lens model settings required to account for them. The numbered subheadings for each system correspond to the labels (1–16) placed on the panels in Figure \ref{fig:overview_16}

\subsection{DESJ0003-3348}

For this lens system, based on the fiducial model, modeling the main deflector light with two elliptical Sérsic profiles significantly improves the fit compared to a single elliptical Sérsic, and is therefore adopted for the reconstruction. In addition, an elliptical Sérsic light component is added to account for a green blob close to the smallest arc.
The source light is modeled using one elliptical Sérsic in each band. 

Three mass models are explored: Model 1 (EPL + external shear), Model 2 (EPL + SIS + external shear, with the SIS component free to vary in position) and Model 3 (EPL + SIS + external shear, with the SIS component constrained to the position of the satellite galaxy as inferred in the r-band). 

\begin{figure*} 
    \centering
    \begin{overpic}[width=1\textwidth]{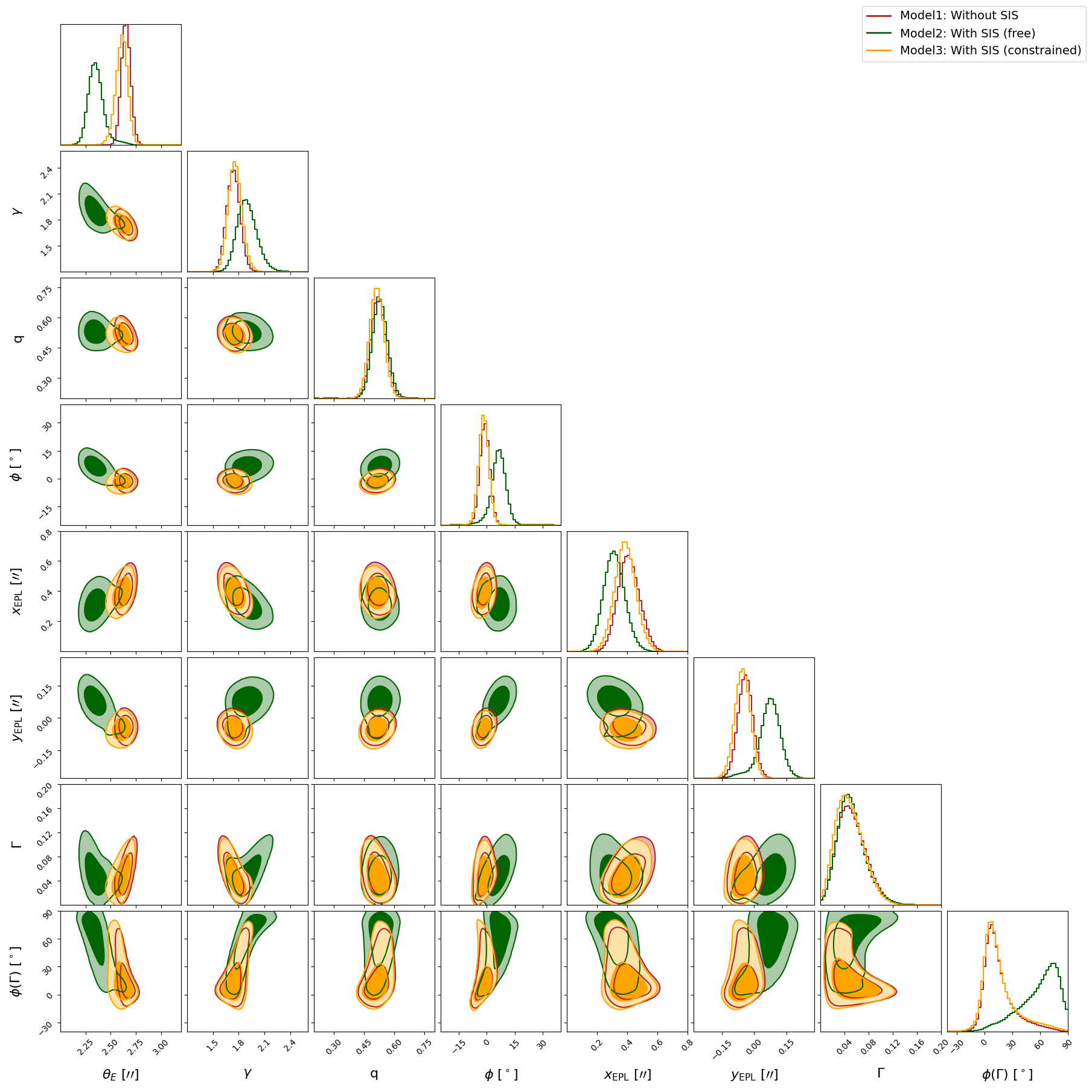}
        \put(290,270){\includegraphics[width=0.4\textwidth]{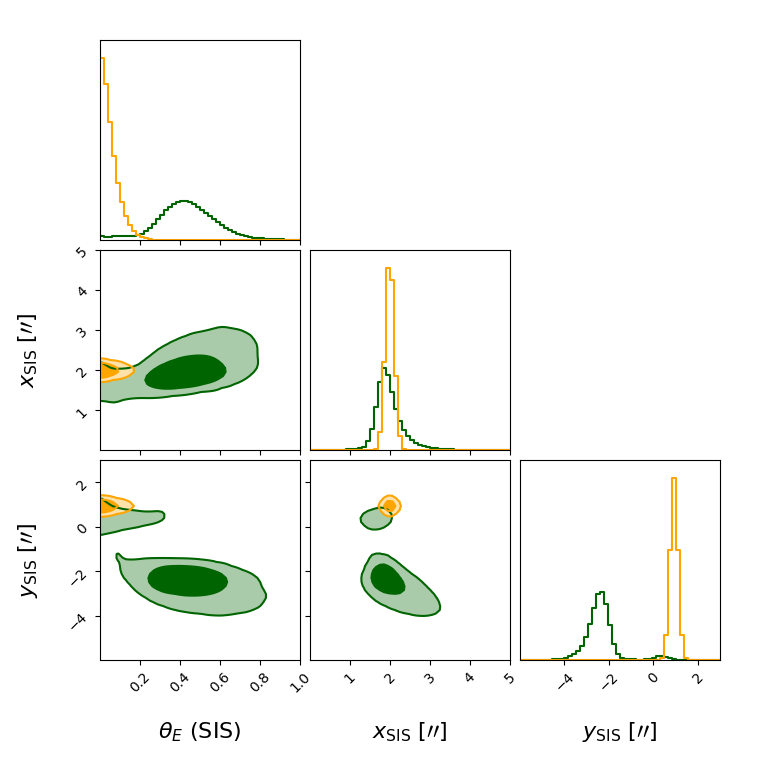}}
    \end{overpic}
    \caption{This figure presents the $1\sigma$ and $2\sigma$ posterior contours of the mass parameters for three lens models for the system DESJ0003-3348. Red corresponds to Model 1 (without an SIS mass), green to Model 2 (with a freely varying SIS component), and orange to Model 3 (with the SIS position constrained to the satellite light). The main 8×8 corner plot shows the posteriors of the EPL component and external shear, while the inset in the upper-right corner displays the posterior distributions of the SIS parameters.  The similarity in the main lensing mass configuration between Model 1 and Model 3, along with the SIS Einstein radius clustering near zero in Model 3, suggests that the data do not strongly support the presence of a satellite mass. In Model 2, the SIS position remains broadly distributed and deviates significantly from the satellite light position, and its Einstein radius is weakly constrained. This suggests that if a dark substructure exists, its parameter is not well constrained by the data.}
    \label{fig:pdf_desj0003}
\end{figure*}

As seen from Figure~\ref{fig:comparison_model_desj0003}, in Model 2, the SIS shifts away from the visible satellite toward a region without significant light, giving only a modest improvement in BIC ($<10$). The posterior distributions (Figure~\ref{fig:pdf_desj0003}) further show that the SIS parameters are poorly constrained, suggesting overfitting rather than detection of a real mass clump. In Model 3, the SIS fixed to the satellite centroid contributes only a weak perturbation and offers no BIC improvement, with $\theta_{E,\mathrm{SIS}}$ sharply peaked near zero ($0.03^{+0.06}_{-0.03}$ at 68\% confidence). We therefore find no strong evidence for a significant satellite mass in this system.

We additionally investigated third and fourth--order multipole terms \citep{1992grle.book.....S, xu2014colddarkmattersubstructuresaccountobserved} and external flexion to introduce higher-order perturbations in the lens mass distribution. While these models yielded BIC improvements of approximately 20–50, 
they also introduced unphysical distortions in the mass distribution. 

Note that the lensed arc signal in the $z$-band for this system is very weak, which can lead to noise-driven colour centroids and large uncertainties in location-related parameters. Therefore, the $z$-band is omitted in the RGB plot (R = $i$, G = $r$, B = $g$) in Figure~\ref{fig:overview_16} and Figure~\ref{fig:comparison_model_desj0003}. All inferred parameters, however, are still based on the full four-band dataset.

\subsection{DESJ0010-4315}
In this system, two extra blue blobs -- one above and one to the lower left of the main lens -- are included in the lens light modeling but not treated as lensing perturbers due to their relatively small sizes and significant angular separation from the lensed source. Additionally, the two satellite galaxies exhibit very blue colors, suggesting they may originate from higher redshifts. 

The reconstructed source appears compact, with a large Sérsic index ($n\sim 6 - 9$ between the i, r and g bands, where the source is bright). Positioned near the fold caustic, it produces a fold quad lens configuration, with the fold pair on the left side of the lens system.

\subsection{DESJ0101-4917}
This system is modeled using the fiducial mass model. 

The fitting result yields a steeper--than--isothermal EPL component with an Einstein radius of $\theta_E = 2.22^{+0.01}_{-0.01} \ ^{\prime \prime}$, while the strength of the external shear $\gamma^{\text{\small ext}} = 0.22^{+0.02}_{-0.02}$ indicates possible influence from galaxies close to the line of sight.

The de-lensed source has a small Sérsic radius ($0.06^{+0.05}_{-0.03}\,^{\prime\prime}$ in r band), and lies across the inner astroid-shaped tangential caustic, which leads to the merging of two images on the left along the tangential critical curve in the lens plane. 

\subsection{DESJ0120-1820}

This system appears as a naked-cusp configuration \citep{1997ApJ...486..681M, Lewis_2002}, where three merging images appear close together on one side of the lens, forming a bright, crescent-shaped arc, while the expected counter-image on the opposite side is either highly demagnified or absent.

\begin{figure}
    \centering
    \includegraphics[width=1\linewidth]{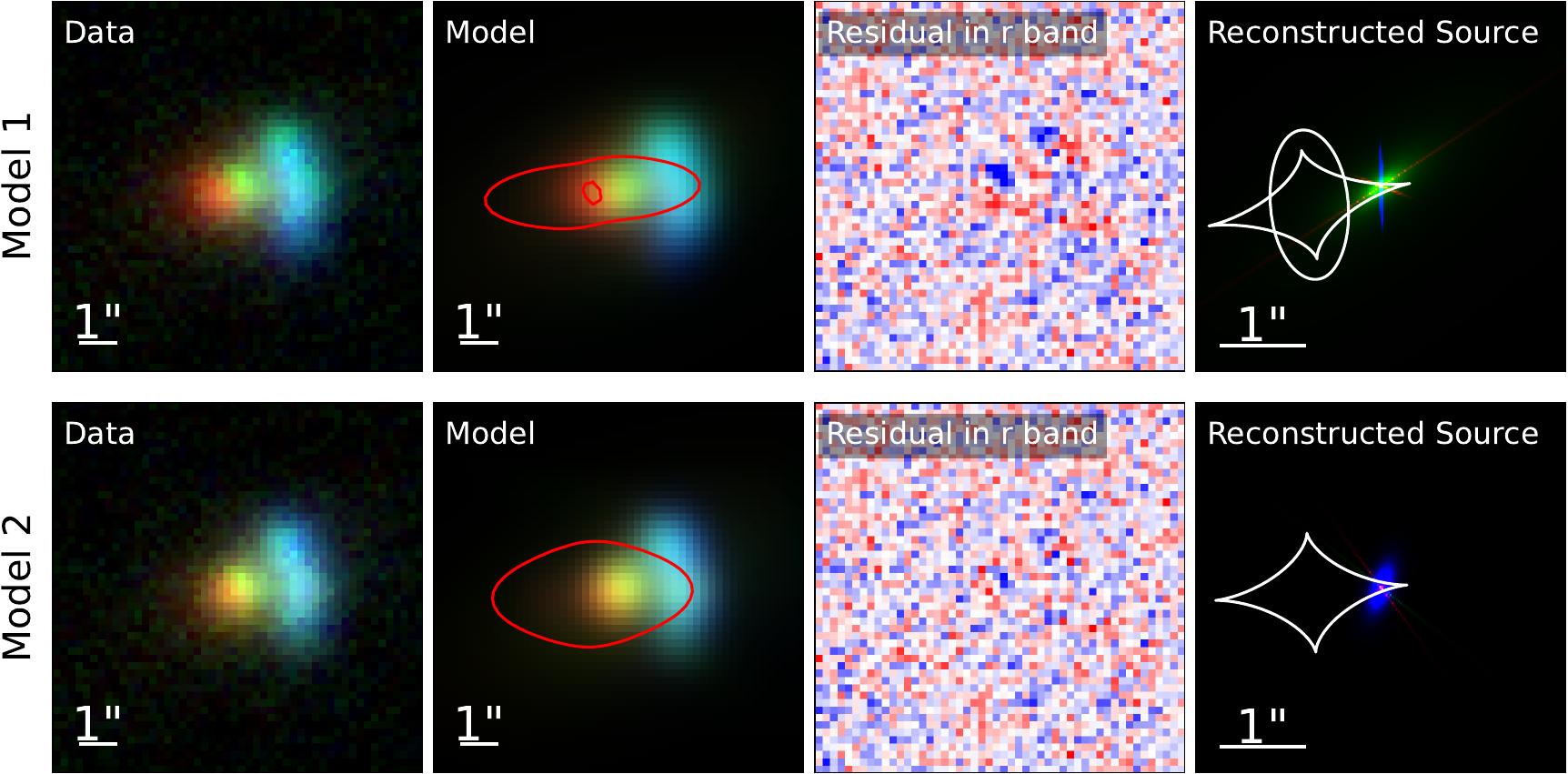}
    \caption{Reconstruction based on the maximum-likelihood results for Model 1 and Model 2 of system DESJ0120-1820. Model 1: the fiducial model, Model 2: adds a second Elliptical Sérsic based on Model 1. From left to right, columns display the observed data, model image, normalized residuals in the r band (at the 3$\sigma$ level), and source-plane reconstruction. Small differences between the data for Model 1 and Model 2 arise from the slightly different shifts and rotations inferred from their respective maximum-likelihood results and applied during band stacking, which also explains the minor variations seen the model comparison of other systems.}
    \label{fig:2models_DESJ0120}
\end{figure}

We reconstruct this lens system starting with the fiducial model, and it leaves noticeable residuals near the crescent-shaped lensed arc in the r-band, which is the band captures significant features of lensed light.
To improve the fit, we adopt Model 2, which retains the same mass and source model but introduces a second elliptical Sérsic profile for the lens light. This significantly reduces residuals (see in Figure~\ref{fig:2models_DESJ0120}) and improves the best-fit with a change in likelihood $\Delta\log\mathcal{L}=177$ , with the BIC changing by $\Delta\text{BIC}=43$ in favour of Model 2.

Posterior comparisons reveal that Model 1 favors a more flattened and elliptical mass distribution and stronger external shear ($\gamma^{\text{\small ext}} \sim 0.23$), whilst Model 2 prefers a heavier, more spherical mass profile and weaker shear ($\gamma^{\text{\small ext}} \sim 0.04$). This suggests that Model 1 might have compensated for an oversimplified light model by artificially increasing the ellipticity in the mass model.

In both models, the source lies near the cusp of the astroid caustic, resulting in a highly magnified, crescent-shaped arc along the critical curve—appearing somewhat fuzzy due to PSF blurring.

\subsection{DESJ0141-1303}
This system includes a nearby satellite galaxy within the cutout, which is modeled using an additional elliptical Sérsic component added to the fiducial model. The data quality is relatively poor, with a signal-to-noise ratio of only $\sim 20$–$30$—substantially lower than the $\gtrsim 50$ typically seen in other systems. This leads to correspondingly larger uncertainties in the inferred parameters.

The reconstruction reveals a diffuse source with a relatively big Sérsic radius of $0.99^{+0.57}_{-0.14}\,\mathrm{arcsec}$ in the r band, located slightly off-center relative to the inner caustic, resulting in an asymmetric ring morphology. The mass distribution is dominated by a nearly circular EPL lens with an Einstein radius of $\theta_E = 1.77^{+0.07}_{-0.07}\,\mathrm{arcsec}$, and moderate external shear of $\gamma^{\text{\small ext}} = 0.12^{+0.07}_{-0.06}$.

\subsection{DESJ0142-1831}
\begin{figure}
    \centering
    \includegraphics[width=1\linewidth]{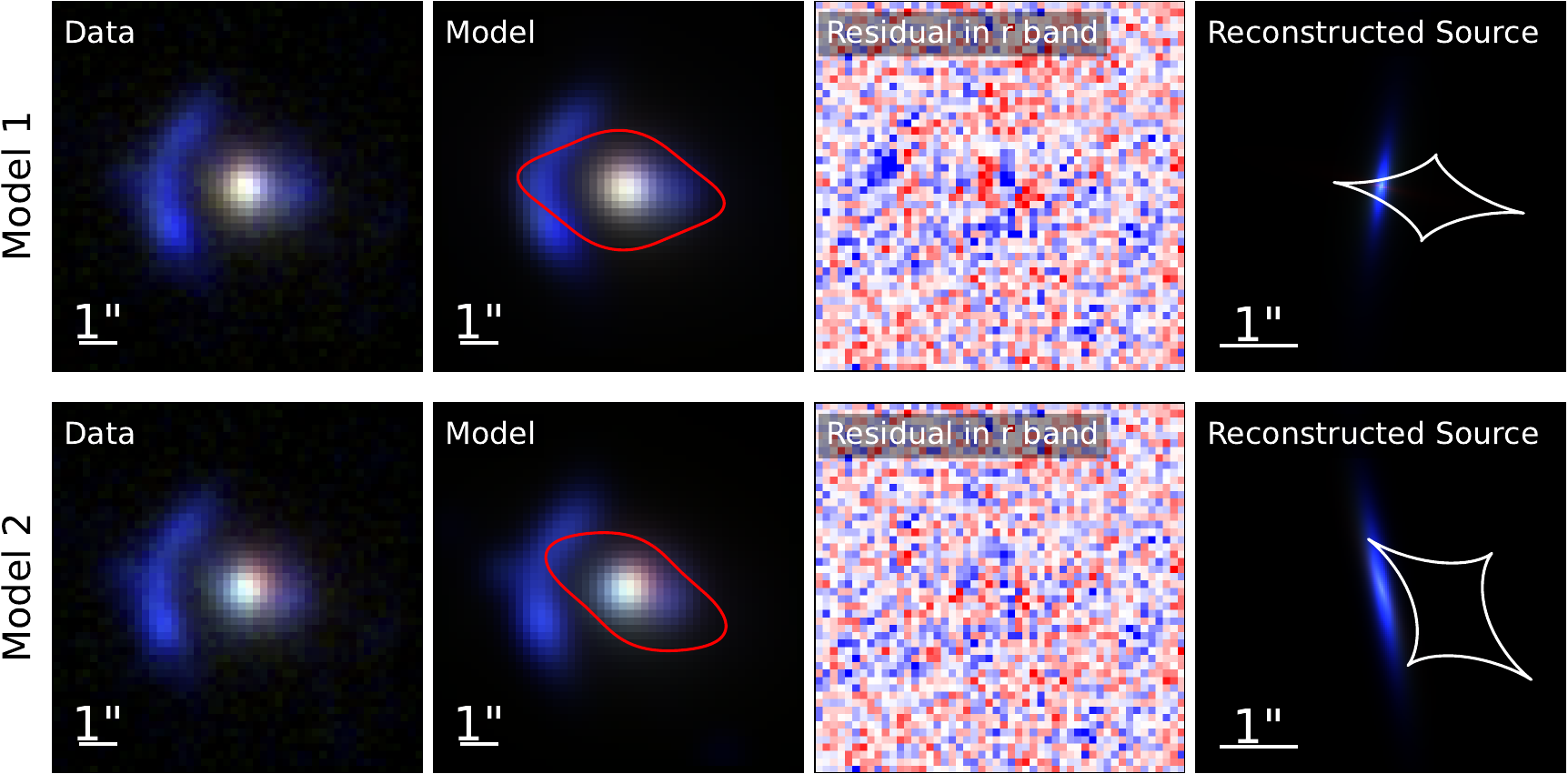}
    \caption{Similar to Figure \ref{fig:2models_DESJ0120}, the model comparison for system DESJ0142-1831. The first row shows Model 1 (fiducial), and the second row shows Model 2 (with additional Sérsic components for the deflector and a satellite near the arc). Model 2 significantly reduces the residuals, particularly the blue residual blob on the left and the central residuals seen in Model 1. In both models, the source lies near a cusp of the inner caustic.}
    \label{fig:2models_DESJ0142}
\end{figure}

For this system, we begin the reconstruction with Model 1 (the fiducial model), where significant residuals persist near the crescent-shaped lensed arc in the r and g bands in the results. These residuals do not match the elongated arc morphology and suggest the presence of a blue satellite. Additionally, a central residual pattern—characterized by inner blue and outer red features—indicates a misrepresentation of the deflector’s light profile, motivating the introduction of an extra Sérsic component for the main deflector light. Model 2 addresses these two limitations and effectively reduces the residuals at aforementioned locations, with the BIC significantly favouring this model by $\Delta\text{BIC}=693$.

Figure~\ref{fig:2models_DESJ0142} illustrates the model comparison, where two models yield noticeably different mass distributions: Model 1 prefers a more extended and elongated profile with a steeper density slope, whereas Model 2 results in a less compact profile with a smaller Einstein radius and stronger external shear, possibly reflecting substructure or environmental tidal effects. 
Despite these differences, both models recover extended sources near the cusp of the inner caustic.
This configuration gives rise to a prominent crescent-shaped arc on one side of the lens and a fainter counter-image on the opposite side.

\subsection{DESJ0150-0304}
\label{app:individual_7}
We explored a series of models to investigate the complex structure of system 7 (DESJ0150-0304) and assess its viability as a strong gravitational lens system. The models differ in both mass and source components, as well as the treatment of the lens light. From top to bottom in Figure~\ref{fig:complexity-DESJ0150}, the models are as follows:

Model (1): Mass is modeled with an elliptical power-law (EPL), a singular isothermal sphere (SIS), and external shear. The source is modeled using a single elliptical Sérsic profile in the \textit{z} and \textit{i} bands, and shapelets in the \textit{r} and \textit{g} bands to improve residuals. The lens light is modeled with three components: two corresponding to the red blobs, and one at the lower-right bright cyan feature on the ring.

Model (2): Shares the same mass and lens light configuration as (1). The source is described by a single elliptical Sérsic profile across all four bands.

Model (3): The mass model consists of only the EPL profile and external shear compared with (2).

Model (4): Similar to Model (2), but without accounting for the lens light contribution from the bright cyan blob on the ring.

Model (5): Same mass and lens light model as Model (1) and (2), The source is modeled with one Sérsic in the \textit{z} and \textit{i} bands, and two Sérsic components in the \textit{r} and \textit{g} bands.

Model (6): Attempts to include a possible counter-image. Uses EPL+SIS+shear for mass, and a single Sérsic source across all bands.

All models shown correspond to best-fit solutions, with Models 1 and 2 further refined using MCMC, while Models 3–6 remain at the PSO best-fit stage. The modeling results reveal several key insights: 

\begin{enumerate}
    \item Two distinct mass clumps are favored: Model (3) with only a single EPL and shear places the EPL between the two red blobs and results in a poorer fit.
    \item The bright cyan blob on the ring is too luminous to be accounted for solely by the lensed source. Model (4), which omits a dedicated lens light component for this feature, leaves significant residuals at its location.
    \item Increasing source complexity (e.g., shapelets in Model 1) reduces residuals, but often produces unphysical source morphologies with spurious holes. A two-Sérsic source (Model 5) slightly yields a modest likelihood improvement but also increases BIC.
    \item Model (6), which attempts to incorporate a potential counter-image in larger field, fails to produce a good fit. This suggests either spurious or incompatible with the assumed mass configuration.
\end{enumerate}

Based on current results, no model can convincingly reproduce all observed features without introducing physical implausibilities in the source. In additional tests not shown here, introducing further mass complexity — such as external flexion, multiple EPL profiles, or extra SIS components — did not improve the fits. 

Spectroscopic redshifts from previous studies \citep[$z_\mathrm{lens}^\dagger = 0.63675$, $z_s^\dagger = 1.390$;][]{tranAGELSurveyStrong2023, baroneAGELSurveyData2025} indicate that the central galaxy and the ring do correspond to physically distinct components, tentatively supporting a strong lens interpretation. However, both measurements have low-quality flags (1 out of 3), which may be due to contamination from nearby components, including the cyan blob on the ring and the red satellite. There remains a small possibility that this system is not a genuine lens. If it is, its morphology likely requires a more flexible mass and source model, and additional spectroscopic data would be invaluable for constraining its true nature.
 
A more detailed discussion of exploring the complexities is presented in Section \ref{subsubsec:complexity in DESJ0150}.

\subsection{DESJ0202-2445}

This lens system exhibits a ring-like morphology, but with a more asymmetric light distribution. We adopted a fiducial model featuring a smooth elliptical power-law (EPL) mass profile and explored additional mass complexity by incorporating third and fourth order multipole and external flexion terms. The addition of multipole moments distorted the critical curve into an unphysical configuration, whereas the external flexion component adjusts the mass distribution into a more asymmetric structure, yielding a better fit. A more detailed discussion of this modeling is provided in Section 
\ref{subsubsec:asymmetries}.

As summarized in Table~\ref{tab:model_summary}, for system 8, Model 1 refers to the fiducial configuration, while Model 2 includes external flexion; the reconstruction presented in Figure~\ref{fig:overview_16} is based on Model 2. Model 2 is strongly favored, with a BIC improvement of more than 200 relative to Model 1, reflecting a substantially better fit.

Under Model 1, the reconstructed source lies near the cusp, and its extended light distribution is lensed into a ring with cusp-like features. The inferred EPL profile is elliptical and vertically aligned, with an aligned external shear with strength 0.18. However, this configuration does not fully reproduce the observed asymmetry in the Einstein ring. In contrast, Model 2 yields improved residuals in the $g$ band and produces enhanced flux in the lower portion of the ring through a lopsided critical curve. 

\subsection{DESJ0212-0852}

By adopting the fiducial model, the arc structure of this lens system is successfully reproduced.   
The source, modeled with a single Sérsic component in each band, exhibits a shallow Sérsic index ($n=0.4 \sim 1.5$), indicative of an extended disk-like galaxy. It is positioned slightly off-center but its light distribution extends across the astroid caustic, resulting in an unevenly magnified Einstein ring with localized brightness variations.

\subsection{DESJ0250-4104}
\label{app:individual_10}
\begin{figure}
    \centering
    \includegraphics[width=1\linewidth]{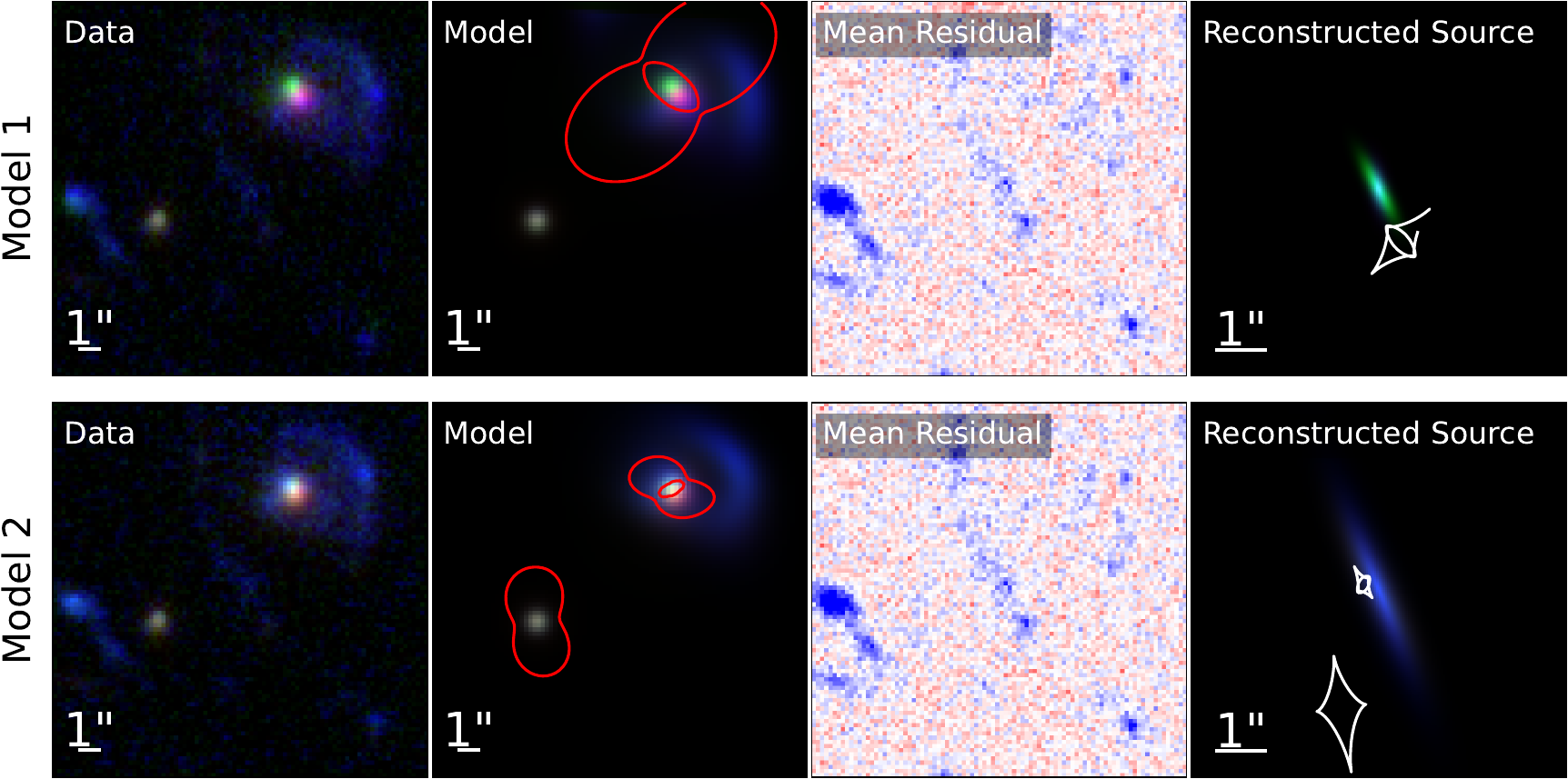}
    \caption{Similar to Figure \ref{fig:2models_DESJ0120}, the model comparison for system DESJ0250-4104. Model 1 uses an EPL+Shear mass model, while Model 2 adds an additional SIS mass component at the location of the satellite in the lower-left region based on Model 1. The long blue trails in the center and lower-left regions of the mean residuals lie outside the modelling mask and correspond to objects within the group or cluster environment. The two models show different mass distributions: Model 1 infers a massive EPL component ($\sim\!10^{12} \mathrm{~M_\odot}$) 
    oriented towards the red satellite, whereas Model 2 yields two comparable mass components (both $\sim\!10^{11} \mathrm{~M_\odot}$). 
    Although Model 2 achieves a higher likelihood, the substantial external shear ($\gamma^{\text{\small ext}} \approx 0.38$) indicates a more complex mass distribution. }
    \label{fig:2models_DESJ0250}
\end{figure}

This is a galaxy-scale strong lens embedded in a group or cluster environment. We explored two mass models: one using a single EPL profile with external shear (Model 1), and the other adding a SIS component at the red satellite (Model 2). When plotting the reconstruction results (Figure~\ref{fig:2models_DESJ0250}), the mask is removed to ensure no spurious light structures appear outside the masked region.
Model 1 infers a relatively large Einstein radius of $3.89^{+0.17}_{-0.20}\, \mathrm{arcsec}$, oriented towards the red satellite. In contrast, Model 2 yields an Einstein radius of $1.41^{+0.09}_{-0.15}\,\mathrm{arcsec}$ for the EPL component (fixed at the main deflector location) and $1.10^{+0.42}_{-0.33}\,\mathrm{arcsec}$ for the SIS component (fixed at the satellite position). In the maximum-likelihood results, Model 2 achieves a higher likelihood and lower BIC than Model 1 ($\Delta\mathrm{BIC}=368$), but requires substantial external shear ($\gamma^{\mathrm{ext}} \approx 0.38$), reflecting the complex environment of the galaxy cluster. Future work will model this system using high-resolution HST imaging.

\subsection{DESJ0305-1024}

Due to the relatively low signal-to-noise ratio ($\sim 15$–$45$ across four bands) for this system, although the fiducial model yields good normalized residuals, the inferred parameters carry large uncertainties. 
The EPL lens mass component is highly elliptical (axis ratio $q = 0.42^{+0.09}_{-0.09}$) with a relatively shallow density slope 
($\gamma^{\text{\tiny EPL}} = 1.69^{+0.13}_{-0.12}$ 
), with the external shear strength ($\gamma^{\text{\small ext}} = 0.17^{+0.05}_{-0.05}$) is moderate to strong.
The extended source overlaps more than half of the caustic structure, producing a nearly complete but slightly irregular Einstein ring.

Notably, the multi--band fitting requires large pixel shifts ($\sim 3$ pixels in $i$ and $g$ band) in the best-fit reconstruction, which significantly improves the cross-band alignment. This mitigates PSF-related distortions and enables a more accurate and self-consistent reconstruction of the system. 

\subsection{DESJ0327-3246}

The reconstruction of lens system, DESJ0327-3246, is based on a fiducial mass model with two additional elliptical Sérsic components to account for two satellite galaxies near the Einstein ring, which are identified by their redder color in multi--band imaging.

The source in the source plane is close to the center of the caustic lines, nearly directly behind the lens, resulting in a nearly complete Einstein ring with quadruply imaged features. The best-fit EPL mass component has the density slope shallower than isothermal, and its strong ellipticity elongates the lensing potential. The external shear, although aligned with the major axis of the EPL, has a relatively mild strength, indicating no significant influence from additional perturbers.

\subsection{DESJ0354-1609}

This lens system is reconstructed using the fiducial model.
The reconstructed source is revealed to cross the fold caustic, leading to the formation of a characteristic fold configuration: two or three lensed images are merging on one side of the lens, while two fainter images appear on the opposite side.
The inferred density slope ($\gamma^{\text{\tiny EPL}} =1.58^{+0.05}_{-0.15}$) deviates from an isothermal profile. This is possibly because the EPL component has a large Einstein radius ($\theta_E = 3.01^{+0.19}_{-0.03}$) and possibly captures the outer regions of the lens plane, where the dark matter halo dominates and flattens the overall mass profile. 
The lensing mass is significantly elliptical, shaping the arc structure. The external shear strength is very large ($\gamma^{\text{\small ext}} = 0.24^{+0.09}_{-0.02}$), suggesting significant lensing contribution from the wider environment. However, it is tightly aligned to the EPL position angle, suggesting more complex lensing structure internal to the lens, and further supporting the hypothesis that a more complex mass distribution than an EPL is required.

\subsection{DESJ0533-2536}

For this lens system, we advance the fiducial model by incorporating a Sérsic elliptical component in the r-band to account for a satellite galaxy in the lower part of the cutout. The system exhibits a distinctive lensed morphology, where three strongly magnified images form an unusual arc-like or ring-like structure on one side of the lens, accompanied by a fainter counter-image on the opposite side.

The inferred Einstein radius ($\theta_E = 3.47^{+0.07}_{-0.07}\ ^{\prime \prime}$) is the largest amongst all the systems in this PISCO sample, while the density slope ($\gamma^{\text{\tiny EPL}} = 1.21^{+0.03}_{-0.03}$) is significantly shallower than the isothermal value, potentially reflecting a transition into the dark matter-dominated regime once again. The external shear is also relatively strong, again suggesting notable tidal perturbations from the surrounding environment or a requirement for more internal mass distribution complexity.

The reconstructed source is compact, with a small Sérsic radius ($0.07^{+0.01}_{-0.01}\ ^{\prime \prime}$). While it lies near a cusp, this is not a typical cusp of the tangential caustic. Rather, the system exhibits a unique caustic structure closely resembling a hyperbolic umbilic (HU) lensing configuration \citep{1992grle.book.....S, 2001stgl.book.....P, Meena_2020, Meena_2023, Meena_2024}, where the radial and tangential caustics meet and interchange a cusp. We highlight and further discuss the significance of this system as the first rare HU galaxy-galaxy lens candidate in Section~\ref{subsec: HU lens}.

\subsection{DESJ2032-5658}

The reconstructed system of DESJ2032-5658 adds two additional Sérsic elliptical components in each band to the fiducial model, to account for two red blobs near the arc. These components are likely satellite galaxies rather than lensed features, as indicated by their distinct color differences compared to the lensed light. We also tested the possibility of these satellites acting as lensing perturbers during the PSO optimization process, but their inclusion did not improve the fit.

The de-lensed source is extended and has a low Sérsic index ($n = 1.05^{+0.47}_{-0.30}$ in r band), spanning the cusp region. This configuration produces a prominent crescent-shaped arc on one side of the lens and a faint counter-image on the other. The lens mass distribution is modeled as a centrally concentrated elliptical power-law (EPL) profile with weak external shear.

\subsection{DESJ2125-6504}
\begin{figure*} 
    \centering
    \begin{overpic}[width=1\textwidth]{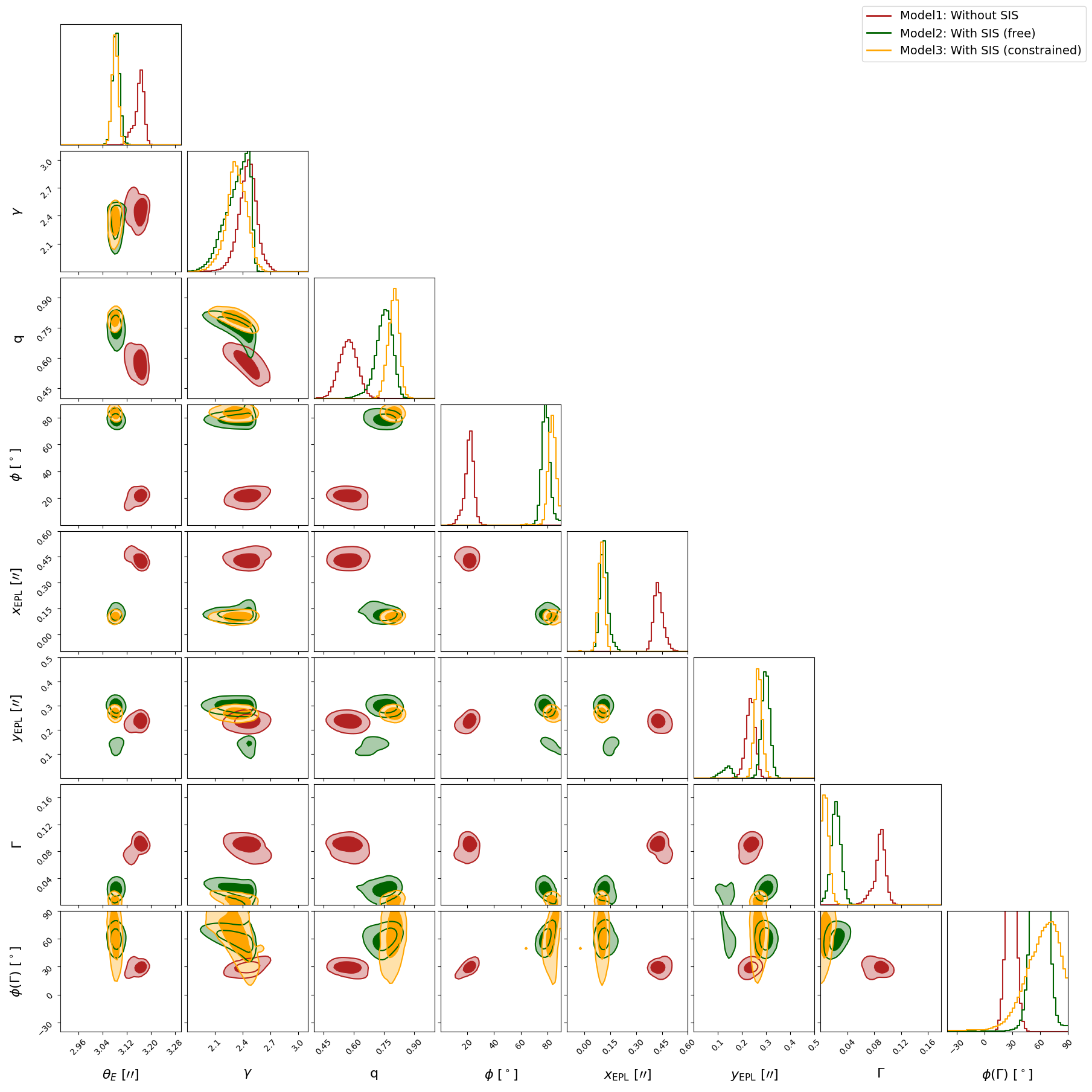}
        \put(290,270){\includegraphics[width=0.4\textwidth]{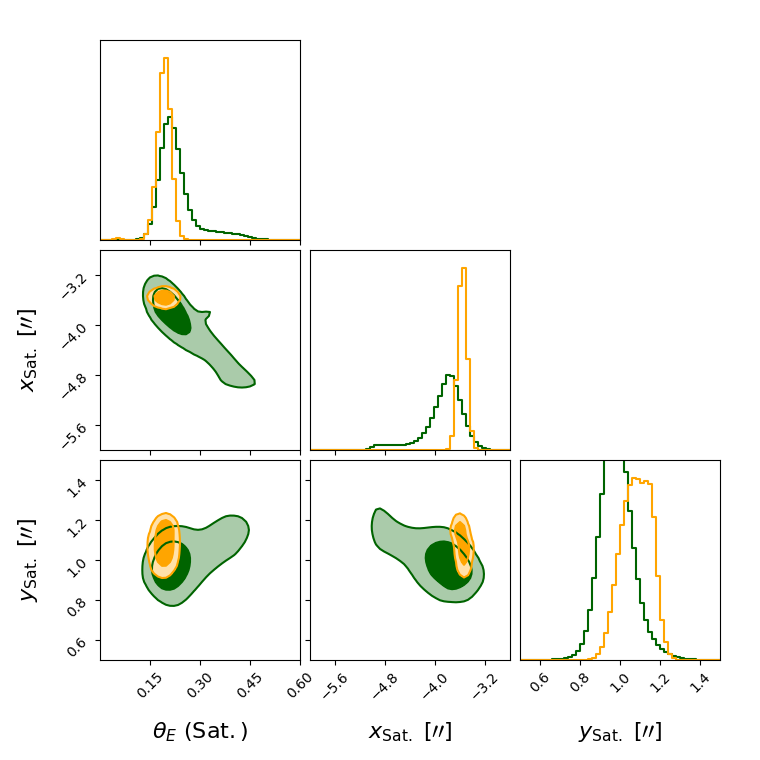}}
    \end{overpic}
    \caption{Same as Figure \ref{fig:pdf_desj0003}, but for the system DESJ2125-6504. The comparison of posteriors highlights differences in mass configurations — particularly in the Einstein radius of the EPL component. Model 1 (without a satellite mass) yields a larger Einstein radius located further to the right and a stronger external shear, compensating for the absence of the satellite mass. For a better visualization of the dominant region of the distribution, we restrict the EPL angle parameter $\phi_{\rm EPL}$ to [0°, 90°] and the external shear angle $\phi_\gamma$ to [–40°, 90°], retaining 97\% of the posterior samples in both cases.}
    \label{fig:pdf_desj2125}
\end{figure*}

This system features a nearly complete Einstein ring. To account for the prominent red blob on the left side, an additional lens light component is included. We investigated the influence of a mass component associated with this satellite in addition to the fiducial model. 
As seen in Figure~\ref{fig:comparison_model_desj2125}, the inferred SIS coincides with the visible satellite, and its inclusion alleviates the need for a displaced EPL center and large external shear otherwise required by Model 1. 
The lower BIC of Model 3 than Model 2 also suggests the utility of imposing informed positional priors in navigating the high-dimensional parameter space ($\sim$ 90 parameters).
From the posterior distributions (Figure~\ref{fig:pdf_desj2125}), the satellite mass is inferred to be $1.8^{+0.9}_{-0.5} \times 10^{10} \mathrm{~M_\odot}$
 solar masses in model2 and $1.5^{+0.2}_{-0.2} \times 10^{10}   \mathrm{~M_\odot}$ in model 3. Correspondingly, the EPL component mass decrease from $3.88^{+0.02}_{-0.05} \times 10^{12} \mathrm{~M_\odot}$ (to Model 1) to $3.687^{+0.019}_{-0.019} \times 10^{12} \mathrm{~M_\odot}$ and $3.679^{+0.012}_{-0.013} \times 10^{12} \mathrm{~M_\odot}$, respectively.

\section{Mock data test}
\label{app:mock_data_test}

\begin{figure}
    \centering
    \includegraphics[width=0.5\linewidth]{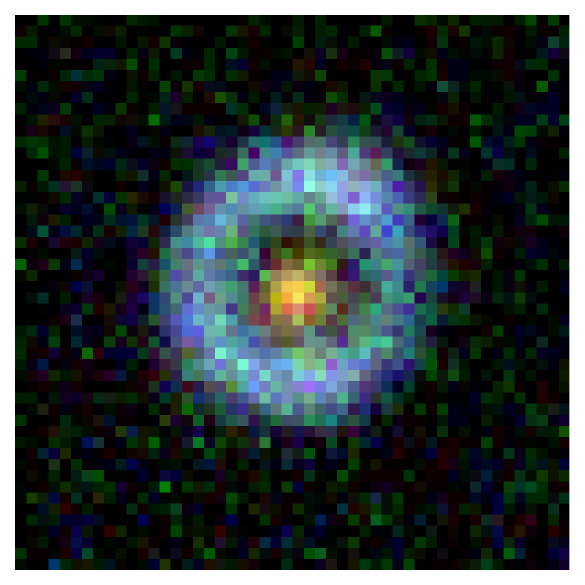}
    \caption{Simulated four-band colour composite of a mock strong lens system used for validating the multi-band modelling pipeline. The mock was generated using the fiducial model described in the main text and includes realistic observational effects chosen to resemble the characteristics of the PISCO imaging.}
    \label{fig:mock system}
\end{figure}

To verify that our multi-band modelling pipeline can reliably recover lens parameters under a high-dimensional configuration, we performed a test on a mock lens system constructed to mimic the properties of our PISCO sample. The mock system was generated using the fiducial lensing configuration adopted in this work: a single elliptical power-law (EPL) mass distribution with external shear, and independent elliptical Sérsic light profiles for the lens and source in each of the four bands. Realistic observational effects were applied, including band-dependent PSFs, noise levels, pixel scales, and inter-band shifts and rotations, chosen to resemble the characteristics of the PISCO imaging. The resulting stacked colour image is shown in Figure~\ref{fig:mock system}.

We ran the full pipeline on this mock system using both single-band and four-band reconstructions. Posterior distributions were obtained using \textsc{Zeus} with $10^5$ steps and a number of walkers set to approximately four times the number of free parameters. Convergence of all MCMC chains was confirmed via autocorrelation analyses. From this mock data test, we summarize the main conclusions as follows:

\begin{enumerate}
    \item \textbf{Parameter Recovery:} As shown in Figure~\ref{fig:MOCK posterior comparison}, the joint four-band reconstruction successfully recovers the input lens mass parameters of the mock system, while single-band fits generally yield broader constraints and, in some cases, mild offsets for individual parameters, reflecting increased degeneracies when fitting each band independently.  
    The mock lens was generated with the following true input parameters (red lines in Figure~\ref{fig:MOCK posterior comparison}):
    \begin{itemize}
        \item Einstein radius $\theta_E = 2.0^{\prime\prime}$
        \item Mass slope $\gamma = 2.2$
        \item Ellipticity components $e_1 = -0.046$, $e_2 = -0.06$
        \item External shear components $\gamma^{\rm ext}_1 = -0.03$, $\gamma^{\rm ext}_2 = 0.03$
    \end{itemize}
    For the multi-band reconstruction with $N=65$ free parameters (including 9 inter-band shift and rotation parameters), the inferred posterior distributions are consistent with the true input values, with the majority of parameters lying within the 68\% credible intervals and the remainder within the 95\% intervals. This test demonstrates that \textsc{Zeus} is able to robustly sample high-dimensional posterior distributions relevant to multi-band strong-lens modelling.

\begin{figure}
    \centering
    \includegraphics[width=1\linewidth]{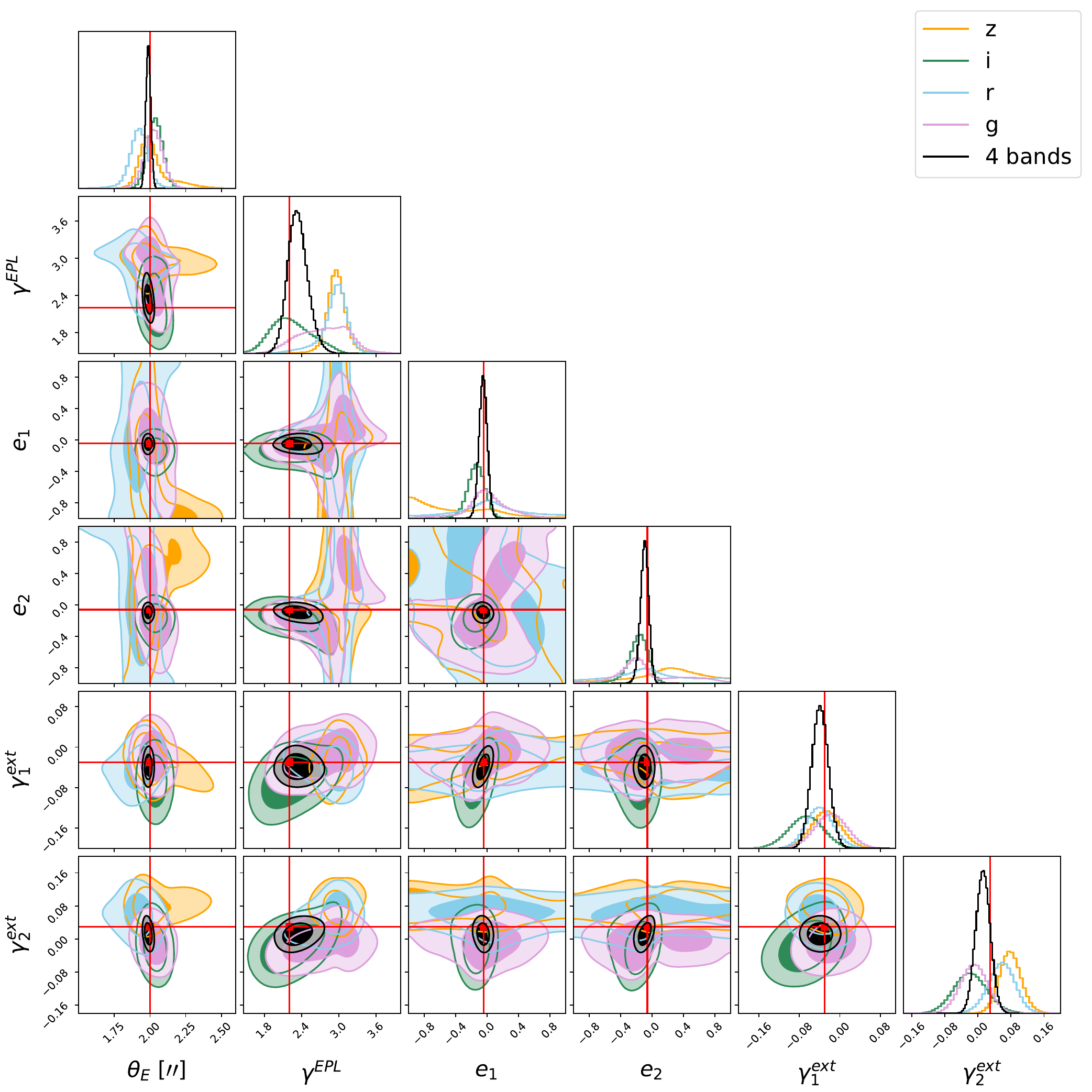}
    \caption{Posterior distributions of the main lens mass parameters for the mock system. Contours show the single-band reconstructions in $z$ (orange), $i$ (seagreen), $r$ (skyblue), and $g$ (plum) bands, and the joint four-band reconstruction in black for comparison. Vertical red lines indicate the true input values used to generate the mock lens system.}
    \label{fig:MOCK posterior comparison}
\end{figure}

    \item \textbf{Likelihood Comparison:} The log-likelihood of the single-band fits was slightly higher than that of the joint four-band fit, reflecting the compromise inherent in enforcing a single mass model across all bands. The lens mass parameters in the joint four-band reconstruction were more stable and physically consistent, with the maximum-likelihood values closely matching the true input parameters. In contrast, the z-band single-band fit exhibits unphysical elliptical lens mass, likely due to the dominance of lens light, highlighting the challenges of single-band modeling in lens-dominated bands for robust mass inference.

    \item \textbf{Computational Cost:} For this mock system, the four-band reconstruction involves substantially more free parameters ($N=65$ compared to $N=20$ for the single-band fits) and  correspondingly requires $\sim 4\times$ the computational time.
    Using the same parallel configuration (20 CPUs) on the Gadi high‑performance computing cluster at NCI Australia, the single-band fits required 10–14 hours of wall-clock time, whereas the four-band reconstruction required $\sim$55 hours.
    Note that for the PISCO samples in this study, the four-band reconstructions require approximately 2–4 times the total CPU usage compared to single-band fits.
\end{enumerate}

This test demonstrates that our pipeline can robustly recover lens parameters in high-dimensional multi-band models, yielding posterior distributions consistent with the true input values and improved constraints that justify the additional computational cost.


\bsp	
\label{lastpage}
\end{document}